\documentclass[oneside,english,reqno]{amsart}
\usepackage[T1]{fontenc}
\usepackage[latin9]{inputenc}
\usepackage{textcomp}
\usepackage{amsbsy}
\usepackage{amstext}
\usepackage{amsthm}
\usepackage{amssymb}
\usepackage{graphicx}
\usepackage{esint}
\usepackage[numbers,sort&compress]{natbib}

\makeatletter
\numberwithin{equation}{section}
\numberwithin{figure}{section}
\theoremstyle{plain}
\newtheorem{thm}{\protect\theoremname}
\theoremstyle{remark}
\newtheorem{rem}[thm]{\protect\remarkname}
\theoremstyle{definition}
\newtheorem{example}[thm]{\protect\examplename}

\@ifundefined{showcaptionsetup}{}{%
 \PassOptionsToPackage{caption=false}{subfig}}
\usepackage{subfig}
\makeatother

\usepackage{babel}
\providecommand{\examplename}{Example}
\providecommand{\remarkname}{Remark}
\providecommand{\theoremname}{Theorem}

\begin{document}
\title[Phase Transformation of Thermoelastic Fluids]{Continuum Thermodynamics of the Phase Transformation of Thermoelastic
Fluids}
\author{Gerard A. Ateshian and Jay J. Shim}
\address{Department of Mechanical Engineering, Columbia University, New York,
New York 10027}
\email{ateshian@columbia.edu}
\address{Department of Mechanical Engineering, Columbia University, New York,
New York 10027}
\curraddr{Siemens Digital Industries Software, Cypress, California}
\email{jjs2215@columbia.edu}
\begin{abstract}
This study uses continuum thermodynamics of pure thermoelastic fluids
to examine their phase transformation. To examine phase transformation
kinetics, a special emphasis is placed on the jump condition for the
axiom of entropy inequality, thereby recovering the conventional result
that stable phase equilibrium coincides with continuity of temperature,
pressure, and free enthalpy across the phase boundary. Moreover, this
jump condition leads to the formulation of a constitutive relation
for the phase transformation mass flux, $-\left[\left[\frac{1}{T}\mathbf{q}\right]\right]\cdot\mathbf{n}^{\Gamma}/\left[\left[s\right]\right]$,
where $\left[\left[\frac{1}{T}\mathbf{q}\right]\right]\cdot\mathbf{n}^{\Gamma}$
is the jump in entropy flux normal to, and across the phase interface
$\Gamma$, and $\left[\left[s\right]\right]$ is the corresponding
jump in entropy. This relation implies that phase transformations
must be accompanied by a jump in temperature across the phase boundary.
Encouraging agreement is found between this formula and limited available
experimental data. Further evidence is needed to conclusively validate
this proposed constitutive model. This continuum framework is well
suited for implementation in a computational framework, such as the
finite element method.
\end{abstract}

\thanks{This study was supported with funds from the National Institute of
General Medical Sciences of the National Institutes of Health (R01GM083925),
and the National Science Foundation, Division of Graduate Education
(Grant No. NSF GRFP DGE-16-44869). The content is solely the responsibility
of the authors and does not necessarily represent the official views
of the National Institutes of Health or the National Science Foundation.}

\maketitle
\global\long\def\tr{\operatorname{tr}}%
\global\long\def\grad{\operatorname{grad}}%
\global\long\def\Grad{\operatorname{Grad}}%
\global\long\def\divg{\operatorname{div}}%
\global\long\def\Divg{\operatorname{Div}}%
\global\long\def\curl{\operatorname{curl}}%
\global\long\def\lap{\operatorname{lap}}%

\section{Introduction}

The primary objective of this study is to formulate phase transformation
kinetics of pure thermoelastic fluids under general `non-equilibrium'
conditions. We present a complete approach for formulating constitutive
models for real gases and liquids, of the type used to populate thermodynamic
tables, using virial expansions. Our approach employs the specific
free energy as the main function of state that needs to be characterized
to completely describe the properties of a thermoelastic fluid. By
adopting an arbitrary reference pressure and reference temperature,
we obtain relatively simple expressions for the specific free energy
of real liquids and gases.

We place a special emphasis on understanding jump conditions across
interfaces, especially those involving a jump in temperature in the
absence or presence of a phase transformation. We show that the concepts
of specific enthalpy and specific free enthalpy emerge naturally from
those jump conditions, albeit using gauge instead of absolute pressure.
To examine phase transformations, we adopt the framework of reactive
mixture theory \citep{Truesdell60,Eringen65,Bowen68,Ateshian07} to
model a phase transformation as a reaction, and specialize it to conditions
prevailing across the phase boundary. We verify that our approach
reproduces the classical result that pressure, temperature and specific
free enthalpy are continuous across the phase boundary at phase equilibrium,
though we also find that continuity of temperature is only sufficient,
not necessary to satisfy the entropy inequality under these conditions.

Finally, we propose a constitutive relation for the reactive (e.g.,
evaporative) mass flux across a phase boundary which satisfies the
entropy inequality jump that regulates the feasibility of interfacial
processes. We find that this reactive mass flux for phase transformation
can only occur in the presence of a non-zero temperature jump and
a non-zero average heat flux across the phase boundary. By also satisfying
the mass, momentum and energy jumps, this constitutive relation produces
a unique solution for the jump in temperature during a phase transformation.
This novel constitutive relation is grounded entirely in the framework
of continuum mechanics, without appeal to the kinetic theory of gases
or statistical rate theory.

A purely continuum-based formulation was adopted here because it represents
a fundamental framework employed by engineers for analyzing thermofluid
processes. We recognize that the axiomatic presentation of conservation
laws for mass, momentum and energy, and the entropy inequality, are
not always favored in other fields of the physical sciences. However,
it is not our purpose here to present a universal, or molecular, or
statistical foundation for thermodynamics of phase transformations.
Instead, for those investigators who favor the continuum approach,
our goal with this study is to fill a missing element of this framework,
namely by providing a method to formulate and constrain a constitutive
model for the evaporation or condensation rates in a phase transformation.

\section{Kinematics and Notation Convention\label{sec:Kinematics-and-Notation}}

We describe the motion of a thermoelastic fluid using the standard
approach in continuum mechanics (for example, see Chapter 2 of \citep{Holzapfel00}).
Let $\mathbf{X}$ denote the position of a material particle in the
reference configuration and let $\mathbf{x}$ represent the spatial
position of this particle at the current time $t$. The motion of
the particle is given by the function $\boldsymbol{\chi}$, such that
$\mathbf{x}=\boldsymbol{\chi}\left(\mathbf{X},t\right)$.\footnote{In fluid mechanics, $\boldsymbol{\chi}\left(\mathbf{X},t\right)$
is commonly described as the equation of the pathline for the particle
starting at $\mathbf{X}$ at time $t=0$.} Any function $f$ associated with the material (such as its temperature
$T$, velocity $\mathbf{v}$, mass density $\rho$, etc.) may be expressed
mathematically in the material frame as $\check{f}\left(\mathbf{X},t\right)$
and in the spatial frame as $\tilde{f}\left(\mathbf{x},t\right)$,
such that $f=\check{f}\left(\mathbf{X},t\right)=\tilde{f}\left(\mathbf{x},t\right)$.
This identity remains valid component-wise when $f$ is a tensor of
any order, as long as $\mathbf{x}$ and $\mathbf{X}$ are expressed
in the same basis. The material time derivative of $f$ is denoted
by $\dot{f}$. It may be evaluated in either frame as
\begin{equation}
\dot{f}=\frac{\partial\check{f}}{\partial t}=\frac{\partial\tilde{f}}{\partial t}+\grad\tilde{f}\cdot\mathbf{v}\,,\label{eq:material-time-derivative}
\end{equation}
where $\grad\tilde{f}=\partial\tilde{f}/\partial\mathbf{x}$ and $\mathbf{v}=\partial\boldsymbol{\chi}/\partial t$
is the material's velocity. In the notation convention adopted here,
we forgo the explicit use of $\check{f}$ and $\tilde{f}$, relying
on $f$ to represent the function in either frame. Thus, $f\left(\mathbf{X},t\right)\equiv\check{f}\left(\mathbf{X},t\right)$
and $f\left(\mathbf{x},t\right)\equiv\tilde{f}\left(\mathbf{x},t\right)$.
For example, $\grad f$ is most conveniently evaluated as $\partial\tilde{f}/\partial\mathbf{x}$
in the spatial frame, but it may then be expressed in the material
frame by substituting the motion $\mathbf{x}=\boldsymbol{\chi}\left(\mathbf{X},t\right)$
into its expression.

The deformation gradient is given by $\mathbf{F}=\partial\boldsymbol{\chi}/\partial\mathbf{X}$.
Its determinant, $J=\det\mathbf{F}$, is the Jacobian of the transformation
from the material to the spatial frame; it represents the ratio of
an elemental volume $dV$ of the material in the current configuration
to the corresponding elemental volume $dV_{r}$ in its reference configuration.
The \emph{volumetric strain}, representing the relative change in
volume of $dV_{r}$, is given by $e=J-1$. Using the chain rule of
differentiation, the material time derivative of the deformation gradient
is $\dot{\mathbf{F}}=\mathbf{L}\cdot\mathbf{F}$ where $\mathbf{L}=\grad\mathbf{v}$.
Using this relation and the fact that $\partial J/\partial\mathbf{F}=J\mathbf{F}^{-T}$,
it follows that
\begin{equation}
\dot{J}=J\divg\mathbf{v}\,.\label{eq:kinematic-constraint}
\end{equation}
This fundamental kinematic constraint between $J$ and $\mathbf{v}$
is rarely used in fluid mechanics and thermodynamics, but we use it
throughout the formulation of our framework. Importantly, this identity
holds in both material and spatial frames.

\section{Single Phase\label{sec:Single-Phase}}

\subsection{Axioms of Conservation and Entropy Inequality\label{subsec:Axioms-of-Conservation}}

Before tackling phase transformations, we first consider the thermodynamics
of a single phase. The axioms of mass, momentum and energy balance
are typically formulated in integral form over a control volume, and
then reduced to differential form following the application of the
divergence theorem. This standard procedure can be found in textbooks
and we omit those details here. In particular, the background for
the material reviewed in this section can be found in the book by
Gurtin et al. \citep{Gurtin10} and in Chapter 4 of \citep{Holzapfel00}.

The differential statement of the axiom of mass balance for a single
constituent (a pure substance in a single phase) is given by
\begin{equation}
\dot{\rho}+\rho\divg\mathbf{v}=0\,,\label{eq:mass-balance}
\end{equation}
where $\rho$ is the mass density of the material (mass per volume
in the current configuration). Substituting $\divg\mathbf{v}=\dot{J}/J$
from the kinematic constraint (\ref{eq:kinematic-constraint}) into
(\ref{eq:mass-balance}) and multiplying across by $J$ produces the
differential statement $\dot{\overline{\rho J}}=0$. This equation
may be integrated to produce the algebraic solution to the equation
of mass balance,
\begin{equation}
\rho=\frac{\rho_{r}}{J}\,,\label{eq:integrated-mass-balance}
\end{equation}
where $\rho_{r}$ is a constant, representing the mass density of
the material in its reference configuration (when $J=1$). This exact
solution is valid in the material and spatial frames.

The relation (\ref{eq:integrated-mass-balance}) is essential for
our formulation of a thermodynamics framework where $J$ is a state
variable. Experimental measurements of the density of a fluid at various
temperatures and pressures are routinely tabulated. To evaluate $J$,
we choose an arbitrary reference configuration for which we know the
absolute temperature $T_{r}$, absolute pressure $P_{r}$, and mass
density $\rho_{r}$. At all other temperatures $T$ and pressures
$P$, we then use (\ref{eq:integrated-mass-balance}) to evaluate
$J=\rho_{r}/\rho$.

The axiom of linear momentum balance requires us to introduce the
Cauchy stress tensor $\boldsymbol{\sigma}$, which is conventionally
taken to be zero in the material's reference configuration. This concept
of setting the stress to zero in some chosen reference configuration
represents an essential element of the presentations of this study
and a deviation from prior continuum thermodynamic treatments, as
will become slowly evident in the developments that follow. An arbitrary
stress-free reference configuration is commonly assumed in solid mechanics,
motivated by the theoretical impossibility of observing residual stresses
in a material domain, even when this domain is seemingly under traction-free
conditions. In a strict sense, this theoretical concept applies to
fluids as well.

The differential statement of the axiom of momentum balance is given
by
\begin{equation}
\rho\dot{\mathbf{v}}=\divg\boldsymbol{\sigma}+\rho\mathbf{b}\,,\label{eq:momentum-balance}
\end{equation}
where $\dot{\mathbf{v}}$ is the acceleration and $\mathbf{b}$ represents
a user-specified specific body force (units of force per mass). For
non-polar media, the axiom of angular momentum balance requires the
stress tensor to be symmetric, $\boldsymbol{\sigma}^{T}=\boldsymbol{\sigma}$.

The differential statement of the axiom of energy balance is
\begin{equation}
\rho\dot{u}=\boldsymbol{\sigma}:\mathbf{D}-\divg\mathbf{q}+\rho r\,,\label{eq:energy-balance}
\end{equation}
where $u$ is the specific internal energy (with units of energy per
mass), $\mathbf{D}$ is the rate of deformation tensor (the symmetric
part of the velocity gradient $\grad\mathbf{v}$), $\mathbf{q}$ is
the heat flux (units of power per area) and $r$ is a user-specified
specific heat supply (units of power per mass) from sources not modeled
explicitly (such as radiative heating in a framework that does not
model electromagnetism, or Joule heating in a framework that does
not model relative motion between charged constituents and their frictional
interactions). The significance of this heat supply term is discussed
in greater detail below.

The functions of state that describe the behavior of specific materials
are the stress $\boldsymbol{\sigma}$, the specific internal energy
$u$, and the heat flux $\mathbf{q}$, which appear in the momentum
and energy balance equations. Experimentally-validated constitutive
relations need to be provided for these functions of state, subject
to constraints imposed by the axiom of entropy inequality \citep{Coleman63}.
The axiom of entropy inequality may be expressed as a differential
statement in the form of the Clausius-Duhem inequality \citep{Truesdell60},
\begin{equation}
\rho\dot{s}+\divg\frac{\mathbf{q}}{T}-\rho\frac{r}{T}\geqslant0\,,\label{eq:entropy-inequality}
\end{equation}
where the specific entropy $s$ represents another function of state.
Since $r$ is a user-specified parameter, we may eliminate it from
the entropy inequality by judiciously combining it with the energy
balance (\ref{eq:energy-balance}) to produce
\begin{equation}
-\rho\left(\dot{a}+s\dot{T}\right)+\boldsymbol{\sigma}:\mathbf{D}-\frac{1}{T}\mathbf{q}\cdot\grad T\geqslant0\,,\label{eq:Clausius-Duhem-inequality}
\end{equation}
where
\begin{equation}
a=u-Ts\label{eq:specific-free-energy}
\end{equation}
is the specific free energy. This important result shows that the
specific free energy is a function of state that emerges naturally
from the axioms of energy balance and entropy inequality. Since the
form of the entropy inequality in (\ref{eq:Clausius-Duhem-inequality})
is free of the user-specified heat supply $r$, it may be used to
place constraints on the functions of state $a$, $s$, $\boldsymbol{\sigma}$
and $\mathbf{q}$.
\begin{rem}
In the classical thermodynamics literature $a$ is named after Helmholtz,
to distinguish it from another function commonly associated with free
energy named after Gibbs. In the framework presented here, $a$ is
the sole measure of free energy. The function of state conventionally
named after Gibbs, which is also called the \emph{free enthalpy},
emerges naturally in a different context, presented further below.
\end{rem}

\subsection{State Variables and Thermodynamic Constraints\label{subsec:State-Variables-Constraints}}

To formulate constitutive relations that satisfy the entropy inequality,
we adapt the approach of Coleman and Noll \citep{Coleman63}. Fundamentally,
we first need to decide which materials and associated mechanisms
we would like to model, then base our choice of state variables on
those constraints. In contrast to the work of these authors, in our
approach we consider that state variables are exclusively observable
variables, derived from measurements of time and space.\footnote{In particular, Coleman and Noll \citep{Coleman63} selected the specific
entropy as a state variable and the temperature as a function of state.
Our approach does not allow this switch because entropy is not observable
whereas temperature can be measured.}

In our case, we would like to limit our choice of materials to thermoelastic
fluids. Thus, to account for the deformation of such materials, we
only need to include the volume ratio $J$ in our list of state variables,
instead of a more complete tensorial measure of strain. To account
for variations of functions of state with temperature, we include
$T$ as a state variable. Finally, to account for the flow of heat,
we also include the temperature gradient $\mathbf{g}=\grad T$. To
help clarify this approach, we note that the rate of deformation $\mathbf{D}$
is excluded from our list of state variables because we choose to
ignore fluid viscosity in our formulation of thermoelastic fluids.

Based on the principle of equipresence \citep{Truesdell60}, all functions
of state are initially assumed to depend on the selected list $\left(T,J,\mathbf{g}\right)$
of state variables. We evaluate the material time derivative of $a$
using the chain rule of differentiation, 
\begin{equation}
\dot{a}=\frac{\partial a}{\partial T}\dot{T}+\frac{\partial a}{\partial J}\dot{J}+\frac{\partial a}{\partial\mathbf{g}}\cdot\dot{\mathbf{g}}\,,\label{eq:adot-chain-rule}
\end{equation}
make use of the kinematic identity (\ref{eq:kinematic-constraint})
in the form $\dot{J}=J\mathbf{I}:\mathbf{D}$ as well as (\ref{eq:integrated-mass-balance}),
substitute the resulting expression into (\ref{eq:Clausius-Duhem-inequality})
and group terms together to produce
\begin{equation}
-\rho\left(\frac{\partial a}{\partial T}+s\right)\dot{T}+\left(\boldsymbol{\sigma}-\rho_{r}\frac{\partial a}{\partial J}\mathbf{I}\right):\mathbf{D}-\rho\frac{\partial a}{\partial\mathbf{g}}\cdot\dot{\mathbf{g}}-\frac{1}{T}\mathbf{q}\cdot\mathbf{g}\geqslant0\,.\label{eq:CD-elastic-fluid}
\end{equation}
In these expressions, $\mathbf{I}$ is the identity tensor. This inequality
must hold for arbitrary processes, which implies arbitrary changes
in the observable variables of state $\dot{T}$, $\mathbf{D}$, $\mathbf{g}$,
and $\dot{\mathbf{g}}$, under our self-imposed constraint that $a$,
$s$, $\boldsymbol{\sigma}$ and $\mathbf{q}$ cannot depend on $\dot{T}$,
$\mathbf{D}$ or $\dot{\mathbf{g}}$.\footnote{In our approach, the arbitrariness of processes is embodied in the
arbitrary variation of observable state variables. Since functions
of state describe the behavior of specific materials, we may not assume
that they can vary arbitrarily. In contrast, Müller \citep{Muller85}
proposed that state variables cannot be arbitrary as they must satisfy
the field equations, so he introduced the field equations into the
entropy inequality using Lagrange multipliers. We do not follow that
procedure here, on the basis that constraints placed on state variables
by the field equations result from the constrained functions of state
appearing in those equations. These alternative views are not contradictory,
as the final results are the same in both approaches.} For example, looking at the first term, which is the only one that
involves $\dot{T}$, we expect that this term must be positive regardless
of the algebraic sign of $\dot{T}$; however, since the coefficient
multiplying $\dot{T}$ is independent of it, this inequality can be
satisfied if and only if the coefficient is zero. Applying the same
reasoning to the terms involving $\mathbf{D}$ and $\dot{\mathbf{g}}$,
we conclude that 
\begin{equation}
s=-\frac{\partial a}{\partial T}\,,\label{eq:entropy-constraint}
\end{equation}
\begin{equation}
\boldsymbol{\sigma}=-p\mathbf{I}\,,\quad p\equiv-\rho_{r}\frac{\partial a}{\partial J}\,,\label{eq:thermodynamic-pressure}
\end{equation}
and
\begin{equation}
\frac{\partial a}{\partial\mathbf{g}}=\mathbf{0}\,.\label{eq:a-independent-g}
\end{equation}
The thermodynamic constraint (\ref{eq:thermodynamic-pressure}) indicates
that the state of stress in a thermoelastic fluid is hydrostatic,
represented by the pressure $p$. The remaining term in (\ref{eq:CD-elastic-fluid})
involves a function of $\mathbf{g}$, preventing us from simplifying
this expression further. Thus, we are left with the \emph{residual
dissipation statement},
\begin{equation}
\mathbf{q}\left(T,J,\mathbf{g}\right)\cdot\mathbf{g}\le0\,.\label{eq:residual-dissipation-compressible}
\end{equation}
Equations (\ref{eq:entropy-constraint}) and (\ref{eq:thermodynamic-pressure})
show that the entropy inequality imposes constraints on the specific
entropy and the stress, making them entirely dependent on the specific
free energy. These relations are sometimes described as basic thermodynamic
relations, but our approach shows that they are a direct result of
applying the entropy inequality to constrain the behavior of a specific
material model (a thermoelastic fluid here).

Equation (\ref{eq:a-independent-g}) indicates that, contrary to our
\emph{a priori} assumption, the free energy cannot depend on the temperature
gradient; it follows from Eqs.(\ref{eq:specific-free-energy}) and
(\ref{eq:entropy-constraint}) that $u$, $s$ and $p$ must also
be independent of those state variables,
\begin{equation}
a=a\left(T,J\right)\,,\quad u=u\left(T,J\right)\,,\quad s=s\left(T,J\right)\,,\quad p=p\left(T,J\right).\label{eq:energy-entropy-state-variables}
\end{equation}

The residual dissipation statement in (\ref{eq:residual-dissipation-compressible})
provides necessary and sufficient constraints on the constitutive
models we may adopt for the heat flux $\mathbf{q}$. Thus, any constitutive
model for $\mathbf{q}$ must satisfy (\ref{eq:residual-dissipation-compressible}),
as will be illustrated below. As we shall also discuss in greater
detail, terms that remain in the residual dissipation represent the
irreversible processes for the material being modeled, whereas terms
from the entropy inequality that reduce to zero (such as the coefficients
of $\dot{T}$, $\mathbf{D}$ and $\dot{\mathbf{g}}$) may be associated
with reversible processes.

The relations obtained in this section are classical relations of
continuum thermodynamics. In particular, the derivation of the relations
of (\ref{eq:entropy-constraint})-(\ref{eq:thermodynamic-pressure})
can be found in \citep{Truesdell84}. The residual dissipation statement
of (\ref{eq:residual-dissipation-compressible}) can be found in \citep{Coleman63}.

\subsection{Implications for Energy Balance\label{subsec:Implications-Energy-Balance}}

Given the constraint on $s$ in (\ref{eq:entropy-constraint}) and
the relation between $a$ and $u$ in Eq.(\ref{eq:specific-free-energy}),
we find that the specific internal energy may be derived entirely
from the specific free energy,
\begin{equation}
u=a-T\frac{\partial a}{\partial T}\,,\label{eq:internal-energy-constraint}
\end{equation}
where $u$ is only a function of $\left(T,J\right)$. Since the material
time derivative of $u$ appears in the energy balance (\ref{eq:energy-balance}),
we may use the chain rule to expand it as
\begin{equation}
\dot{u}=\frac{\partial u}{\partial T}\dot{T}+\frac{\partial u}{\partial J}\dot{J}\,.\label{eq:sie-chain-rule}
\end{equation}
The coefficient of $\dot{T}$ in this expression is denoted by $c_{v}$,
where
\begin{equation}
c_{v}\left(T,J\right)\equiv\frac{\partial u}{\partial T}=T\frac{\partial s}{\partial T}=-T\frac{\partial^{2}a}{\partial T^{2}}\label{eq:specific-heat-v}
\end{equation}
is conventionally defined as the \emph{isochoric specific heat capacity}
\citep{Haase69}.

Similarly, we may evaluate $\partial u/\partial J$ from (\ref{eq:internal-energy-constraint})
while also employing (\ref{eq:thermodynamic-pressure}),
\begin{equation}
\frac{\partial u}{\partial J}=\frac{\partial a}{\partial J}-T\frac{\partial}{\partial T}\left(\frac{\partial a}{\partial J}\right)=-\frac{p}{\rho_{r}}+\frac{T}{\rho_{r}}\frac{\partial p}{\partial T}\,.\label{eq:du-dJ}
\end{equation}
We may substitute (\ref{eq:specific-heat-v}) and (\ref{eq:du-dJ})
into (\ref{eq:sie-chain-rule}) and the resulting expression into
the energy balance (\ref{eq:energy-balance}) to produce the axiom
of energy balance for a thermoelastic fluid,
\begin{equation}
\rho c_{v}\dot{T}=-T\frac{\partial p}{\partial T}\frac{\dot{J}}{J}-\divg\mathbf{q}+\rho r\,.\label{eq:energy-balance-fluid}
\end{equation}
It becomes apparent from this relation that $-T\frac{\partial p}{\partial T}\frac{\dot{J}}{J}$
is the heat supply density resulting from the rate of change of the
fluid volume ratio, whereas $-\divg\mathbf{q}$ is the heat supply
density resulting from a converging heat flux.

Yet another way to rearrange the energy balance is to express $\dot{u}=\dot{a}+\dot{T}s+T\dot{s}$
based on (\ref{eq:specific-free-energy}), then use (\ref{eq:entropy-constraint})-(\ref{eq:a-independent-g})
to find $\dot{a}=-s\dot{T}-p\dot{J}/\rho_{r}$, so that $\dot{u}=T\dot{s}-p\dot{J}/\rho_{r}$.
Substituting this result into (\ref{eq:energy-balance}), and making
use of (\ref{eq:kinematic-constraint}) produces an alternative form
of the energy balance in terms of the material time derivative of
the specific entropy,
\begin{equation}
\dot{s}=\frac{1}{T}\left(-\frac{1}{\rho}\divg\mathbf{q}+r\right)\,.\label{eq:energy-balance-entropy}
\end{equation}
This expression shows that the entropy changes over time if, and only
if, heat is supplied from a user-specified source ($r\ne0$), or provided
in the form of a diverging heat flux ($\divg\mathbf{q}\ne\mathbf{0}$),
such that neither combines with the other to produce $\dot{s}=0$.

As a final note on this subtopic, if we now subtract (\ref{eq:energy-balance-entropy})
from the entropy inequality (\ref{eq:entropy-inequality}), we recover
the residual dissipation inequality in (\ref{eq:residual-dissipation-compressible}).
This outcome emphasizes that the mechanisms responsible for irreversible
processes are well defined, given a set of constitutive assumptions
embodied in the choice of state variables.

The relations presented in this section are also common in the continuum
mechanics literature. For example, the form of equations (\ref{eq:energy-balance-fluid})-(\ref{eq:energy-balance-entropy})
is consistent with those presented in section 57.5 of \citep{Gurtin10}.

\subsection{Heat Conduction\label{sec:Heat-Conduction}}

To satisfy the constraint on $\mathbf{q}$ in (\ref{eq:residual-dissipation-compressible})
unconditionally for any $\mathbf{g}$, we recognize that $\mathbf{q}$
must be proportional to the temperature gradient $\mathbf{g}$,
\begin{equation}
\mathbf{q}\left(T,J,\mathbf{g}\right)=-k\left(T,J,\mathbf{g}\right)\mathbf{g}\label{eq:Fourier's-Law}
\end{equation}
where $k$ is a scalar function of state (since fluids are isotropic)
known as the \emph{thermal conductivity}. Now, the residual dissipation
(\ref{eq:residual-dissipation-compressible}), which reduces to $k\mathbf{g}\cdot\mathbf{g}\ge0$,
is satisfied if and only if $k$ is positive for all $\left(T,J,\mathbf{g}\right)$.
The constitutive relation in (\ref{eq:Fourier's-Law}) is a generalization
of \emph{Fourier's law} of heat conduction; in most materials and
processes, it is observed experimentally that $k$ is negligibly dependent
on the temperature gradient $\mathbf{g}$.

Substituting Fourier's law into the energy balance (\ref{eq:energy-balance-fluid})
produces
\begin{equation}
\rho c_{v}\dot{T}=-T\frac{\partial p}{\partial T}\frac{\dot{J}}{J}+\divg\left(k\grad T\right)+\rho r\,.\label{eq:energy-balance-Fourier}
\end{equation}
This equation is generally called the \emph{heat equation}. This is
a standard procedure for finding a constitutive model for $\mathbf{q}$
that satisfies the residual dissipation statement \citep{Coleman63}.

\subsection{Reversible and Irreversible Processes\label{sec:Irreversible-and-Reversible}}

We found two types of constraints resulting from the Clausius-Duhem
inequality: Constraints that reduce the corresponding terms to zero,
such as those in (\ref{eq:entropy-constraint})-(\ref{eq:a-independent-g}),
and constraints that persist in the residual dissipation statement,
as in (\ref{eq:residual-dissipation-compressible}). We deduced that
the processes associated with terms that vanish from the residual
dissipation are \emph{reversible}, since they produce no dissipation,
whereas processes that persist in the residual dissipation are \emph{irreversible}.
In the derivations presented above it becomes apparent that processes
which only alter the temperature ($\dot{T}\ne0$), or the volume ($\dot{J}\ne0$)
while maintaining zero dissipation, $-\mathbf{q}\cdot\mathbf{g}/T=0$,
are reversible processes. Therefore, a necessary and sufficient condition
for reversibility in this material model is to satisfy $\mathbf{q}\cdot\mathbf{g}=0$.
In real materials, where $\mathbf{q}$ must be proportional to $\mathbf{g}$
with $k>0$, this reversibility condition is satisfied if and only
if $\mathbf{g}=\mathbf{0}$ (uniform temperature throughout a process).
Therefore, any process that generates a temperature gradient ($\mathbf{g}\ne\mathbf{0}$)
is an irreversible process. (For materials idealized as perfect heat
insulators, the thermal conductivity is $k=0$, in which case $\mathbf{q}=\mathbf{0}$
even if $\mathbf{g}\ne\mathbf{0}$; however, no real materials exist
for which $k=0$.)

In fluid mechanics and thermodynamics we often consider \emph{isentropic
flows} or \emph{isentropic processes}; these are processes that keep
the entropy uniform in space and constant in time, thus $\dot{s}=0$.
Examining the energy balance given in the form of (\ref{eq:energy-balance-entropy}),
we note that an isentropic process must satisfy $\rho r-\divg\mathbf{q}=0$.
A sufficient condition is to satisfy $\mathbf{q}=\mathbf{0}$ and
$r=0$, which represents an \emph{adiabatic process}. Therefore, adiabatic
processes in a single constituent material are always isentropic and
reversible (since $\mathbf{q}=\mathbf{0}$). However, processes may
exist where $\mathbf{q}\ne\mathbf{0}$ (irreversible) even though
$\rho r-\divg\mathbf{q}=0$ and thus $\dot{s}=0$ (such as one-dimensional
steady-state heat conduction). Therefore, isentropic processes are
not necessarily reversible processes.

In introductory engineering thermodynamics courses it is generally
assumed that the temperature is uniform within each domain under consideration
in a process, thus $\mathbf{g}=\mathbf{0}$ implying $\mathbf{q}=\mathbf{0}$
according to (\ref{eq:Fourier's-Law}). The only mechanism by which
heat may be exchanged in such isothermal processes is via a non-zero
user-specified heat supply $r$. In that case (\ref{eq:energy-balance-entropy})
reduces to $\dot{s}=r/T$. Hence, an isentropic process ($\dot{s}=0$)
must be adiabatic ($r=0$) and reversible (since $\mathbf{q}=\mathbf{0}$),
in which case these terms become interchangeable (isentropic$\Leftrightarrow$adiabatic$\Leftrightarrow$reversible).
The derivations provided here show that this is only a special, idealized
case. Generally, isentropic processes are not always reversible.

Recall that we eliminated the heat supply $r$ from the entropy inequality
(\ref{eq:entropy-inequality}) by combining it with the energy balance
(\ref{eq:energy-balance}). As a result, $r$ does not appear in the
residual dissipation statement (\ref{eq:entropy-jump-latent-1}) and
we may wonder whether this user-specified heat supply term is dissipative
or not. This apparent ambiguity arises from the fact that we introduced
$r$ to simulate various potential sources of heat supplies, such
as microwave heating, Joule heating, exothermic or endothermic chemical
reactions, etc., without explicitly accounting for the mechanisms
that give rise to these phenomena. In reality, all these illustrative
phenomena are dissipative, as shown for reactive processes as well
as frictional interactions between electrically neutral or charged
species in \citep{Ateshian07}. Therefore, the absence of $r$ in
the residual dissipation statement for a single constituent (here,
a thermoelastic fluid) represents a simplifying idealization of actual
dissipative processes.
\begin{example}
\label{exa:viscous-dissipation}So far we have idealized fluids to
be inviscid. In this example we briefly review how fluid viscosity
contributes to the residual dissipation. Since viscous stresses are
generated when a fluid is sheared, we need to introduce the rate of
deformation tensor, $\mathbf{D}$, as an additional state variable
in our list, $\left(T,J,\mathbf{g},\mathbf{D}\right)$. When analyzing
this slightly more general framework using the entropy inequality,
we obtain the same constraints as in (\ref{eq:entropy-constraint})-(\ref{eq:a-independent-g}),
supplemented by $\partial a/\partial\mathbf{D}=\mathbf{0}$, implying
that neither $a$, nor any of its related functions of state ($p$,
$u$ and $s$) can depend on $\mathbf{D}$. In this case, the residual
dissipation includes an additional term,
\begin{equation}
\boldsymbol{\tau}:\mathbf{D}-\frac{1}{T}\mathbf{q}\cdot\mathbf{g}\ge0\,,\label{eq:CD-inequality-viscous}
\end{equation}
where $\boldsymbol{\tau}\equiv\boldsymbol{\sigma}+p\mathbf{I}$, and
$\boldsymbol{\tau}:\mathbf{D}$ describes the heat supply density
generated in the fluid due to viscous dissipation. Since this term
does not vanish from the residual dissipation, we conclude that viscous
dissipation is an irreversible process. Furthermore, when including
viscous stresses, the energy balance equation presented in (\ref{eq:energy-balance-fluid})
has an additional term,
\begin{equation}
\rho c_{v}\dot{T}=-T\frac{\partial p}{\partial T}\frac{\dot{J}}{J}-\divg\mathbf{q}+\rho r+\boldsymbol{\tau}:\mathbf{D}\,,\label{eq:energy-balance-viscous}
\end{equation}
which clearly shows that $\boldsymbol{\tau}:\mathbf{D}$ is a heat
supply density analogous to $\rho r$. This example emphasizes that
$r$ is a generic placeholder for processes that have not been modeled
explicitly via the adoption of suitable state variables (such as neglecting
viscosity in our thermoelastic fluid model). Similarly, the common
assumption that $\mathbf{g}=\mathbf{0}$ in introductory thermodynamic
textbooks implies that the heat flux $\mathbf{q}$ may not be modeled
explicitly in that classical framework. In that case, the only mechanism
by which heat exchanges may be included in the energy balance is to
let $\rho r$ serve as a placeholder for a converging heat flux $-\divg\mathbf{q}$.
\end{example}

The concepts presented and illustrated in this section demonstrate
that the framework of undergraduate engineering thermodynamics textbooks
is fundamentally suited for reversible processes only ($\mathbf{q}=\mathbf{0}$
and $\boldsymbol{\tau}=\mathbf{0}$), as recognized explicitly in
those textbooks (for example, see section 5.3.1 of \citep{Moran88}).

\subsection{Constitutive Relations for the Free Energy\label{subsec:Constitutive-Relations}}

The constraints placed by the entropy inequaliy on the functions of
state demonstrate that the pressure $p$, specific entropy $s$, and
internal energy $u$ may all be derived from the specific free energy
$a$. Thus, the formulation of constitutive relations for $a$ represents
an essential foundation for modeling the thermodynamics of real and
idealized fluids. In our approach we use $J$ as a kinematic state
variable, instead of the more commonly used specific volume $\upsilon=1/\rho$,
and the gauge pressure $p$ as a function of state, instead of the
absolute pressure $P$. We start with ideal gases, as they represent
a canonical problem, then work our way through real gases using the
concept of virial expansions \citep{Wagner02}. We then consider real
liquids, with special consideration of their phase transition curve.
As commonly done in thermodynamics, we work from the assumption that
experimental static measurements exist for the fluid pressure $P$
at various temperatures $T$ and densities $\rho$. We also assume
that the isochoric specific heat capacity $c_{v}$ has been characterized
experimentally over a range of conditions, as specified below \citep{Wagner02}.

\subsubsection{Ideal Gases\label{subsec:Ideal-Gases}}

Ideal gases represent a special case of inviscid compressible fluids.
The absolute pressure in ideal gases may be related to temperature
and density via $P=RT\rho/M$, where $R$ is the universal gas constant
and $M$ is the gas molar mass. The deviation from ideal behavior
occurs at low temperatures and high pressures. From experimental observations,
$c_{v}$ in ideal gases is only a function of temperature; conventionally
its value for ideal gases is represented by $c_{v0}\left(T\right)$.
We now use these relations, in addition to the integrated form (\ref{eq:integrated-mass-balance})
of the mass balance equation, to recover the functions of state $p$
and $a$.

The gauge pressure $p=P-P_{r}$ is taken to be zero in the reference
state, when $T=T_{r}$ and $J=1$, thus $\rho=\rho_{r}$. The reference
pressure is $P_{r}=RT_{r}\rho_{r}/M$ and the gauge pressure $p$
of an ideal gas has the form
\begin{equation}
p\left(T,J\right)=\frac{RT_{r}}{M}\rho_{r}\left(\frac{T}{JT_{r}}-1\right)\,.\label{eq:IG-elastic-pressure}
\end{equation}
Substituting this expression into the thermodynamic constraint of
(\ref{eq:thermodynamic-pressure}) and integrating with respect to
$J$ produces
\begin{equation}
a\left(T,J\right)=-\frac{RT_{r}}{M}\left[\frac{T}{T_{r}}\left(1-\ln\frac{T}{JT_{r}}\right)-J\right]+a_{0}\left(T\right)\,,\label{eq:IG-a-a0}
\end{equation}
where $a_{0}\left(T\right)$ is an integration function, carefully
chosen to produce $a=a_{0}$ when $p=0$ (i.e., when $J=T/T_{r}$
according to (\ref{eq:IG-elastic-pressure})). Then, $a-a_{0}$ represents
the part of the free energy that varies with pressure. We substitute
this expression into (\ref{eq:specific-heat-v}) to find that
\begin{equation}
c_{p0}\left(T\right)=-Ta_{0}^{\prime\prime}\left(T\right)\,,\label{eq:IG-cp}
\end{equation}
where
\begin{equation}
c_{p0}\left(T\right)\equiv c_{v0}\left(T\right)+\frac{R}{M}\label{eq:specific-heat-p}
\end{equation}
is conventionally called the \emph{isobaric specific heat capacity}
\citep{Moran88}, since it is evaluated from that part of the specific
free energy, $a_{0}\left(T\right)$, which does not vary with pressure.
Since $c_{v0}$ and $c_{p0}$ are only functions of temperature for
an ideal gas, we may integrate this expression once, subject to the
condition $a_{0}^{\prime}\left(T_{r}\right)=-s_{r}$, to produce
\begin{equation}
a_{0}^{\prime}\left(T\right)=-s_{r}-s^{\circ}\left(T\right)\,,\quad s^{\circ}\left(T\right)\equiv\int_{T_{r}}^{T}\frac{c_{p0}\left(\vartheta\right)}{\vartheta}d\vartheta\,,\label{eq:IG-so-def}
\end{equation}
where $s_{r}$ is the entropy in the reference configuration. The
function $s^{\circ}\left(T\right)$, which satisfies $s^{\circ}\left(T_{r}\right)=0$,
is often tabulated. Integrating this expression a second time, subject
to the condition $a_{0}\left(T_{r}\right)=a_{r}$ where $a_{r}$ is
the specific free energy in the reference configuration, produces
\begin{equation}
a_{0}\left(T\right)=a_{r}+a^{\circ}\left(T\right)-s_{r}\left(T-T_{r}\right)\,,\quad a^{\circ}\left(T\right)\equiv-\int_{T_{r}}^{T}s^{\circ}\left(\vartheta\right)\,d\vartheta\,,\label{eq:IG-psi0}
\end{equation}
where the function $a^{\circ}\left(T\right)$ satisfies $a^{\circ}\left(T_{r}\right)=0$.
Therefore, the final expression for the specific free energy of an
ideal gas is
\begin{equation}
a\left(T,J\right)=\frac{R}{M}\left(JT_{r}-T+T\ln\frac{T}{JT_{r}}\right)+a^{\circ}\left(T\right)+a_{r}-s_{r}\left(T-T_{r}\right)\,.\label{eq:IG-free-energy}
\end{equation}
This result shows that a complete characterization of the specific
free energy for a thermoelastic fluid requires experimental measurements
of $p\left(T,J\right)$ as well as an experimental estimation of $c_{v0}\left(T\right)$.
(In practice, it is common for $c_{v0}\left(T\right)$ to be calculated
using statistical thermodynamics.) It also shows that, in a continuum
framework, the functions of state are given within arbitrary reference
values $a_{r}$ and $s_{r}$.

This expression for $a\left(T,J\right)$ may be substituted into (\ref{eq:entropy-constraint})
to produce the specific entropy
\begin{equation}
s\left(T,J\right)=-\frac{R}{M}\ln\frac{T}{JT_{r}}+s^{\circ}\left(T\right)+s_{r}\,,\label{eq:IG-entropy}
\end{equation}
and into (\ref{eq:internal-energy-constraint}) to produce the specific
internal energy,
\begin{equation}
u\left(T,J\right)=\frac{R}{M}\left(JT_{r}-T\right)+u^{\circ}\left(T\right)+u_{r}\,,\label{eq:IG-internal-energy}
\end{equation}
where $u^{\circ}\left(T\right)\equiv a^{\circ}\left(T\right)+Ts^{\circ}\left(T\right)$
and $u_{r}\equiv a_{r}+T_{r}s_{r}$. We may use these relations to
evaluate the specific gauge enthalpy $h$ of an ideal gas as
\begin{equation}
h\left(T,J\right)\equiv u+\frac{p}{\rho}=u^{\circ}\left(T\right)+u_{r}\,.\label{eq:IG-enthalpy}
\end{equation}
Notably the gauge enthalpy is independent of the volume ratio $J$,
as it only varies with temperature. Importantly, the enthalpy evaluated
here uses the gauge pressure, not the absolute pressure.

The relation of (\ref{eq:IG-free-energy}) for the specific free energy
of an ideal gas is not given in undergraduate engineering thermodynamics
textbooks, despite its simplicity. In fact, the concept of free energy
is not typically mentioned in textbooks for mechanical engineering
students, as can be deduced from the absence of this term from the
subject index of popular coursebooks \citep{Van-Wylen78,Moran88}.
Moreover, this relation cannot be found in compressible flow textbooks,
such as \citep{Liepmann57,Schreier82}, nor in edited books on continuum
thermodynamics \citep{Domingos73,Truesdell84}. It is also not given
in papers that report methods for formulating thermodynamic properties
of water \citep{Wagner02}. The closest analogy to this relation can
be found in the book by Müller \citep{Muller85}, who derived an expression
in a similar manner, but assumed that $p$ is the absolute pressure;
the resulting expression (equation 6.71 in that book) is necessarily
not the same as (\ref{eq:IG-free-energy}) above. Indeed, based on
our review of the classical literature, we believe that the relation
presented in (\ref{eq:IG-free-energy}) is original.
\begin{example}
\label{exa:IG-constant-cv}We may specialize ideal gas relations to
the case where $c_{v0}$ and $c_{p0}$ are constants. Then it can
be shown that the specific free energy is
\begin{equation}
\begin{aligned}a\left(T,J\right) & =\frac{RT_{r}}{M}\left(J-1-\frac{T}{T_{r}}\ln J\right)-c_{v0}T\ln\frac{T}{T_{r}}\\
 & +\left(c_{v0}-s_{r}\right)\left(T-T_{r}\right)+a_{r}
\end{aligned}
\,,\label{eq:IG-constC-a}
\end{equation}
the specific entropy is
\begin{equation}
s\left(T,J\right)=\frac{R}{M}\ln J+c_{v0}\ln\frac{T}{T_{r}}+s_{r}\,,\label{eq:IG-constC-s}
\end{equation}
the specific internal energy is
\begin{equation}
u\left(T,J\right)=\frac{RT_{r}}{M}\left(J-1\right)+c_{v0}\left(T-T_{r}\right)+u_{r}\,,\label{eq:IG-constC-u}
\end{equation}
and the specific gauge enthalpy is
\begin{equation}
h\left(T,J\right)=c_{p0}\left(T-T_{r}\right)+u_{r}\,.\label{eq:IG-constC-h}
\end{equation}
\end{example}

While equation (\ref{eq:IG-constC-a}) is not given elsewhere, the
relation of (\ref{eq:IG-constC-h}) for the enthalpy, and the relation
of (\ref{eq:IG-constC-u}) for the internal energy, evaluated in the
special case when $T_{r}=0$, reproduce classical textbook relations
\citep{Moran88}. It is also easy to show that setting $\dot{s}=0$
in (\ref{eq:IG-entropy}) and (\ref{eq:IG-constC-s}) can reproduce
classical relations between temperature and volume for isentropic
processes in ideal gases \citep{Van-Wylen78,Moran88}.

\subsubsection{Real Gases\label{sec:Real-Gases}}

The behavior of a real gas can be modeled by assuming that its absolute
pressure $P$ relates to the pressure $P^{\text{IG}}$ of an ideal
gas according to
\begin{equation}
P\left(T,J\right)=Z\left(T,J\right)P^{\text{IG}}\left(T,J\right)\,,\label{eq:compressibility-factor-def}
\end{equation}
where $Z\left(T,J\right)$ is known as the \emph{compressibility factor},
which measures the deviation of the pressure response from ideal gas
behavior ($Z=1$). The compressibility factor may be easily evaluated
for any state from the measurement of pressure and its calculation
from the ideal gas relation. We now propose that the behavior of real
gases can be modeled as a function of $\left(T,J\right)$ using the
virial expansion
\begin{equation}
p\left(T,J\right)=P_{r}\sum_{k=1}^{m}A_{k}\left(T\right)\left(\frac{T}{JT_{r}}-1\right)^{k}\,,\label{eq:rg-gage-pressure}
\end{equation}
where $P_{r}$ is the referential pressure, and $A_{k}\left(T\right)$
are virial coefficients (all unitless) and the expansion is truncated
after $m$ terms.

Using the thermodynamic constraint between $p$ and $a$ in (\ref{eq:thermodynamic-pressure}),
we may integrate the gauge pressure in (\ref{eq:rg-gage-pressure})
once with respect to $J$ to yield
\begin{equation}
\begin{aligned}a\left(T,J\right) & =\frac{P_{r}}{\rho}A_{1}\left(T\right)\left[\frac{T}{JT_{r}}\left(\ln\frac{T}{JT_{r}}-1\right)+1\right]\\
 & +\frac{P_{r}}{\rho}A_{2}\left(T\right)\left(\frac{T}{JT_{r}}\left(\frac{T}{JT_{r}}-2\ln\frac{T}{JT_{r}}\right)-1\right)\\
 & +\frac{P_{r}}{\rho}A_{3}\left(T\right)\left(\frac{1}{2}\frac{T}{JT_{r}}\left(\frac{T^{2}}{J^{2}T_{r}^{2}}-6\frac{T}{JT_{r}}+6\ln\frac{T}{JT_{r}}+3\right)+1\right)\\
 & +a_{0}\left(T\right)+\cdots
\end{aligned}
\label{eq:rg-a-final}
\end{equation}
for $m=3$ as an illustration, such that $a=a_{0}\left(T\right)$
when $p=0$ ($J=T/T_{r}$). Here again, $a-a_{0}$ represents the
part of the specific strain energy that varies with pressure. For
an ideal gas we set $A_{1}\left(T\right)=1$ and $A_{k}\left(T\right)=0$
for $k=2$ to $m$ to recover the expression in (\ref{eq:IG-a-a0}).

We may then evaluate $c_{v}\left(T,J\right)=-T\partial^{2}a/\partial T^{2}$
based on (\ref{eq:specific-heat-v}), which is a cumbersome expression,
and simplify the resulting expression to the case when $p=0$ to find
the isochoric specific heat capacity at the reference pressure $P_{r}$,
\begin{equation}
c_{vr}\left(T\right)\equiv c_{v}\left(T,\frac{T}{T_{r}}\right)=-\frac{P_{r}}{\rho_{r}T_{r}}A_{1}\left(T\right)-Ta_{0}^{\prime\prime}\left(T\right)\,.\label{eq:rg-cvr}
\end{equation}
This formula is valid for any number of virial coefficients $m$.
We may evaluate $a_{0}\left(T\right)$ by integrating this expression
twice with respect to temperature, just as we did for ideal gases.
As a side note, this relation shows why $c_{p0}$ as defined in (\ref{eq:specific-heat-p})
is not valid for real gases. Indeed, for real gases we may use (\ref{eq:rg-cvr})
to define $c_{pr}$ at the reference pressure $P_{r}$ as
\begin{equation}
c_{pr}\left(T\right)\equiv c_{vr}\left(T\right)+Z_{r}A_{1}\left(T\right)\frac{R}{M}=-Ta_{0}^{\prime\prime}\left(T\right)\,,\label{eq:rg-cp}
\end{equation}
where $Z_{r}R/M=P_{r}/\rho_{r}T_{r}$ is the compressibility factor
in the reference state. Thus, the factor $Z_{r}A_{1}\left(T\right)$
influences the relation between $c_{pr}$ and $c_{vr}$.

The relation (\ref{eq:rg-cp}) shows that $a_{0}\left(T\right)$ may
be evaluated from experimental measurements of $c_{pr}\left(T\right)$
at the reference pressure $P_{r}$ and by integrating the resulting
expression with respect to $T$ once,
\begin{equation}
s_{0}\left(T\right)=-a_{0}^{\prime}\left(T\right)=s^{\circ}\left(T\right)+s_{r}\,,\quad s^{\circ}\left(T\right)=\int_{T_{r}}^{T}\frac{c_{pr}\left(\tau\right)}{\tau}\,d\tau\,,\label{eq:rg-s-circle}
\end{equation}
and then again to get
\begin{equation}
a_{0}\left(T\right)=a^{\circ}\left(T\right)+a_{r}-s_{r}\left(T-T_{r}\right)\,,\quad a^{\circ}\left(T\right)=-\int_{T_{r}}^{T}s^{\circ}\left(\tau\right)\,d\tau\,.\label{eq:rg-a-circle}
\end{equation}
Here again, the referential entropy $s_{r}=-a_{0}^{\prime}\left(T_{r}\right)$
and free energy $a_{r}=a_{0}\left(T_{r}\right)$ are arbitrary constants.
Once we have obtained the complete solution for $a\left(T,J\right)$,
we may differentiate it suitably to evaluate $s$ from (\ref{eq:entropy-constraint}),
$u$ from (\ref{eq:specific-free-energy}), and $h=u+p/\rho$. We
present an illustration of this type of constitutive relation for
water vapor in Example~\ref{exa:Vapor-water-properties} below, after
our presentation of phase transformations.

As for the case of ideal gases, we believe that the expression of
(\ref{eq:rg-a-final}) for the specific free energy of a real gas
is original, as we have not been able to find an equivalent expression
in the prior literature.

\subsubsection{Liquids\label{sec:Liquids}}

We propose that the gauge pressure in a liquid be given by the virial
expansion
\begin{equation}
p\left(T,J\right)=p_{\sigma}\left(T\right)+P_{r}\sum_{k=1}^{n}B_{k}\left(T\right)\left(J-J_{\sigma}\left(T\right)\right)^{k}\,,\label{eq:liquid-pressure}
\end{equation}
where $p_{\sigma}\left(T\right)$ is the gauge pressure, and $J_{\sigma}\left(T\right)$
is the volume ratio, of the liquid on the liquid-vapor saturation
curve.\footnote{Numerically, it may be more convenient to evaluate the volumetric
strain on the saturation curve, $e_{\sigma}\left(T\right)=J_{\sigma}\left(T\right)-1$,
and replace $J-J_{\sigma}\left(T\right)$ with $e-e_{\sigma}\left(T\right)$
in (\ref{eq:liquid-pressure}), where $e=J-1$.} Here, the virial coefficients $B_{k}\left(T\right)$ are all unitless.
These functions are constructed such that $p_{\sigma}\left(T_{r}\right)=0$
and $J_{\sigma}\left(T_{r}\right)=1$, in order to satisfy $p\left(T_{r},1\right)=0$
in the reference configuration. The relation of (\ref{eq:liquid-pressure})
also satisfies $p\left(T,J_{\sigma}\left(T\right)\right)=p_{\sigma}\left(T\right)$,
implying that we automatically recover the temperature dependence
of the pressure on the saturation curve. The most obvious choice of
a reference configuration for a pure liquid substance is its triple
point, since it represents the lowest pressure and temperature at
which the liquid phase can exist in stable form.\footnote{In the absence of prior knowledge of the saturation curve, the constitutive
model in (\ref{eq:liquid-pressure}) may be substituted with $p\left(T,J\right)/P_{r}=C_{0}\left(T\right)+\sum_{k=1}^{n}C_{k}\left(T\right)\left(J-1\right)^{k}$,
where $C$'s represent the unitless virial coefficients, satisfying
$C_{0}\left(T_{r}\right)=0$. Then, the saturation curve may be obtained
from the constitutive models for the liquid and vapor phases of a
pure substance, using phase equilibrium conditions discussed in Section~\ref{sec:Phase-Transformations}.}

Using the thermodynamic constraint (\ref{eq:thermodynamic-pressure}),
we may integrate (\ref{eq:liquid-pressure}) once with respect to
$J$ to obtain
\begin{equation}
a\left(T,J\right)=\frac{J_{\sigma}\left(T\right)-J}{\rho_{r}}\left(p_{\sigma}\left(T\right)+P_{r}\sum_{k=1}^{n}\frac{1}{k+1}B_{k}\left(T\right)\left(J-J_{\sigma}\left(T\right)\right)^{k}\right)+a_{\sigma}\left(T\right)\,.\label{eq:liquid-free-energy}
\end{equation}
Here, the integration function $a_{\sigma}\left(T\right)$ represents
the value of the specific free energy on the saturation curve, where
$J=J_{\sigma}\left(T\right)$. In particular, we note that $a\left(T_{r},1\right)=a_{\sigma}\left(T_{r}\right)\equiv a_{r}$
is the specific free energy in the reference state.

It follows from (\ref{eq:entropy-constraint}) that the specific entropy
of this liquid is given by
\begin{equation}
\begin{aligned}s\left(T,J\right) & =\frac{J_{\sigma}^{\prime}\left(T\right)}{\rho_{r}}\left(p_{\sigma}\left(T\right)-p\left(T,J\right)\right)+\frac{1}{\rho_{r}}p_{\sigma}^{\prime}\left(T\right)\left(J-J_{\sigma}\left(T\right)\right)\\
 & +s_{\sigma}\left(T\right)+\frac{P_{r}}{\rho_{r}}\sum_{k=1}^{n}\frac{1}{k+1}B_{k}^{\prime}\left(T\right)\left(J-J_{\sigma}\left(T\right)\right)^{k+1}
\end{aligned}
\,,\label{eq:liquid-entropy}
\end{equation}
where the specific entropy $s_{\sigma}\left(T\right)$ on the saturation
curve is
\begin{equation}
s_{\sigma}\left(T\right)=-a_{\sigma}^{\prime}\left(T\right)-\frac{J_{\sigma}^{\prime}\left(T\right)}{\rho_{r}}p_{\sigma}\left(T\right)\,,\label{eq:liquid-entropy-sat}
\end{equation}
and its value in the reference configuration is given by $s\left(T_{r},1\right)=-a_{\sigma}^{\prime}\left(T_{r}\right)\equiv s_{r}$.
The specific internal energy may be evaluated from $u=a+Ts$ and the
specific gauge enthalpy from $h=u+p/\rho$.

We may also evaluate the isochoric specific heat capacity from $c_{v}=T\partial s/\partial T$,
see (\ref{eq:specific-heat-v}). On the saturation curve ($J=J_{\sigma}\left(T\right)$),
we find that the isochoric specific heat capacity satisfies
\begin{equation}
\begin{aligned}\frac{1}{T}c_{v\sigma}\left(T\right) & =-2\frac{J_{\sigma}^{\prime}\left(T\right)}{\rho_{r}}p_{\sigma}^{\prime}\left(T\right)-\frac{J_{\sigma}^{\prime\prime}\left(T\right)}{\rho_{r}}p_{\sigma}\left(T\right)\\
 & +\frac{P_{r}}{\rho_{r}}\left(J_{\sigma}^{\prime}\left(T\right)\right)^{2}B_{1}\left(T\right)-a_{\sigma}^{\prime\prime}\left(T\right)
\end{aligned}
\,.\label{eq:liquid-cv-sat}
\end{equation}
Therefore, experimental estimation of $c_{v\sigma}\left(T\right)$,
along with experimental measurements of $p\left(T,J\right)$, make
it possible to evaluate $a_{\sigma}^{\prime\prime}\left(T\right)$
using the above formula. Integrating $a_{\sigma}^{\prime\prime}\left(T\right)$
with respect to temperature twice, using $a_{\sigma}^{\prime}\left(T_{r}\right)=-s_{r}$
and $a_{\sigma}\left(T_{r}\right)=a_{r}$, provides a determination
of $s_{\sigma}\left(T\right)$ and $a_{\sigma}\left(T\right)$. The
integration constants $s_{r}$ and $a_{r}$ remain arbitrary. Conventionally
for liquids, we set $s_{r}=0$ and $a_{r}=0$ at the triple point,
which serves as a convenient reference configuration for the liquid.
\begin{example}
\label{exa:Liquid-water-properties}The thermodynamic properties of
liquid water may be obtained from NIST (https://webbook.nist.gov/chemistry/fluid/).
We can download these properties for a range of pressures and temperatures
and treat $T$, $P$, $\rho$ and $c_{v\sigma}\left(T\right)$ as
experimental measurements, then reconstruct the rest of the NIST thermodynamic
table entries using the relations presented in this section to validate
our approach. At the triple point of water the reference state is
$T_{r}=273.16\,\text{K}$, $P_{r}=0.61165\,\text{kPa}$, and $\rho_{r}=999.79\,\text{kg}/\text{m}^{3}$.
Using downloaded values of $P$ and $\rho$ on the saturation curve
(at $1\,\text{K}$ intervals from $T_{r}$ to $646.16\,\text{K}$),
we may evaluate $J_{\sigma}\left(T\right)$ from (\ref{eq:integrated-mass-balance})
and $p_{\sigma}\left(T\right)$ from $P-P_{r}$ (\figurename~\ref{fig:liquid-J-p-sigma}),
\begin{figure}
\begin{centering}
\includegraphics[width=3.13in]{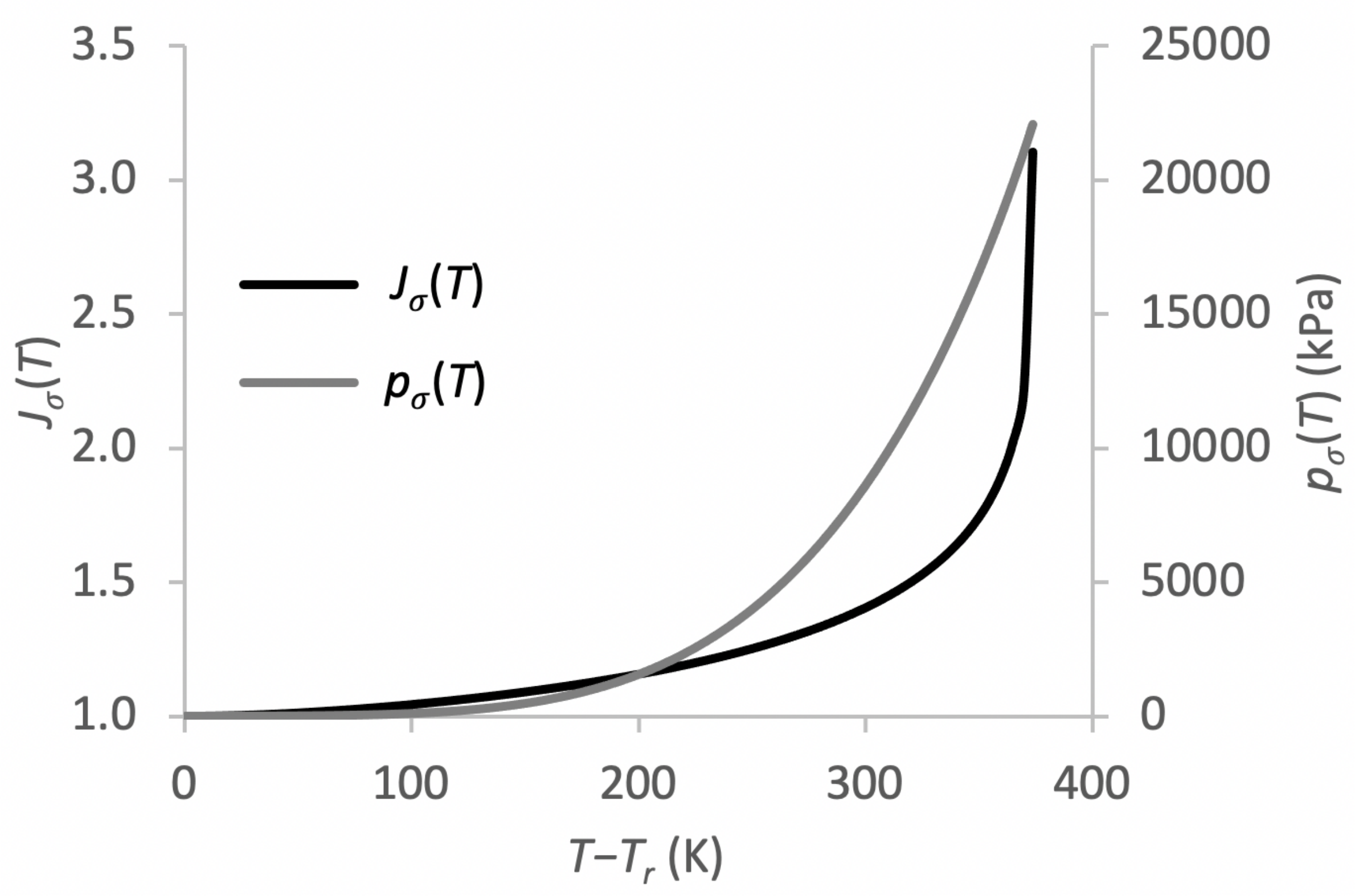}
\par\end{centering}
\caption{Plots of the volume ratio $J_{\sigma}\left(T\right)$ and gauge pressure
$p_{\sigma}\left(T\right)$ for liquid water on its saturation curve,
using NIST data.\label{fig:liquid-J-p-sigma}}
\end{figure}
 then interpolate them using piecewise cubic polynomials and differentiate
them when necessary, using a suitable software package such as Mathematica
(Wolfram Research Inc.). Using downloaded properties over a broad
range of $P$ at selected values of $T$ in the range $T_{r}\le T\le T_{c}$
(where $T_{c}=647.1\,\text{K}$ is the critical temperature), we may
fit (\ref{eq:liquid-pressure}) to $P-P_{r}$ versus $J-J_{\sigma}\left(T\right)$
at each $T$ to obtain $B_{1}\left(T\right)$ (with coefficients of
determination in the range $0.99997\le R^{2}\le1.00000$ for the various
values of $T$) and $B_{2}\left(T\right)$ (with $0.99991\le R^{2}\le1.00000$)
for a virial expansion with $m=2$. Then, the discrete set of coefficients
$B_{1}$ and $B_{2}$ may be fitted to polynomial functions of $T$
(\figurename~\ref{fig:Virial-liquid-H2O}) to produce
\begin{figure}
\begin{centering}
\subfloat[]{\begin{centering}
\includegraphics[width=3.13in]{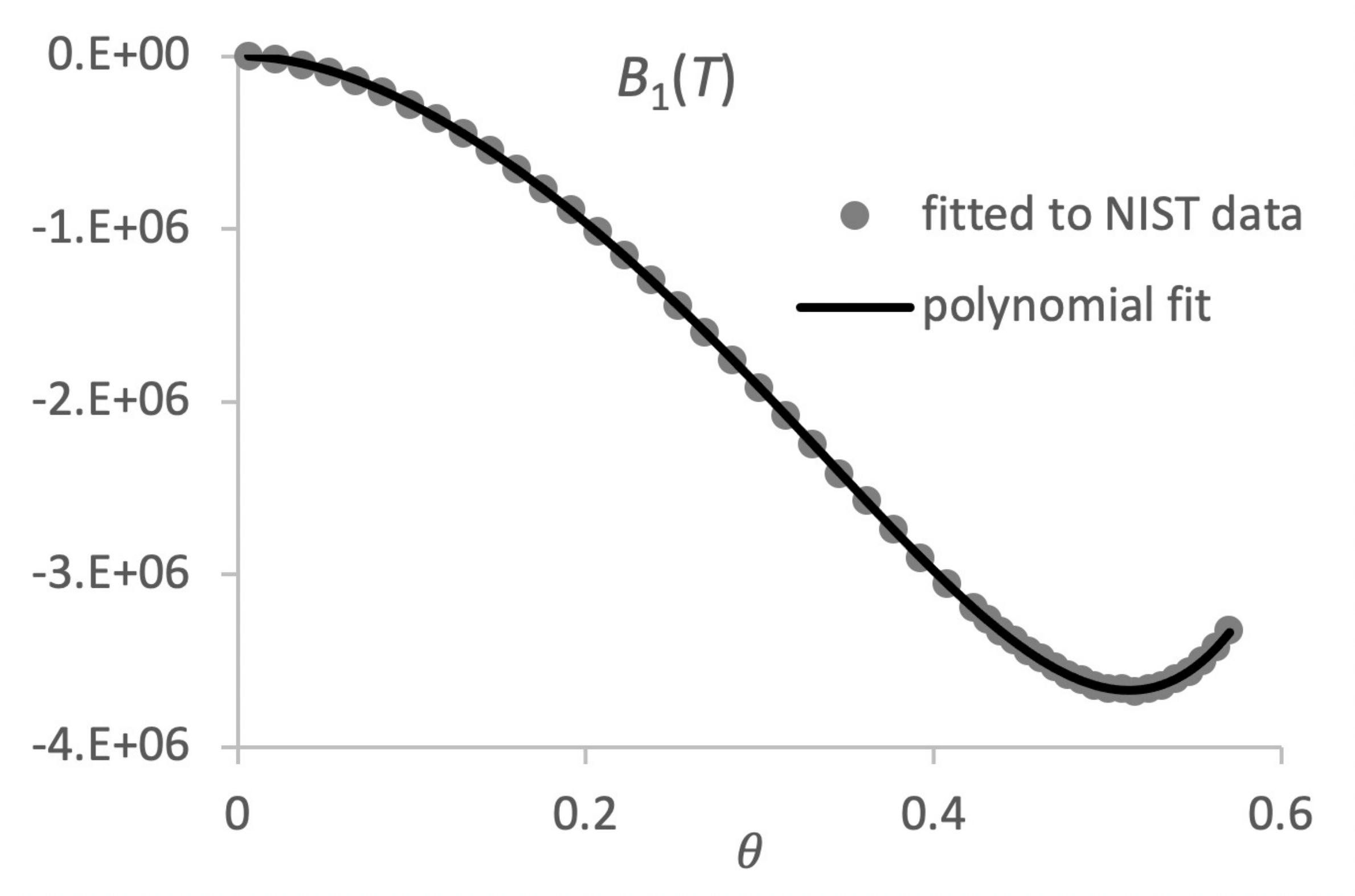}
\par\end{centering}
}
\par\end{centering}
\begin{centering}
\subfloat[]{\begin{centering}
\includegraphics[width=3.13in]{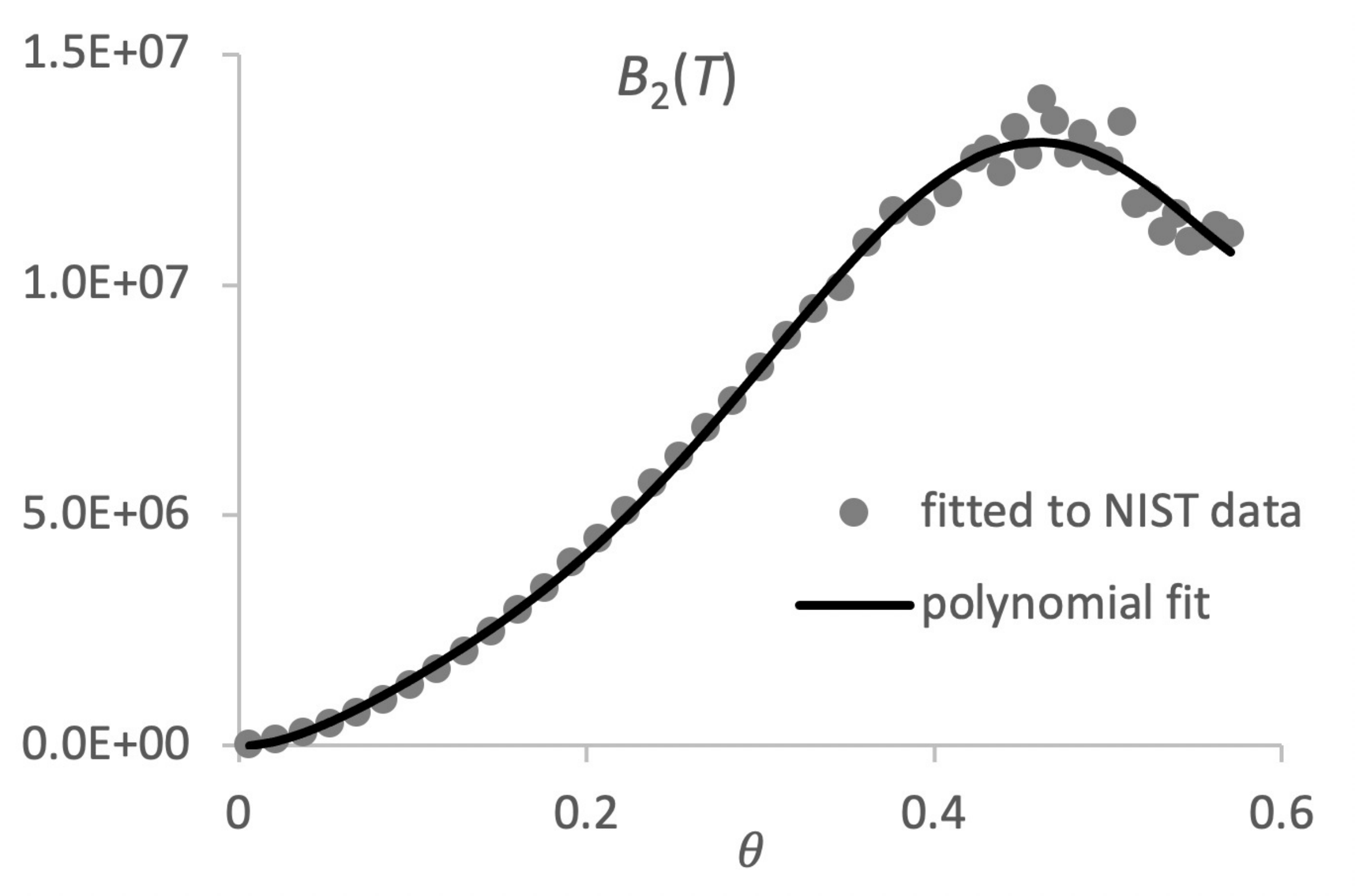}
\par\end{centering}
}
\par\end{centering}
\caption{Virial coefficients $B_{1}\left(T\right)$ and $B_{2}\left(T\right)$
for liquid water, plotted against $\theta=1-T/T_{c}$. Symbols represent
values obtained by fitting $p$ in equation (\ref{eq:liquid-pressure})
versus $J-J_{\sigma}\left(T\right)$ at selected increments of temperature
in the range $T_{r}\le T\le T_{c}$. Solid lines represents polynomial
approximations of those values as given in equations (\ref{eq:liquid-B1})
and (\ref{eq:liquid-B2}).\label{fig:Virial-liquid-H2O}}
\end{figure}
 
\begin{equation}
\begin{aligned}B_{1}\left(T\right) & =9.59435\times10^{7}\,\theta^{6}+8.95496\times10^{7}\,\theta^{5}-1.37737\times10^{8}\,\theta^{4}\\
 & +7.36381\times10^{7}\,\theta^{3}-3.42044\times10^{7}\,\theta^{2}
\end{aligned}
\,,\label{eq:liquid-B1}
\end{equation}
with $R^{2}=1.00000$ and
\begin{equation}
\begin{aligned}B_{2}\left(T\right) & =9.24187\times10^{9}\,\theta^{6}-1.43932\times10^{10}\,\theta^{5}+7.77799\times10^{9}\,\theta^{4}\\
 & -1.88532\times10^{9}\,\theta^{3}+2.70511\times10^{8}\,\theta^{2}
\end{aligned}
\,,\label{eq:liquid-B2}
\end{equation}
with $R^{2}=0.99885$, where $\theta=1-T/T_{c}$ and $B_{1}\left(T\right)$
and $B_{2}\left(T\right)$ are unitless. The scatter observed in $B_{2}\left(T\right)$
for lower values of $T$ (higher values of $\theta$) arises from
the greater uncertainty in this parameter for this range of temperatures.
These expressions may be differentiated as needed in (\ref{eq:liquid-entropy}).
Using downloaded values of $c_{v\sigma}\left(T\right)$ on the saturation
curve, we then evaluate $a_{\sigma}^{\prime\prime}\left(T\right)$
from (\ref{eq:liquid-cv-sat}) and integrate it twice to solve for
$s_{\sigma}\left(T\right)$ and $a_{\sigma}\left(T\right)$, subject
to the condition that $s_{r}=0$ and $a_{r}=0$ at the triple point
(\figurename~\ref{fig:liquid-a-s-sigma}). A comparison of these
results against NIST data entries yields errors ranging from $-0.14\%$
to $0\%$ for $a_{\sigma}\left(T\right)$ and from $-1.30\%$ to $0\%$
for $s_{\sigma}\left(T\right)$, with the largest magnitudes occurring
in the vicinity of $T_{c}$. With these functions, we may now calculate
$a$ and $s$ for any $\left(T,J\right)$ using (\ref{eq:liquid-free-energy})
and (\ref{eq:liquid-entropy}), allowing us to further verify the
accuracy of our constitutive model (\ref{eq:liquid-free-energy})
for liquid water against the NIST values as obtained from standard
methods \citep{Wagner02}. For $T_{r}\le T\le643.15\,\text{K}$ and
$0.9\le p\le22\,000\,\text{kPa}$, we find that the error against
NIST data ranges from $-0.74\%$ to $0.56\%$ for $a\left(T,J\right)$
($0.00\pm0.06\,\%$ mean\textpm standard deviation) and from $-1.52\%$
to $0.05\%$ for $s\left(T,J\right)$ ($-0.02\pm0.08\,\%)$. Similarly
good agreement is found for the specific internal energy $u\left(T,J\right)$
($-2.12\%$ to $0.05\%$, $-0.02\pm0.11\:\%$) and specific enthalpy
$h\left(T,J\right)$ ($-2.06\%$ to $0.41\%$, $-0.02\pm0.12\,\%$).
These small errors imply that the formulation of the constitutive
model for liquids in this study is valid; small discrepancies with
NIST data arise most likely from the simplistic choices of interpolation
and approximation functions adopted in this illustrative example.
\begin{figure}
\begin{centering}
\subfloat[]{\begin{centering}
\includegraphics[width=3.13in]{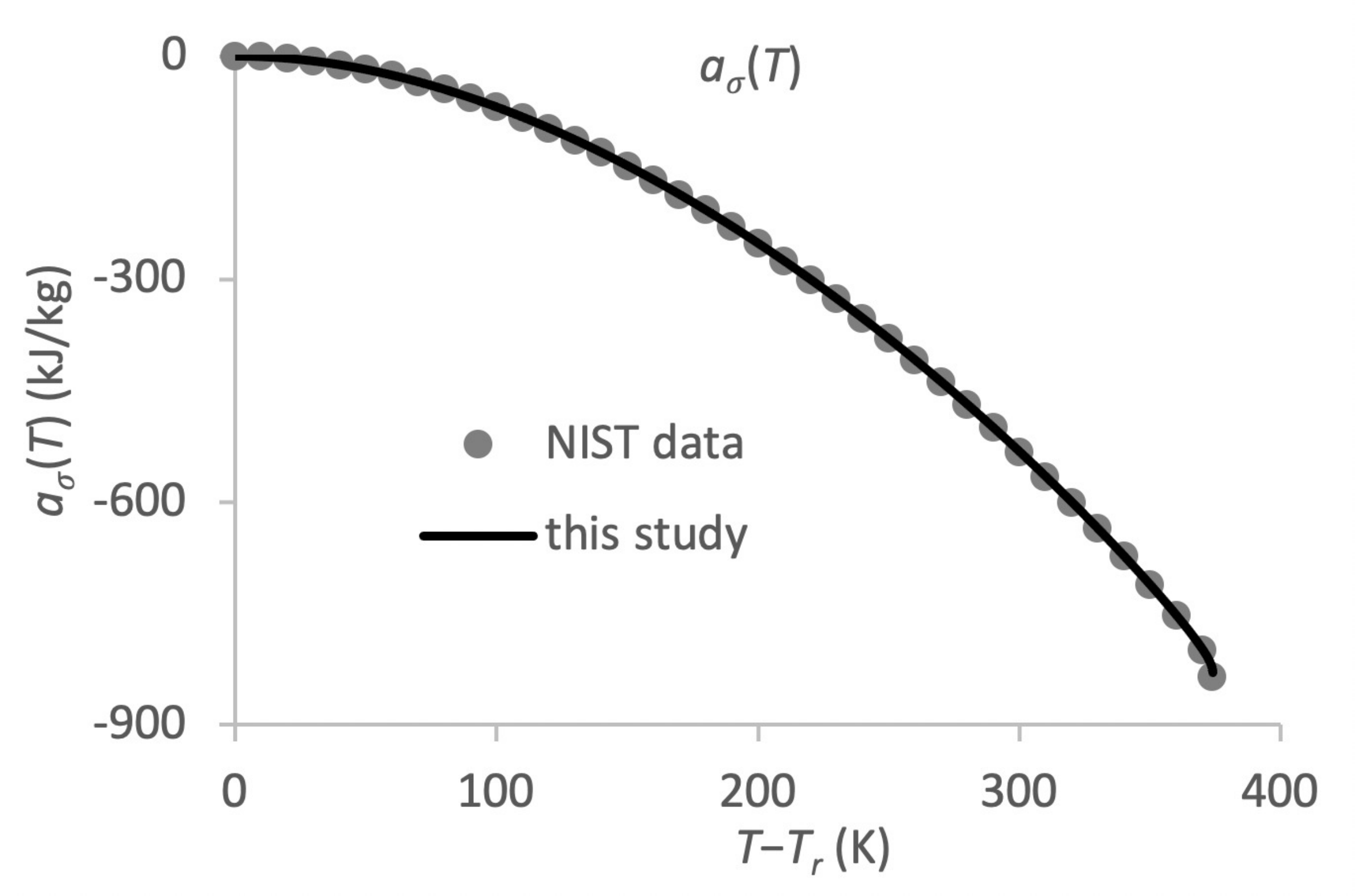}
\par\end{centering}
}
\par\end{centering}
\begin{centering}
\subfloat[]{\begin{centering}
\includegraphics[width=3.13in]{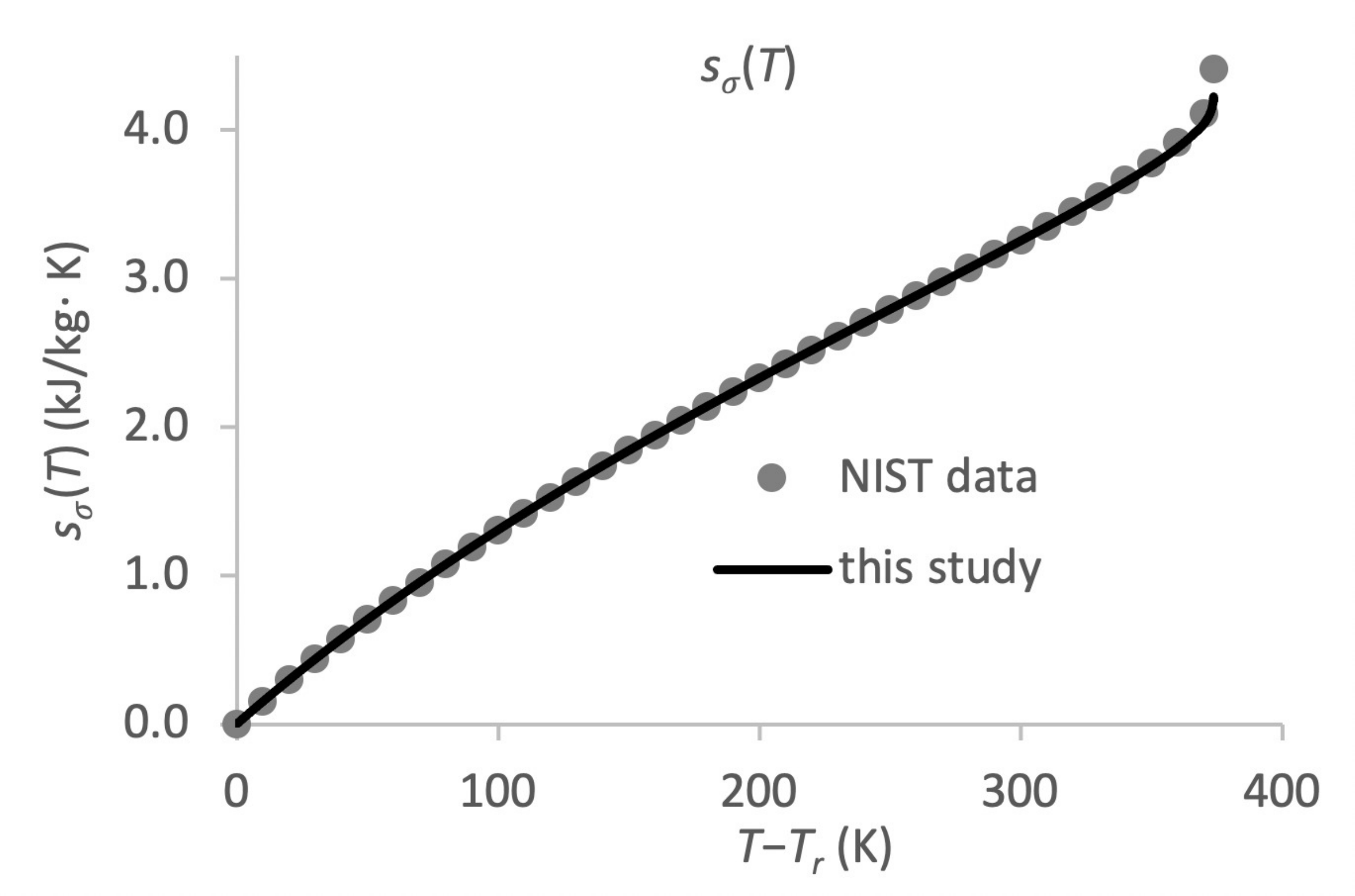}
\par\end{centering}
}
\par\end{centering}
\caption{(a) Specific free energy $a_{\sigma}\left(T\right)$ and (b) specific
entropy $s_{\sigma}\left(T\right)$ of liquid water on the saturation
curve, evaluated from the method of this study (solid lines) and compared
to NIST values (symbols).\label{fig:liquid-a-s-sigma}}
\end{figure}
\end{example}

\subsection{Interface Jump Conditions\label{subsec:Interface-Jump-Conditions}}

Interface jump conditions are needed to impose valid boundary conditions
in any boundary value problem in continuum mechanics \citep{Kelly64,Muller85,Hutter04}.
These jump conditions are derived from the same balance axioms that
produce the governing differential equations. In some applications,
the interface represents a material region, such as a membrane or
shell domain separating two fluids, while in other cases it is an
immaterial surface, such as the phase boundary between liquid and
gas phases of a fluid, a shock wave in a compressible fluid, or an
imaginary section through a material. In Appendix~\ref{sec:Appendix},
we derive interface jump conditions for immaterial interfaces that
may move and deform, and we neglect certain effects, such as surface
tension and its rate of work, that may occur due to disruption in
intermolecular bonds introduced by the presence of such interfaces.
These jump conditions are derived under the assumption that the interface
separates two domains that each contains a single phase of the same
substance. The phase may be the same on both sides of the interface,
in which case the interface is permeable to the material. If the phases
are different, the interface is assumed to be impermeable. Later,
we generalize interface jump conditions to account for phase transformations
on a phase boundary.

\subsubsection{Summary of Jump Conditions}

Jump conditions are derived under the assumption that the interface
$\Gamma$ separates two domains denoted by $V_{+}$ and $V_{-}$.
The velocity of $\Gamma$ is $\mathbf{v}^{\Gamma}$ and the unit normal
on $\Gamma$ is $\mathbf{n}^{\Gamma}$, pointing away from $V_{+}$.
For any quantity $f$, the expression $\left[\left[f\right]\right]\equiv f_{+}-f_{-}$
represents the jump in $f$ across $\Gamma$, with $f_{\pm}$ representing
the values of $f$ in $V_{\pm}$, on $\Gamma$. The jump condition
derived from the axiom of mass balance is
\begin{equation}
\left[\left[\rho\mathbf{u}^{\Gamma}\right]\right]\cdot\mathbf{n}^{\Gamma}=0\,,\label{eq:mass-balance-jump}
\end{equation}
where
\begin{equation}
\mathbf{u}^{\Gamma}\equiv\mathbf{v}-\mathbf{v}^{\Gamma}\,.\label{eq:interface-relative-velocity}
\end{equation}
Evidently, the mass balance jump in (\ref{eq:mass-balance-jump})
simply enforces continuity of the mass flux across and normal to the
interface $\Gamma$. It is noteworthy that the mass balance jump does
not impose any constraint on the tangential component of the mass
flux or the velocity. Indeed, any tangential constraint would need
to depend on the specifics of a particular analysis. For example,
if $\Gamma$ separates two solid materials that slide past each other,
there is no requirement for enforcing continuity of tangential mass
flux or velocity. If the two solids are glued together however, the
nature of this problem provides a specific additional interface condition,
namely the continuity of the velocity component tangential to the
interface. For viscous fluids this is known as the no-slip condition;
it is recognized to be an empirically validated condition.

For thermoelastic fluids, the jump condition on the axiom of momentum
balance reduces to
\begin{equation}
\left[\left[p\mathbf{I}+\rho\mathbf{u}^{\Gamma}\otimes\mathbf{u}^{\Gamma}\right]\right]\cdot\mathbf{n}^{\Gamma}=\mathbf{0}\,.\label{eq:momentum-jump-redux}
\end{equation}
This jump condition tells us that the sum of pressure and linear momentum
flux across and normal to $\Gamma$ is conserved. The jump condition
on the axiom of energy balance takes the form
\begin{equation}
\left[\left[\rho\left(h+\frac{1}{2}\mathbf{u}^{\Gamma}\cdot\mathbf{u}^{\Gamma}\right)\mathbf{u}^{\Gamma}+\mathbf{q}\right]\right]\cdot\mathbf{n}^{\Gamma}=0\,,\label{eq:energy-jump-redux}
\end{equation}
where $h\equiv u+p/\rho$. Thus, the concept of enthalpy emerges naturally
from the jump condition on the energy. Since $p$ represents a gauge
pressure, we refer to $h$ as the \emph{specific gauge enthalpy}.
Note that $\frac{1}{2}\rho\mathbf{u}^{\Gamma}\cdot\mathbf{u}^{\Gamma}$
is the kinetic energy density of the material relative to $\Gamma$;
we may refer to it as the diffusive kinetic energy density across
$\Gamma$. The jump condition (\ref{eq:energy-jump-redux}) tells
us that the flux of enthalpy and diffusive kinetic energy plus the
heat flux across and normal to $\Gamma$ is conserved.

Finally, the jump condition on the axiom of entropy inequality is
given by
\begin{equation}
\left[\left[\rho s\mathbf{u}^{\Gamma}+\frac{1}{T}\mathbf{q}\right]\right]\cdot\mathbf{n}^{\Gamma}\le0\,.\label{eq:entropy-jump}
\end{equation}
The entropy inequality jump enforces a constraint on the sum of entropy
fluxes carried by mass convection and heat conduction, across and
normal to $\Gamma$. Since we have already formulated the constraints
(\ref{eq:entropy-constraint}) and (\ref{eq:residual-dissipation-compressible})
on the functions of state $s$ and $\mathbf{q}$ on either side of
$\Gamma$, the entropy inequality jump (\ref{eq:entropy-jump}) serves
to place a constraint on the feasibility of processes across $\Gamma$,
as illustrated next.

These jump conditions are standard in the continuum mechanics literature
\citep{Kelly64,Eringen65,Muller85,Buratti03,Hutter04,Ateshian07}.

\subsubsection{Normal Shock Wave\label{subsec:Normal-Shock-Wave}}

This normal shock wave problem illustrates how jump conditions may
be used to solve for one-dimensional steady flow of an ideal gas with
constant specific heat capacity across a \emph{stationary shock wave}.
We apply the jump conditions for mass, momentum and energy across
a non-reactive interface (the shock wave) which is permeable to the
substance ($\mathbf{u}^{\Gamma}\cdot\mathbf{n}^{\Gamma}\ne0$).

In our framework, $\rho_{r}$ and $T_{r}$ are invariant reference
values for the gas on either side of $\Gamma$, therefore $\left[\left[\rho_{r}\right]\right]=0$
and $\left[\left[T_{r}\right]\right]=0$. Similar constraints apply
to invariants $R$ and $M$. Substituting the ideal gas relations
in (\ref{eq:IG-elastic-pressure}) and Example~\ref{exa:IG-constant-cv}
into the jump conditions of (\ref{eq:mass-balance-jump}), (\ref{eq:momentum-jump-redux})
and (\ref{eq:energy-jump-redux}) produces
\begin{equation}
\left[\left[\frac{1}{J}\mathbf{u}^{\Gamma}\right]\right]\cdot\mathbf{n}^{\Gamma}=0\,,\label{eq:sw-mass-jump}
\end{equation}
\begin{equation}
\left[\left[\frac{1}{J}\left(\frac{RT}{M}\mathbf{I}+\mathbf{u}^{\Gamma}\otimes\mathbf{u}^{\Gamma}\right)\right]\right]\cdot\mathbf{n}^{\Gamma}=\mathbf{0}\,,\label{eq:sw-momentum-jump}
\end{equation}
\begin{equation}
\left[\left[\frac{\rho_{r}}{J}\left(c_{p0}\left(T-T_{r}\right)+\frac{1}{2}\mathbf{u}^{\Gamma}\cdot\mathbf{u}^{\Gamma}\right)\mathbf{u}^{\Gamma}+\mathbf{q}\right]\right]\cdot\mathbf{n}^{\Gamma}=0\,.\label{eq:sw-energy-jump}
\end{equation}
Consider one-dimensional steady flow across a stationary shock wave
$\Gamma$ ($\mathbf{v}^{\Gamma}=\mathbf{0}$, so that $\mathbf{u}^{\Gamma}=\mathbf{v}$),
with uniform temperatures $T_{+}$ upstream and $T_{-}$ downstream,
such that $\mathbf{q}=\mathbf{0}$ (adiabatic conditions) in both
domains, leading to $\left[\left[\mathbf{q}\right]\right]\cdot\mathbf{n}^{\Gamma}=0$.
Let $T_{+}$, $v_{+}\equiv\mathbf{v}_{+}\cdot\mathbf{n}^{\Gamma}$
and $J_{+}$ be given and solve for the corresponding downstream variables
$T_{-}$, $v_{-}$ and $J_{-}$. From the mass jump (\ref{eq:sw-mass-jump})
we find that
\begin{equation}
\frac{1}{J_{+}}v_{+}=\frac{1}{J_{-}}v_{-}\,,\label{eq:sw-mass-redux}
\end{equation}
where $v_{-}=\mathbf{v}_{-}\cdot\mathbf{n}^{\Gamma}$. From the momentum
jump (\ref{eq:sw-momentum-jump}), we find that
\begin{equation}
\frac{1}{J_{+}}\left(\frac{RT_{+}}{M}+v_{+}^{2}\right)=\frac{1}{J_{-}}\left(\frac{RT_{-}}{M}+v_{-}^{2}\right)\,.\label{eq:sw-momentum-redux}
\end{equation}
From the energy jump (\ref{eq:sw-energy-jump}), making use of (\ref{eq:sw-mass-redux})
and (\ref{eq:IG-cp}), 
\begin{equation}
\frac{R\gamma}{M\left(\gamma-1\right)}T_{+}+\frac{1}{2}v_{+}^{2}=\frac{R\gamma}{M\left(\gamma-1\right)}T_{-}+\frac{1}{2}v_{-}^{2}\,,\label{eq:sw-energy-redux}
\end{equation}
where $\gamma\equiv c_{p0}/c_{v0}$ is the specific heat capacity
ratio. This system of quadratic equations admits two solutions, namely
the trivial solution $J_{-}=J_{+}$, $T_{-}=T_{+}$ and $v_{-}=v_{+}$
(implying that there is no shock wave), and the following solution,
\begin{equation}
\begin{aligned}\frac{J_{-}}{J_{+}} & =\frac{\text{Ma}_{+}^{2}\left(\gamma-1\right)+2}{\text{Ma}_{+}^{2}\left(1+\gamma\right)}\\
\frac{T_{-}}{T_{+}} & =\frac{\left(2\gamma\text{Ma}_{+}^{2}-\gamma+1\right)\left(\text{Ma}_{+}^{2}\left(\gamma-1\right)+2\right)}{\text{Ma}_{+}^{2}\left(\gamma+1\right)^{2}}\\
\frac{v_{-}}{v_{+}} & =\frac{\text{Ma}_{+}^{2}\left(\gamma-1\right)+2}{\text{Ma}_{+}^{2}\left(1+\gamma\right)}
\end{aligned}
\,,\label{eq:sw-solution}
\end{equation}
where
\begin{equation}
\text{Ma}_{+}^{2}=\frac{Mv_{+}^{2}}{\gamma RT_{+}}\,,\quad\text{Ma}_{-}^{2}=\frac{Mv_{-}^{2}}{\gamma RT_{-}}\,,\label{eq:sw-Mach-numbers}
\end{equation}
are the squares of Mach numbers upstream and downstream, respectively.
This result, which reproduces the classical solution for a normal
shock wave \citep{Shapiro53}, illustrates the application of the
jump conditions derived above to a canonical problem involving an
interface. As is evident from the solution in (\ref{eq:sw-solution}),
the temperature is not continuous across $\Gamma$. This observation
is important because it clearly establishes that jump conditions do
not necessarily enforce temperature continuity. We revisit this concept
below when we analyze phase transformations across an interface $\Gamma$.

\begin{figure}
\begin{centering}
\includegraphics[width=3.13in]{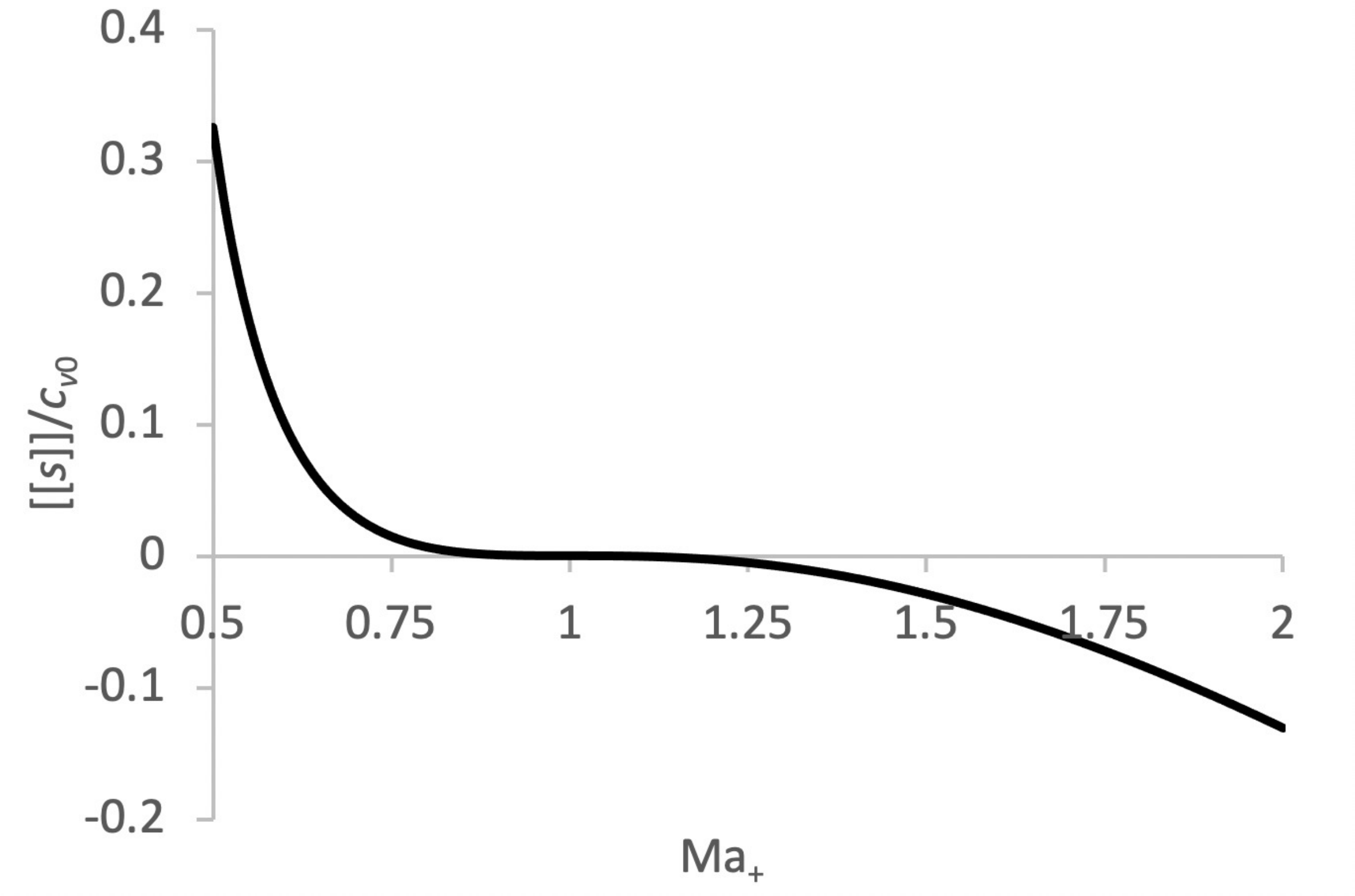}
\par\end{centering}
\caption{Plot of $\left[\left[s\right]\right]/c_{v0}$ across a normal shock
wave, as a function of the upstream Mach number $\text{Ma}_{+}$ ($\gamma=1.4$).
The entropy inequality jump (\ref{eq:entropy-jump}) is only satisfied
when $\text{Ma}_{+}\ge1$.\label{fig:Jump-in-entropy}}
\end{figure}

Using the function of state for the specific entropy $s$ of an ideal
gas as given in (\ref{eq:IG-constC-s}), we find that the change in
$s$ across the shock wave is given by
\begin{equation}
s_{+}-s_{-}=\left[\left[s\right]\right]=c_{v0}\left(\left(1-\gamma\right)\ln\frac{J_{-}}{J_{+}}-\ln\frac{T_{-}}{T_{+}}\right)\,.\label{eq:sw-entropy-jump}
\end{equation}
In particular, when $\text{Ma}_{+}=1$ (i.e., when there is no shock
wave) we find that $\left[\left[s\right]\right]=0$. Furthermore,
a plot of $\left[\left[s\right]\right]/c_{v0}$ versus $\text{Ma}_{+}$
shows that it is positive when $\text{Ma}_{+}<1$ and negative when
$\text{Ma}_{+}>1$ (\figurename~\ref{fig:Jump-in-entropy}). Combining
the mass jump condition for the mixture (\ref{eq:mass-balance-jump})
with the entropy inequality jump (\ref{eq:entropy-jump}) under our
given assumption that $\left[\left[\mathbf{q}\right]\right]\cdot\mathbf{n}^{\Gamma}=0$
shows that we must satisfy $\left[\left[s\right]\right]\le0$ whenever
the mass flux across the shock wave is positive ($\rho\mathbf{u}^{\Gamma}\cdot\mathbf{n}>0$).
This means that the entropy must increase at the downstream side of
the shock wave. We conclude that the entropy inequality requires that
shock waves occur only when the upstream flow is supersonic, $\text{Ma}_{+}\ge1$.
In other words, \emph{the flow cannot spontaneously jump from subsonic
to supersonic} under the adiabatic 1D flow conditions of this problem,
as has been demonstrated in the classical compressible flow literature
\citep{Liepmann57}. This application of the jump condition for entropy
inequality across a normal shock wave illustrates the significance
of this constraint on interfacial processes.

\section{Reactive Mixtures\label{sec:Reactive-Mixtures}}

Most thermodynamic applications deal with mixtures of multiple constituents,
such as mixtures of gases, mixtures of the liquid and gas phases of
a fluid, or solutions containing a solvent and solutes. To analyze
such mixtures we may use the framework of mixture theory \citep{Truesdell60,Eringen65,Bowen68,Williams84,Ateshian07,Bothe15}.
This general framework can encompass a wide range of complexities,
including phase changes, chemical reactions, diffusion of neutral
or charged solutes, etc. In this study we choose to only consider
phase transformations of a pure fluid substance across a boundary.

To formulate the jump condition for mass balance across a reactive
interface $\Gamma$, we first need to present the axiom of mass balance
for mixture constituents in a control volume. The basic concept of
mixture theory is that any number of mixture constituents may be present
in an elemental material region. In the context of a continuum framework,
it is assumed mathematically that the size of this material region
is infinitesimal; in other words, mathematically, all these materials
coexist at a single point. In reality, the continuum model is a valid
approximation only down to a representative scale of the microstructure.
Thus, the elemental region should contain a sufficient number of molecules
of the mixture constituents to obey the governing equations of the
continuum framework.

In mixture theory we distinguish each mixture constituent $\alpha$
using a superscript, such as $\rho^{\alpha}$ for representing its
apparent mass density, which represents the mass of constituent $\alpha$
per volume of the mixture. The general framework of mixture theory
allows each constituent to follow a motion independent of others;
thus, each constituent may have its own motion $\boldsymbol{\chi}^{\alpha}\left(\mathbf{X}^{\alpha},t\right)$
and velocity $\mathbf{v}^{\alpha}=\partial\boldsymbol{\chi}^{\alpha}/\partial t$.

\subsection{Axiom of Mass Balance for Reactive Constituents\label{subsec:Mass-jump-reactive}}

The material presented in this section can be found in classical mixture
theory studies of reacting media \citep{Truesdell60,Kelly64,Eringen65,Bowen68,Truesdell84,Muller85}
and the more recent literature (such as \citep{Ateshian07,Bothe15}).
Since a mixture may involve reactions that exchange mass among its
constituents (e.g., phase transformations, chemical reactions, etc.),
the axiom of mass balance for each constituent $\alpha$ must account
for this mass supply. The integral statement of mass balance in a
fixed control volume $V$ takes the form
\begin{equation}
\frac{d}{dt}\int_{V}\rho^{\alpha}\,dV=-\int_{\partial V}\rho^{\alpha}\mathbf{v}^{\alpha}\cdot\mathbf{n}\,dS+\int_{V}\hat{\rho}^{\alpha}\,dV\,,\label{eq:integral-mass-alpha}
\end{equation}
where $\mathbf{n}$ is the outward unit normal to the boundary $\partial V$
of $V$ and $\hat{\rho}^{\alpha}$ is the mass density supply to constituent
$\alpha$ due to reactions with all other mixture constituents. This
mass density supply is a function of state that requires a constitutive
model, which may take a different form for chemical reactions, phase
transformations, etc. It has units of mass per volume, per time, and
it is positive when mass is added to constituent $\alpha$, negative
when mass is lost from that constituent, and zero in the absence of
reactions involving constituent $\alpha$. Using the divergence theorem
to convert the surface integral in (\ref{eq:integral-mass-alpha})
to a volume integral, we obtain the differential statement of mass
balance for each constituent $\alpha$,
\begin{equation}
\frac{\partial\rho^{\alpha}}{\partial t}+\divg\left(\rho^{\alpha}\mathbf{v}\right)=\hat{\rho}^{\alpha}\,.\label{eq:diff-mass-alpha}
\end{equation}
A fundamental principle of mixture theory is that the mixture should
obey the axioms of mass, momentum and energy balance of a single constituent
(a pure substance in a single phase). Taking the sum of (\ref{eq:diff-mass-alpha})
over all $\alpha$ and equating it to (\ref{eq:mass-balance}) shows
that the mixture density is $\rho=\sum_{\alpha}\rho^{\alpha}$, the
mixture velocity is $\mathbf{v}=\sum_{\alpha}\rho^{\alpha}\mathbf{v}^{\alpha}/\rho$,
and the mass density supplies must satisfy the constraint
\begin{equation}
\sum_{\alpha}\hat{\rho}^{\alpha}=0\,.\label{eq:mass-supply-constraint}
\end{equation}
Thus, mass gained by products of a reaction must balance the mass
lost from reactants to conserve the mass of the mixture.

We may now apply the methodology described in Appendix~\ref{sec:Appendix}
to evaluate the mass balance jump condition for each constituent $\alpha$,
recognizing that $\hat{\rho}^{\alpha}$ in $V_{\epsilon}$ does not
reduce to zero as $\epsilon\to0$. In that case, the jump condition
becomes
\begin{equation}
\left[\left[\rho^{\alpha}\left(\mathbf{v}^{\alpha}-\mathbf{v}^{\Gamma}\right)\right]\right]\cdot\mathbf{n}^{\Gamma}=-\bar{\rho}^{\alpha}\,.\label{eq:mass-jump-constituent}
\end{equation}
Here, $\bar{\rho}^{\alpha}$ is the area density supply of mass to
constituent $\alpha$ due to reactions taking place on $\Gamma$,
such that
\[
\lim_{\epsilon\to0}\int_{V_{\epsilon}\left(t\right)}\hat{\rho}^{\alpha}\,dV=\lim_{\epsilon\to0}\int_{\Gamma}\hat{\rho}^{\alpha}\,\epsilon\,dA=\int_{\Gamma}\bar{\rho}^{\alpha}\,dA\,.
\]
Just as $\hat{\rho}^{\alpha}$ is a function of state in $V$, so
is $\bar{\rho}^{\alpha}$ a function of state on $\Gamma$; it has
units of mass flux (mass per area per time) and may also be described
as the \emph{reactive mass flux} of $\alpha$.

The mass balance jump is applied separately to each constituent $\alpha$.
Moreover, the reactive mass flux $\bar{\rho}^{\alpha}$ satisfies
the same type of constraint as $\hat{\rho}^{\alpha}$ in (\ref{eq:mass-supply-constraint}),
namely
\begin{equation}
\sum_{\alpha}\bar{\rho}^{\alpha}=0\,.\label{eq:reactive-mass-flux-constraint}
\end{equation}

\subsection{Reaction Kinetics\label{subsec:Reaction-Kinetics}}

Reaction kinetics and the stoichiometry of reacting mixtures have
been described in the prior literature \citep{Bowen68a,Truesdell84,Prudhomme10}.
Reactions may occur among the constituents of a mixture which result
in a temporal evolution of the mass content of reactants and products.
A forward reaction between mixture constituents may be expressed as
\citep{Prudhomme10}
\begin{equation}
\sum_{\alpha}\nu_{R}^{\alpha}\mathcal{E}^{\alpha}\to\sum_{\alpha}\nu_{P}^{\alpha}\mathcal{E}^{\alpha}\,,\label{eq:forward-reaction}
\end{equation}
where $\mathcal{E}^{\alpha}$ is the material associated with constituent
$\alpha$, $\nu_{R}^{\alpha}$ represents the stoichiometric coefficient
of reactant $\alpha$ and $\nu_{P}^{\alpha}$ is that of the corresponding
product. Similarly, a reversible reaction may be written as
\begin{equation}
\sum_{\alpha}\nu_{R}^{\alpha}\mathcal{E}^{\alpha}\rightleftharpoons\sum_{\alpha}\nu_{P}^{\alpha}\mathcal{E}^{\alpha}\,.\label{eq:reversible-reaction}
\end{equation}
The summations are taken over all mixture constituents, though constituents
that are not reactants in that particular reaction will have $\nu_{R}^{\alpha}=0$,
and those that are not products will have $\nu_{P}^{\alpha}=0$.

Since the stoichiometric coefficients count the number of moles of
reactants and products, we may relate the mass concentration $\rho^{\alpha}$
to the molar concentration $c^{\alpha}$ of constituent $\alpha$
via $\rho^{\alpha}=M^{\alpha}c^{\alpha}$, where $M^{\alpha}$ is
the molar mass of $\alpha$. Similarly, we may define the molar density
supply $\hat{c}^{\alpha}$ such that $\hat{\rho}^{\alpha}=M^{\alpha}\hat{c}^{\alpha}$.
The stoichiometry of the reaction imposes constraints on the molar
density supplies which may be represented by $\hat{c}^{\alpha}=\nu^{\alpha}\hat{\zeta}$,
where $\hat{\zeta}$ is the \emph{molar production rate} for the reaction
(units of mole per volume, per time), and $\nu^{\alpha}=\nu_{P}^{\alpha}-\nu_{R}^{\alpha}$
is the net stoichiometric coefficient of $\alpha$ in the reaction.
Substituting these relations into the constraint (\ref{eq:mass-supply-constraint})
on mass supplies produces $\sum_{\alpha}M^{\alpha}\nu^{\alpha}=0$.
This relation may be recognized as the classical requirement to balance
the molar mass of reactants and products in a reaction.

For reactions taking place on an interface $\Gamma$, we may similarly
define the reactive molar flux $\bar{\zeta}$ such that
\begin{equation}
\bar{\rho}^{\alpha}=\nu^{\alpha}M^{\alpha}\bar{\zeta}\,.\label{eq:reactive-mass-molar-flux}
\end{equation}
Substituting this relation into (\ref{eq:reactive-mass-flux-constraint})
reproduces the same stoichiometric balance equation given in the previous
paragraph. The implication of (\ref{eq:reactive-mass-molar-flux})
is that only one constitutive relation is needed for $\bar{\zeta}$
to uniquely define the reactive mass supplies $\bar{\rho}^{\alpha}$
for all constituents $\alpha$. In chemical kinetics, it is common
to adopt the `law of mass action' for the constitutive model of
$\bar{\zeta}$ (or $\hat{\zeta}$).

\section{Phase Transformations\label{sec:Phase-Transformations}}

Given our framework of reactive mixtures, we can model phase transitions
across a phase boundary modeled as an interface $\Gamma$ under the
assumption that the reactive mixture on either side of $\Gamma$ only
contains one phase of the pure substance. The relations derived in
this section are valid under general conditions of phase transformation
kinetics. Phase equilibrium is only considered as a special, clearly
identified case.

Let a phase transformation take place across an interface $\Gamma$
that separates two phases, $\mathcal{S}^{a}$ and $\mathcal{S}^{b}$,
of the same pure substance $\mathcal{S}$, with the unit normal $\mathbf{n}^{\Gamma}$
of $\Gamma$ pointing away from the domain of $\mathcal{S}^{a}$ (i.e.,
phase $\mathcal{S}^{a}$ resides in $V_{+}$ and phase $\mathcal{S}^{b}$
resides in $V_{-}$, \figurename~\ref{fig:Interface-jump-conditions}).
Since we have limited our derivations to thermoelastic fluids, $a$
and $b$ may interchangeably denote the liquid and vapor phases. Consider
that the phase transformation reaction may proceed in either direction,
\begin{equation}
\mathcal{S}^{a}\rightleftharpoons\mathcal{S}^{b}\,.\label{eq:reversible-phase-transformation}
\end{equation}
Our goal is to apply the jump conditions presented in Sections~\ref{subsec:Interface-Jump-Conditions}
and \ref{subsec:Mass-jump-reactive}, under the general assumption
that each side of $\Gamma$ is a mixture of both phases, while recognizing
that the apparent density of phase $\mathcal{S}^{a}$ is zero in the
domain of $\mathcal{S}^{b}$ ($\rho_{+}^{a}=\rho^{a}$, $\rho_{-}^{a}=0$),
and that of phase $\mathcal{S}^{b}$ is zero in the domain of $\mathcal{S}^{a}$
($\rho_{-}^{b}=\rho^{b}$, $\rho_{+}^{b}=0$). Thus, the mixture density
and velocity on the side of $\mathcal{S}^{a}$ are $\rho=\rho^{a}$
and $\mathbf{v}=\mathbf{v}^{a}$, respectively, and those on the side
of $\mathcal{S}^{b}$ are $\rho=\rho^{b}$ and $\mathbf{v}=\mathbf{v}^{b}$,
and we may forgo the use of subscripts $+$ and $-$.

Note that the derivations presented in Sections~\ref{subsec:Mass-Balance-Jump},
\ref{subsec:Momentum-Balance-Jump-1} and \ref{subsec:Energy-Balance-Jump-1}
below revisit and extend the classical Stefan problem \citep{Stefan91}
as applied to fluid phase transformations. More recent investigations
have also proposed to use jump conditions to analyze phase transformations
in thermoelastic fluids, however some of those studies proved or assumed
\emph{a priori} that the temperature is continuous across the phase
interface, therefore they recovered the classical relations for phase
equilibrium \citep{Muller85,Fried95,Gray97,Hutter04}, whereas our
presentation investigates general conditions for phase transformations,
without assuming temperature continuity as explained in Section~\ref{subsec:Entropy-Inequality-Jump-1}
below. Later studies in the recent literature \citep{Svendsen96,Fried99,Buratti03,Danescu04}
did account for a temperature jump, though they did not follow the
approach we present here, nor did they derive an expression for the
phase transition mass flux or investigate experimental validations.

\subsection{Mass Balance Jump\label{subsec:Mass-Balance-Jump}}

Since $\Gamma$ is the phase boundary, we treat it as an immaterial
surface. Based on the stoichiometry of this reaction ($\nu^{a}=-1$,
$\nu^{b}=+1$ in (\ref{eq:reversible-phase-transformation})), reactive
mass fluxes $\bar{\rho}^{\alpha}$ ($\alpha=a,\,b$) must satisfy
$\bar{\rho}^{a}+\bar{\rho}^{b}=0$, or $\bar{\rho}^{b}=-\bar{\rho}^{a}=M\bar{\zeta}$
according to (\ref{eq:reactive-mass-molar-flux}), where $M$ is the
molar mass of substance $\mathcal{S}$ and $\bar{\zeta}$ is the reactive
molar flux, such that $M\bar{\zeta}$ is the net reactive mass flux
from the forward reaction $\mathcal{S}^{a}\to\mathcal{S}^{b}$ and
the reverse reaction $\mathcal{S}^{a}\leftarrow\mathcal{S}^{b}$.
Thus, $\bar{\zeta}$ may be positive if the forward reaction dominates,
negative when the reverse reaction dominates, or zero when the reversible
reaction (\ref{eq:reversible-phase-transformation}) has reached phase
equilibrium. Here, we used the reactive molar flux $\bar{\zeta}$
to emphasize that one mole of phase $a$ has been exchanged with one
mole of phase $b$ on $\Gamma$. Now, the mass balance jump for each
phase is reduced from (\ref{eq:mass-jump-constituent}) to
\begin{equation}
\rho^{a}\left(\mathbf{v}^{a}-\mathbf{v}^{\Gamma}\right)\cdot\mathbf{n}^{\Gamma}=\rho^{b}\left(\mathbf{v}^{b}-\mathbf{v}^{\Gamma}\right)\cdot\mathbf{n}^{\Gamma}=M\bar{\zeta}\,.\label{eq:PT-mass-jump}
\end{equation}
The mixture densities $\rho^{a}$ and $\rho^{b}$ may be very different
from each other (liquid versus vapor phases), implying that $\mathbf{v}^{a}$
may also be very different from $\mathbf{v}^{b}$. The phase boundary
moves with normal velocity $\mathbf{v}^{\Gamma}\cdot\mathbf{n}^{\Gamma}$.
By solving for $\mathbf{v}^{\Gamma}\cdot\mathbf{n}^{\Gamma}$ in both
equations,
\begin{equation}
\mathbf{v}^{\Gamma}\cdot\mathbf{n}^{\Gamma}=\mathbf{v}^{a}\cdot\mathbf{n}^{\Gamma}-\frac{M\bar{\zeta}}{\rho^{a}}=\mathbf{v}^{b}\cdot\mathbf{n}^{\Gamma}-\frac{M\bar{\zeta}}{\rho^{b}}\,,\label{eq:PT-interface-velocity}
\end{equation}
we also find that the normal component of the relative velocity between
the two phases is
\begin{equation}
\left(\mathbf{v}^{a}-\mathbf{v}^{b}\right)\cdot\mathbf{n}^{\Gamma}=M\bar{\zeta}\left(\frac{1}{\rho^{a}}-\frac{1}{\rho^{b}}\right)\,,\label{eq:PT-relative-velocity}
\end{equation}
which may also be written as
\begin{equation}
\left[\left[\mathbf{v}\right]\right]\cdot\mathbf{n}^{\Gamma}=M\bar{\zeta}\left[\left[\frac{1}{\rho}\right]\right]\,.\label{eq:PT-relvel-redux}
\end{equation}

\begin{example}
\label{exa:water-evaporates}Consider that $\mathcal{S}^{a}$ is liquid
water ($a=\ell$) and $\mathcal{S}^{b}$ is water vapor ($b=v$),
such that the reaction represents the evaporation of water (hence,
$\bar{\zeta}>0$ by assumption). Assume that the liquid is stationary
($\mathbf{v}^{\ell}=\mathbf{0}$) even while its boundary $\Gamma$
is receding as the water evaporates. From (\ref{eq:PT-interface-velocity})
we can solve for the receding velocity, $\mathbf{v}^{\Gamma}\cdot\mathbf{n}^{\Gamma}=-M\bar{\zeta}/\rho^{\ell}$.
We also find from (\ref{eq:PT-relative-velocity}) that $\mathbf{v}^{v}\cdot\mathbf{n}^{\Gamma}=\left(\frac{1}{\rho^{v}}-\frac{1}{\rho^{\ell}}\right)M\bar{\zeta}$;
since $\rho^{\ell}>\rho^{v}$ for liquid and vapor, the latter result
shows that $\mathbf{v}^{v}\cdot\mathbf{n}^{\Gamma}>0$, i.e., the
vapor water moves away from the liquid to make up for the increase
in volume as liquid water transforms into vapor.
\end{example}

\subsection{Momentum Balance Jump\label{subsec:Momentum-Balance-Jump-1}}

For our choice of mixtures on both sides of the phase boundary $\Gamma$,
the momentum jump condition (\ref{eq:momentum-jump-redux}) becomes
\[
\left(p^{a}-p^{b}\right)\mathbf{n}^{\Gamma}+\underbrace{\rho^{a}\left(\mathbf{v}^{a}-\mathbf{v}^{\Gamma}\right)\cdot\mathbf{n}^{\Gamma}}_{=M\bar{\zeta}}\left(\mathbf{v}^{a}-\mathbf{v}^{\Gamma}\right)-\underbrace{\rho^{b}\left(\mathbf{v}^{b}-\mathbf{v}^{\Gamma}\right)\cdot\mathbf{n}^{\Gamma}}_{=M\bar{\zeta}}\left(\mathbf{v}^{b}-\mathbf{v}^{\Gamma}\right)=\mathbf{0}\,.
\]
Taking the dot product of this expression with $\mathbf{n}^{\Gamma}$
and using the general relation of (\ref{eq:PT-relative-velocity})
produces the jump condition on the pressure,
\begin{equation}
\left[\left[p\right]\right]=-\left(M\bar{\zeta}\right)^{2}\left[\left[\frac{1}{\rho}\right]\right]\,.\label{eq:PT-pressure-jump}
\end{equation}
This result shows that the jump in pressure results from differences
in the densities of phases $\mathcal{S}^{a}$ and $\mathcal{S}^{b}$
of $\mathcal{S}$. It is also proportional to the square of the mass
flux from the reaction. When $\rho^{b}<\rho^{a}$, we find that $p^{a}>p^{b}$.
Thus, if $a$ is the liquid phase and $b$ is the vapor phase, the
liquid pressure is higher than the vapor pressure regardless of the
direction of the phase transformation (i.e., evaporation or condensation).
At phase equilibrium ($M\bar{\zeta}=0$), there is no jump in pressure
across $\Gamma$.

\subsection{Energy Balance Jump\label{subsec:Energy-Balance-Jump-1}}

So far, no constraint has emerged on the continuity of $T$ across
$\Gamma$. The same argument applies to the gradient $\mathbf{g}$
of $T$, implying that the heat flux $\mathbf{q}$ may exhibit a discontinuity
across $\Gamma$. Now, the energy jump may be reduced from (\ref{eq:energy-jump-redux})
to
\begin{equation}
M\bar{\zeta}\left[\left[h+\frac{1}{2}\mathbf{u}^{\Gamma}\cdot\mathbf{u}^{\Gamma}\right]\right]+\left[\left[\mathbf{q}\right]\right]\cdot\mathbf{n}^{\Gamma}=0\,,\label{eq:phase-transformation-energy}
\end{equation}
where $\left[\left[h\right]\right]=h^{a}-h^{b}$ and $\left[\left[\mathbf{q}\right]\right]=\mathbf{q}^{a}-\mathbf{q}^{b}$.
This form of the energy jump condition shows that when $M\bar{\zeta}=0$
there can be no jump in the heat flux, $\left[\left[\mathbf{q}\right]\right]\cdot\mathbf{n}^{\Gamma}=0$.
In other words, the heat flux is continuous across $\Gamma$, $\mathbf{q}^{a}\cdot\mathbf{n}^{\Gamma}=\mathbf{q}^{b}\cdot\mathbf{n}^{\Gamma}$,
only at phase equilibrium. This is consistent with our expectation
that a phase transformation requires the exchange of energy in the
form of heat.

Since $h=u+p/\rho$ in general, we may also write
\begin{equation}
\left[\left[h\right]\right]=\left[\left[u\right]\right]+\left[\left[\frac{p}{\rho}\right]\right]=\left[\left[u\right]\right]+\frac{1}{\rho^{\Gamma}}\left[\left[p\right]\right]+p^{\Gamma}\left[\left[\frac{1}{\rho}\right]\right]\,,\label{eq:h-jump-split}
\end{equation}
where
\begin{equation}
\frac{1}{\rho^{\Gamma}}\equiv\frac{1}{2}\left(\frac{1}{\rho^{a}}+\frac{1}{\rho^{b}}\right)\,,\quad p^{\Gamma}\equiv\frac{1}{2}\left(p^{a}+p^{b}\right)\,\label{eq:interface-density-pressure}
\end{equation}
may be viewed as the respective values of $1/\rho$ and $p$ on $\Gamma$.
Using the jump condition on the pressure found in (\ref{eq:PT-pressure-jump}),
this expression for $\left[\left[h\right]\right]$ may be rewritten
as
\begin{equation}
\left[\left[h\right]\right]=\left[\left[u\right]\right]+\left(p^{\Gamma}-\frac{1}{\rho^{\Gamma}}\left(M\bar{\zeta}\right)^{2}\right)\left[\left[\frac{1}{\rho}\right]\right]\,.\label{eq:phase-transformation-enthalpy}
\end{equation}

Now consider that there is no slip between the two phases $\mathcal{S}^{a}$
and $\mathcal{S}^{b}$ on $\Gamma$, so that $\mathbf{u}^{\Gamma}=\left(\mathbf{u}^{\Gamma}\cdot\mathbf{n}^{\Gamma}\right)\mathbf{n}^{\Gamma}$
(i.e., the tangential component of $\mathbf{u}^{\Gamma}$ on $\Gamma$
is zero). In that case, we may use the mass jump condition (\ref{eq:PT-mass-jump})
and the relations of (\ref{eq:interface-density-pressure}) to get
\begin{equation}
\left[\left[\frac{1}{2}\mathbf{u}^{\Gamma}\cdot\mathbf{u}^{\Gamma}\right]\right]=\frac{1}{\rho^{\Gamma}}\left(M\bar{\zeta}\right)^{2}\left[\left[\frac{1}{\rho}\right]\right]\,.\label{eq:PT-diffusive-KE-jump}
\end{equation}
It follows from this result and (\ref{eq:PT-latent-heat-def}) that
\begin{equation}
\left[\left[h+\frac{1}{2}\mathbf{u}^{\Gamma}\cdot\mathbf{u}^{\Gamma}\right]\right]=\left[\left[u\right]\right]+p^{\Gamma}\left[\left[\frac{1}{\rho}\right]\right]\,,\label{eq:PT-outer-enthalpy-jump}
\end{equation}
so that the jump condition (\ref{eq:phase-transformation-energy})
on the energy balance may be rewritten in its final form as
\begin{equation}
M\bar{\zeta}\left(\left[\left[u\right]\right]+p^{\Gamma}\left[\left[\frac{1}{\rho}\right]\right]\right)+\left[\left[\mathbf{q}\right]\right]\cdot\mathbf{n}^{\Gamma}=0\,.\label{eq:PT-energy-redux}
\end{equation}

By definition, the \emph{specific latent hea}t $L$ for the phase
transformation of $\mathcal{S}^{a}$ into $\mathcal{S}^{b}$ is the
ratio of the net heat flux across the phase boundary, $\left[\left[\mathbf{q}\right]\right]\cdot\mathbf{n}^{\Gamma}$,
to the mass flux $M\bar{\zeta}$ produced in the phase transformation,
\begin{equation}
L=-\left(\left[\left[u\right]\right]+p^{\Gamma}\left[\left[\frac{1}{\rho}\right]\right]\right)=\frac{\left[\left[\mathbf{q}\right]\right]\cdot\mathbf{n}^{\Gamma}}{M\bar{\zeta}}\,.\label{eq:PT-latent-heat-def}
\end{equation}
The units of $L$ are those of energy per mass. Alternatively, we
may define the molar latent heat of transformation as $\bar{L}=ML$.
An important observation is that the definition of $L$ in (\ref{eq:PT-latent-heat-def})
is valid for arbitrary processes.
\begin{example}
Consider a phase transformation which takes place at a very slow,
but non-zero rate $M\bar{\zeta}$. According to the momentum jump
condition (\ref{eq:PT-pressure-jump}), in the limit when $M\bar{\zeta}\approx0$,
we find that
\begin{equation}
\left(M\bar{\zeta}\right)^{2}\left[\left[\frac{1}{\rho}\right]\right]=\left[\left[p\right]\right]\approx0\,,\label{eq:slow-phase-transformation}
\end{equation}
implying that $p^{a}\approx p^{b}\approx p^{\Gamma}$. In this limiting
case it follows from (\ref{eq:PT-latent-heat-def}) that
\begin{equation}
L_{0}=\lim_{M\bar{\zeta}\to0}L=-\left[\left[h\right]\right]\,.\label{eq:tabulated-latent-heat}
\end{equation}
Thus, in the limit of phase transformation equilibrium, the specific
latent heat is equal to the jump in enthalpy across $\Gamma$. Indeed,
the specific latent heat of transformation is often called the \emph{specific
enthalpy of formation}. Generally, it is $L_{0}$ (not $L$) which
is tabulated along with other thermodynamic properties of a substance.
In this limiting case, we may estimate the slow reactive mass flux
from
\begin{equation}
M\bar{\zeta}\approx-\frac{\left[\left[\mathbf{q}\right]\right]\cdot\mathbf{n}^{\Gamma}}{\left[\left[h\right]\right]}\,,\label{eq:PT-slow-mass-flux}
\end{equation}
where $\left[\left[\mathbf{q}\right]\right]\cdot\mathbf{n}^{\Gamma}$
is evaluated from the knowledge of the temperature gradient and thermal
conductivities in the domains on either side of $\Gamma$. The expression
of (\ref{eq:PT-slow-mass-flux}) is the classical Stefan condition
\citep{Stefan91,Ward04}.
\begin{example}
Let $\mathcal{S}^{a}$ represent liquid water and $\mathcal{S}^{b}$
represent vapor water at phase equilibrium ($T^{\ell}=T^{v}$) and
atmospheric pressure. The jump in enthalpy for water at these conditions
is $\left[\left[h\right]\right]=h^{\ell}-h^{v}=-2256\,\text{kJ}/\text{kg}$.
It follows from (\ref{eq:tabulated-latent-heat}) that the specific
enthalpy of evaporation is $L_{0}=2256\,\text{kJ}/\text{kg}$ at atmospheric
pressure.
\end{example}

\end{example}

\subsection{Entropy Inequality Jump\label{subsec:Entropy-Inequality-Jump-1}}

Using the above result (\ref{eq:PT-mass-jump}) for the mass jump
condition, the jump condition (\ref{eq:entropy-jump}) on the entropy
inequality across $\Gamma$ reduces to
\begin{equation}
M\bar{\zeta}\left[\left[s\right]\right]+\left[\left[\frac{1}{T}\mathbf{q}\right]\right]\cdot\mathbf{n}^{\Gamma}\le0\,,\label{eq:entropy-jump-latent-1}
\end{equation}
where $\left[\left[s\right]\right]=s^{a}-s^{b}$. Since the ratio
$\mathbf{q}/T$ emerges naturally in the entropy inequality, we may
call it the \emph{thermal entropy flux}. This constraint informs us
of the circumstances under which a phase transformation may occur.
Since $T$ and $\mathbf{q}$ may both be discontinuous across $\Gamma$,
we can express the jump in thermal entropy flux in the direction normal
to $\Gamma$ as
\begin{equation}
\left[\left[\frac{1}{T}\mathbf{q}\right]\right]\cdot\mathbf{n}^{\Gamma}=\frac{1}{T^{\Gamma}}\left[\left[\mathbf{q}\right]\right]\cdot\mathbf{n}^{\Gamma}+\left[\left[\frac{1}{T}\right]\right]\mathbf{q}^{\Gamma}\cdot\mathbf{n}^{\Gamma}\,,\label{eq:split-jump-product}
\end{equation}
where
\begin{equation}
\frac{1}{T^{\Gamma}}\equiv\frac{1}{2}\left(\frac{1}{T^{a}}+\frac{1}{T^{b}}\right)\,,\quad\mathbf{q}^{\Gamma}\equiv\frac{1}{2}\left(\mathbf{q}^{a}+\mathbf{q}^{b}\right)\,\label{eq:split-jump-variables}
\end{equation}
represent the respective values of $1/T$ and $\mathbf{q}$ on $\Gamma$.
Then (\ref{eq:entropy-jump-latent-1}) may be rewritten as
\begin{equation}
T^{\Gamma}M\bar{\zeta}\left[\left[s\right]\right]+\left[\left[\mathbf{q}\right]\right]\cdot\mathbf{n}^{\Gamma}+\left[\left[\frac{1}{T}\right]\right]T^{\Gamma}\mathbf{q}^{\Gamma}\cdot\mathbf{n}^{\Gamma}\le0\,,\label{eq:entropy-jump-latent-2}
\end{equation}
where we have used the fact that $T^{\Gamma}>0$. Negating this inequality
and adding it to the energy jump (\ref{eq:PT-energy-redux}) to eliminate
$\left[\left[\mathbf{q}\right]\right]\cdot\mathbf{n}^{\Gamma}$, yields
the final general form of the constraint placed by the entropy inequality
on phase transformations on the interface $\Gamma$,
\begin{equation}
M\bar{\zeta}\left(\left[\left[u\right]\right]+p^{\Gamma}\left[\left[\frac{1}{\rho}\right]\right]-T^{\Gamma}\left[\left[s\right]\right]\right)-\left[\left[\frac{1}{T}\right]\right]T^{\Gamma}\mathbf{q}^{\Gamma}\cdot\mathbf{n}\ge0\,.\label{eq:entropy-jump-latent-3}
\end{equation}

To better understand this non-trivial constraint it is beneficial
to examine additional limiting cases. When the phase transformation
reaction equilibrates ($M\bar{\zeta}=0$, implying $\left[\left[\mathbf{q}\right]\right]\cdot\mathbf{n}^{\Gamma}=0$
from the energy jump (\ref{eq:PT-energy-redux})), the constraint
of (\ref{eq:entropy-jump-latent-2}) simplifies to
\begin{equation}
\left[\left[\frac{1}{T}\right]\right]\mathbf{q}^{\Gamma}\cdot\mathbf{n}^{\Gamma}\le0\,,\quad\text{at phase equilibrium}\label{eq:entropy-jump-no-transformation}
\end{equation}
where $\mathbf{q}^{\Gamma}\cdot\mathbf{n}^{\Gamma}=\mathbf{q}^{a}\cdot\mathbf{n}^{\Gamma}=\mathbf{q}^{b}\cdot\mathbf{n}^{\Gamma}$.
Thus, if a jump in $T$ exists, it must satisfy this constraint whenever
the heat flux $\mathbf{q}^{\Gamma}\cdot\mathbf{n}^{\Gamma}$ is non-zero;
for example, if $\mathbf{q}^{\Gamma}\cdot\mathbf{n}^{\Gamma}>0$ (net
heat flowing from the domain of $\mathcal{S}^{a}$ to the domain of
$\mathcal{S}^{b}$) we must have $T^{a}>T^{b}$. However, when $\mathbf{q}^{\Gamma}\cdot\mathbf{n}^{\Gamma}=0$
(e.g., when $\mathbf{q}^{a}\cdot\mathbf{n}=\mathbf{q}^{b}\cdot\mathbf{n}=0$
as in the normal shock wave problem in Section~\ref{subsec:Normal-Shock-Wave}),
the sign of $\left[\left[1/T\right]\right]$ is inconsequential. Conversely,
when the temperature is continuous across $\Gamma$ ($T^{a}=T^{b}$),
the sign of $\mathbf{q}^{\Gamma}\cdot\mathbf{n}^{\Gamma}$ is inconsequential
(i.e., the net heat flux may occur in either direction). We also conclude
that \emph{continuity of the temperature across }$\Gamma$ (i.e.,
$\left[\left[1/T\right]\right]=0$) is sufficient to satisfy the entropy
inequality \emph{at phase equilibrium}, but it\emph{ is not a necessary
condition}. As is evident from (\ref{eq:entropy-jump-no-transformation})
and the examples described in this paragraph, phase equilibrium may
exist over a broad range of conditions.

Now, consider that a phase transformation occurs at very slow but
non-zero rate $M\bar{\zeta}$ as defined in (\ref{eq:PT-slow-mass-flux}).
It follows from the momentum jump condition (\ref{eq:PT-pressure-jump})
that $\left[\left[p\right]\right]\approx0$. Consider that the phase
transformation is taking place very close to the phase equilibrium
state of the reversible reaction (\ref{eq:reversible-phase-transformation})
and that $\left[\left[1/T\right]\right]\approx0$ (i.e., assume that
$T$ is continuous across $\Gamma$, consistent with the common assumption
of phases transitions in the limit of reversibility). Now, (\ref{eq:entropy-jump-latent-3})
simplifies to
\begin{equation}
M\bar{\zeta}\left[\left[g\right]\right]\ge0\,,\quad\text{near phase equilibrium}\label{eq:Gibbs-inequality-constraint}
\end{equation}
where $g\equiv h-Ts$ is the definition of the \emph{specific Gibbs
function} (also called the \emph{specific free enthalpy}, since $g=a+p/\rho$).
In other words, the Gibbs function emerges naturally from the entropy
inequality constraint during phase transformations on an interface
$\Gamma$, \emph{in the limit as $M\bar{\zeta}\to0$. }The inequality
(\ref{eq:Gibbs-inequality-constraint}) represents the thermodynamic
constraint that must be satisfied by the constitutive models for $g^{a}$
and $g^{b}$ (and thus, the specific free energies $a^{a}$ and $a^{b}$)
in order for the phase transformation to occur on $\Gamma$, at sufficiently
slow rates $M\bar{\zeta}$, in the neighborhood of phase equilibrium.
For the forward reaction $\mathcal{S}^{a}\to\mathcal{S}^{b}$ (with
$\bar{\zeta}>0$), we must have $g^{a}\ge g^{b}$ according to (\ref{eq:Gibbs-inequality-constraint});
for the reverse reaction $\mathcal{S}^{b}\to\mathcal{S}^{a}$ ($\bar{\zeta}<0$),
we must have $g^{a}\le g^{b}$. In order for the phase transformation
reaction to proceed slowly in either direction starting from the equilibrium
state, we may argue that $\left[\left[g\right]\right]=0$ at phase
equilibrium (even though this constraint is not strictly needed to
satisfy (\ref{eq:Gibbs-inequality-constraint}) when $M\bar{\zeta}=0$
exactly). Therefore, continuity of the specific Gibbs function across
the phase boundary is a sufficient condition for allowing phase transformations
to initiate in either direction (forward or reverse), when the temperature
is continuous across $\Gamma$. This thermodynamic requirement implies
that the saturation curve between $\mathcal{S}^{a}$ and $\mathcal{S}^{b}$
may be identified as the line along which $\left[\left[g\right]\right]=0$
is satisfied simultaneously with $\left[\left[1/T\right]\right]=0$.
This is a classical result of the thermodynamics of phase transformations;
this presentation emphasizes the special set of conditions under which
this result holds.

We may now argue that (\ref{eq:entropy-jump-no-transformation}) describes
the general spectrum of metastable phase equilibrium conditions, of
which one special case, corresponding to $\left[\left[g\right]\right]=0$
and $\left[\left[1/T\right]\right]=0$, represents a stable phase
equilibrium state, based on the reasoning outlined in the previous
paragraph. It also bears to emphasize that the condition (\ref{eq:Gibbs-inequality-constraint})
that lets us identify the saturation curve is sufficient but not necessary
for phase transformations to occur, as it does not represent the general
case of allowable phase transformations, which are governed by (\ref{eq:entropy-jump-latent-1})
or its equivalent forms (\ref{eq:entropy-jump-latent-2}) and (\ref{eq:entropy-jump-latent-3}).
A complete description of the kinetics of phase transformations requires
the formulation of a constitutive model for the reactive mass flux
$M\bar{\zeta}$, which must rely on experimental validation and simultaneously
satisfy (\ref{eq:entropy-jump-latent-1}). Such a constitutive model
is proposed further below. Prior to that however, now that we have
established that $\left[\left[g\right]\right]=0$ and $\left[\left[1/T\right]\right]=0$
at stable phase equilibrium, we first need to clarify the relations
between the reference free energies and entropies of the liquid and
vapor phases.

To the best of our knowledge, the derivations presented in this section
are original, as we could not find a similar presentation in the prior
literature. They represent one of the major findings of this study.

\subsection{Reference Configurations for Liquid and Vapor Phases\label{subsec:Reference-Configurations}}

We are now ready to figure out how to set the reference configuration
of liquid and vapor phases of a pure substance, taking into account
the phase equilibrium conditions outlined in the previous section.
Recall that we previously proposed constitutive relations for the
specific free energy $a\left(T,J\right)$ of real gases (Section~\ref{sec:Real-Gases})
and liquids (Section~\ref{sec:Liquids}). In both cases we explained
that we could evaluate $a\left(T,J\right)$ as well as the specific
entropy $s\left(T,J\right)$ to within arbitrary constants $a_{r}$
and $s_{r}$, respectively. Now consider that we use these constitutive
relations for the liquid and vapor phases of a pure substance, denoting
their specific free energies as $a^{\ell}$ and $a^{v}$, respectively.

The proper way to set up a common reference configuration for the
vapor and liquid phases of a pure substance is to select the same
reference pressure $P_{r}$ and temperature $T_{r}$ for both phases.
The most logical choice for a common reference is the triple point
of the pure substance, where vapor, liquid and solid phases coexist,
since this also allows us to account for phase transformations between
the solid and liquid or the solid and vapor phases, as needed. We
now need to determine the relation between $a_{r}^{v}$ and $a_{r}^{\ell}$
at the triple point. To figure it out, we use the phase equilibrium
condition $\left[\left[g\right]\right]=0$ for the reversible phase
transformation $\mathcal{S}^{\ell}\rightleftharpoons\mathcal{S}^{v}$
which was deduced in Section~\ref{sec:Phase-Transformations}, evaluated
specifically at the triple point. Recall that the specific Gibbs function
is related to the specific free energy via $g=h-Ts=a+p/\rho$ where,
in our approach, $p$ represents the gauge pressure. Since $p=0$
in the reference configuration (by definition), the jump condition
$\left[\left[g_{r}\right]\right]=0$ reduces to $\left[\left[a_{r}\right]\right]=0$
at the triple point, which is equivalent to stating that
\begin{equation}
a_{r}^{\ell}=a_{r}^{v}\,.\label{eq:liquid-gas-a-ref}
\end{equation}
Since this reference configuration is arbitrary, we can select $a_{r}^{\ell}=a_{r}^{v}=0$.
\begin{rem}
In conventional thermodynamics where the absolute pressure $P$ is
used instead of the gauge pressure, using $g=a+P/\rho$ and $\left[\left[g_{r}\right]\right]=0$
at the triple point would produce
\begin{equation}
a_{r}^{\ell}+\frac{P_{r}}{\rho_{r}^{\ell}}=a_{r}^{v}+\frac{P_{r}}{\rho_{r}^{v}}\,\text{conventional.}\label{eq:g-jump-classical}
\end{equation}
Combining this result with the conventional choice of $a_{r}^{\ell}=0$
produces
\begin{equation}
a_{r}^{v}=P_{r}\left(\frac{1}{\rho_{r}^{\ell}}-\frac{1}{\rho_{r}^{v}}\right)\,\text{conventional.}\label{eq:a-vapor-classical}
\end{equation}
Thus, contrary to our gauge pressure approach, the referential free
energies of liquid and vapor phases are not equal to each other in
the classical framework, $\left[\left[a_{r}\right]\right]\ne0$. In
practice, this means that the specific free energy from our real gas
constitutive model in Section~\ref{sec:Real-Gases} will deviate
from the conventionally tabulated values $a^{v}=u^{v}-Ts^{v}$ by
an amount comparable to or smaller in magnitude than the value of
$a_{r}^{v}$ obtained from standard tables.
\end{rem}

To determine the reference values $s_{r}^{v}$ and $s_{r}^{\ell}$
for the specific entropies, we examine the jump condition on the energy
balance in the limit as the phase transformation approaches equilibrium.
As shown in (\ref{eq:tabulated-latent-heat}), we find that the latent
heat of transformation, in the limit as the phase transformation reaction
equilibrates, is equal to the jump in enthalpy, $\left[\left[h\right]\right]=-L_{0}$.
Using the gauge enthalpy $h=u+p/\rho$, we can examine this jump condition
at the triple point reference configuration, where $p=0$, so that
\begin{equation}
L_{r}=-\left[\left[u_{r}\right]\right]=u_{r}^{v}-u_{r}^{\ell}=\underbrace{a_{r}^{v}-a_{r}^{\ell}}_{=-\left[\left[a_{r}\right]\right]=0}+T_{r}\underbrace{\left(s_{r}^{v}-s_{r}^{\ell}\right)}_{=-\left[\left[s_{r}\right]\right]}\,,\label{eq:latent-heat-s-jump}
\end{equation}
where $L_{r}$ is the latent heat of evaporation near phase equilibrium
at the triple point. If we let $s_{r}^{\ell}=0$ as is the convention,
then $s_{r}^{v}=L_{r}/T_{r}$. Thus, unlike the referential specific
free energy, the jump in referential specific entropy is not generally
equal to zero in our gauge pressure approach. The evaluation of $s_{r}^{v}$
is contingent on the experimental measurement of $L_{r}$. Importantly,
a single measurement of the latent heat of evaporation (e.g., $L_{r}$
at the triple point) is sufficient to predict $L_{0}$ along the entire
saturation curve.
\begin{rem}
In classical thermodynamics we use $h=u+P/\rho$ to evaluate the specific
enthalpy. Thus, at the triple point reference configuration we would
have
\begin{equation}
L_{r}=u_{r}^{v}-u_{r}^{\ell}+P_{r}\left(\frac{1}{\rho_{r}^{v}}-\frac{1}{\rho_{r}^{\ell}}\right)=\underbrace{a_{r}^{v}-a_{r}^{\ell}+P_{r}\left(\frac{1}{\rho_{r}^{v}}-\frac{1}{\rho_{r}^{\ell}}\right)}_{=-\left[\left[g_{r}\right]\right]=0}+T_{r}\left(s_{r}^{v}-s_{r}^{\ell}\right)\,,\,\text{conventional}\label{eq:latent-heat-s-classical}
\end{equation}
where we used the phase equilibrium condition $\left[\left[g_{r}\right]\right]=0$
at the triple point, with the classical definition of $g_{r}=a_{r}+P_{r}/\rho_{r}$.
Comparing (\ref{eq:latent-heat-s-jump}) to (\ref{eq:latent-heat-s-classical}),
it becomes evident that the result from our gauge pressure approach
for the jump $\left[\left[s\right]\right]$ across the saturation
curve is the same as in the conventional approach. It follows from
a similar argument using (\ref{eq:tabulated-latent-heat}) that the
jump in enthalpy $\left[\left[h\right]\right]=h^{\ell}-h^{v}$ across
the saturation curve has the same value in conventional thermodynamics
(where absolute pressure is used) and our approach (where gauge pressure
is used).
\begin{figure}
\begin{centering}
\subfloat[]{\begin{centering}
\includegraphics[width=3.13in]{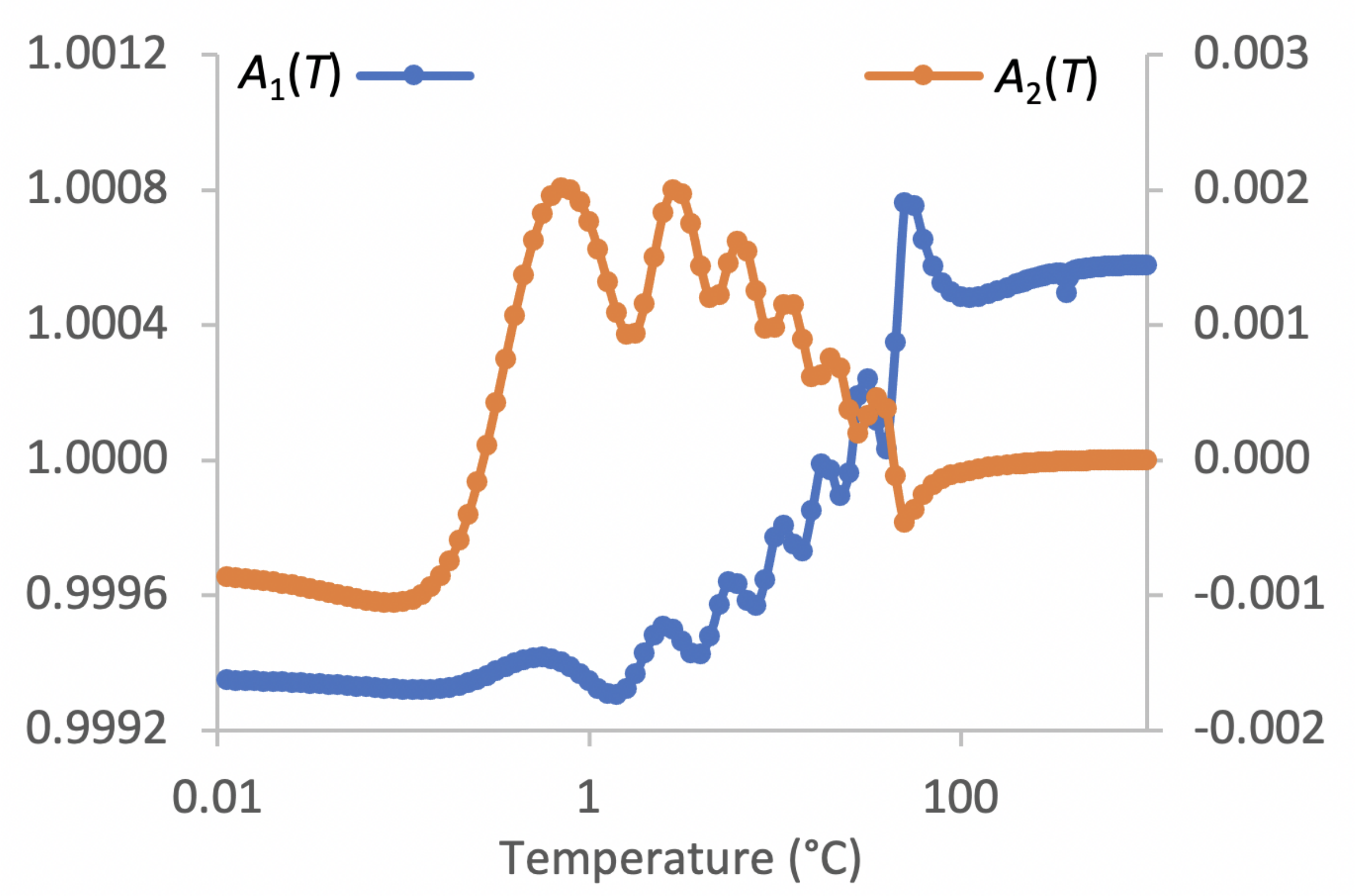}
\par\end{centering}
}
\par\end{centering}
\begin{centering}
\subfloat[]{\begin{centering}
\includegraphics[width=3.13in]{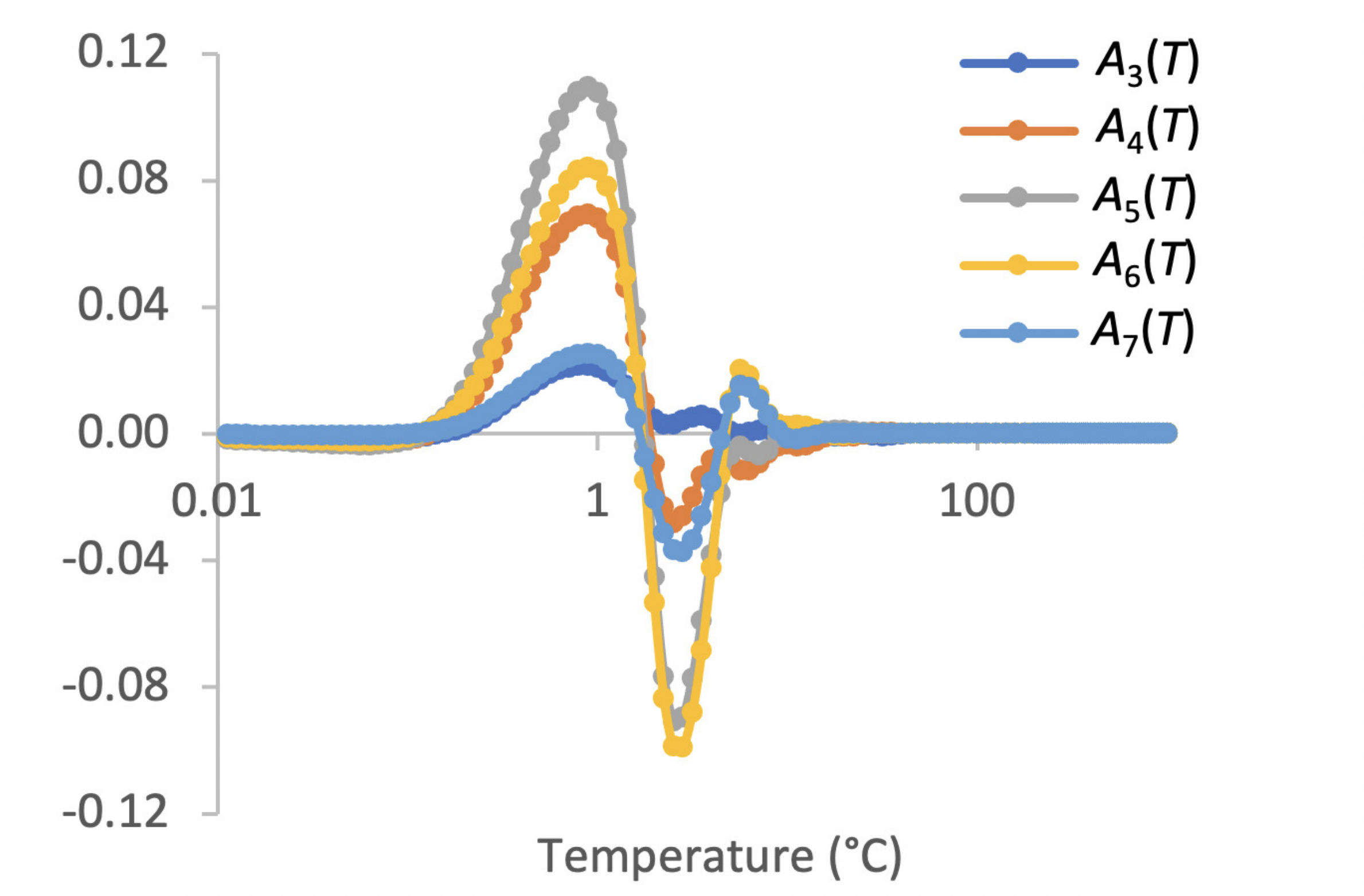}
\par\end{centering}
}
\par\end{centering}
\caption{Virial coefficients $A_{1-7}\left(T\right)$ for water vapor, plotted
against the temperature in $\text{°C}$ on a logarithmic scale. Symbols
represent values obtained by fitting $p/P_{r}$ in equation (\ref{eq:rg-gage-pressure})
versus $\frac{T}{JT_{r}}-1$ to pressure data downloaded from NIST,
at selected temperatures in the range $0.01$ to $1000\,\text{°C}$.
Solid curves represent interpolations of those coefficients. \label{fig:H2Ovapor-virial-coeffs}}
\end{figure}
\end{rem}

\begin{example}
\label{exa:Vapor-water-properties}As done in Example~\ref{exa:Liquid-water-properties}
with liquid water, we download the properties of water vapor at various
pressures and temperatures (https://webbook.nist.gov/chemistry/fluid/)
and fit the virial expansion (\ref{eq:rg-gage-pressure}) for the
pressure in real gases to this data set to obtain discrete values
of the virial coefficients $A_{k}\left(T\right)$ at selected values
of $T$ in the range $0.01-1000\,{^\circ}\text{C}$. These discrete
sets are then interpolated using piecewise cubic polynomials and differentiated
as needed. By trial and error, it is determined that seven virial
coefficients ($m=7$ in (\ref{eq:rg-gage-pressure})) can accurately
reproduce the response of water vapor (\figurename~\ref{fig:H2Ovapor-virial-coeffs}).
Similarly, we download the isobaric specific heat capacity $c_{pr}\left(T\right)$
at the triple-point pressure $P_{r}$ for the same range of temperatures
and use those values to integrate $a_{0}^{\prime\prime}\left(T\right)$
in (\ref{eq:rg-cp}) twice with respect to $T$, assuming that $a_{r}^{v}=0$
and $s_{r}^{v}=L_{r}/T_{r}=9.15549341\,\text{kJ}/\text{kg}$, to obtain
$s_{0}^{v}\left(T\right)$ in (\ref{eq:rg-s-circle}) and $a_{0}^{v}\left(T\right)$
in (\ref{eq:rg-a-circle}). We then evaluate $a^{v}\left(T,J\right)$
from (\ref{eq:rg-a-final}) for any $\left(T,J\right)$, such as those
on the saturation curve, and compare those values to the NIST-tabulated
specific free energy (where $a_{r}^{v}=-125.994900\,\text{kJ}/\text{kg}$),
showing that they initially deviate from each other, then become nearly
identical at temperatures above $50\,\text{°C}$ (\figurename~\ref{fig:H2Ovapor-properties-1}a).
We also evaluate $s^{v}\left(T,J\right)$ by substituting (\ref{eq:rg-a-final})
into (\ref{eq:entropy-constraint}) and verify that its values along
the saturation curve agree with the NIST-tabulated values (\figurename~\ref{fig:H2Ovapor-properties-1}b).
We similarly compare our calculated specific internal energy $u^{v}=a^{v}+Ts^{v}$
on the saturation curve to the NIST-tabulated values, showing an initial
deviation at temperatures below $50\,\text{°C}$ consistent with the
difference noted in $a^{v}$ (\figurename~\ref{fig:H2Ovapor-properties-2}a).
Finally, the specific gauge enthalpy $h^{v}=u^{v}+p^{v}/\rho^{v}$
is evaluated in our approach and compared to the tabulated value of
the specific enthalpy, confirming that these measures are identical
(\figurename~\ref{fig:H2Ovapor-properties-2}b).
\begin{figure}
\begin{centering}
\subfloat[]{\includegraphics[width=3.25in]{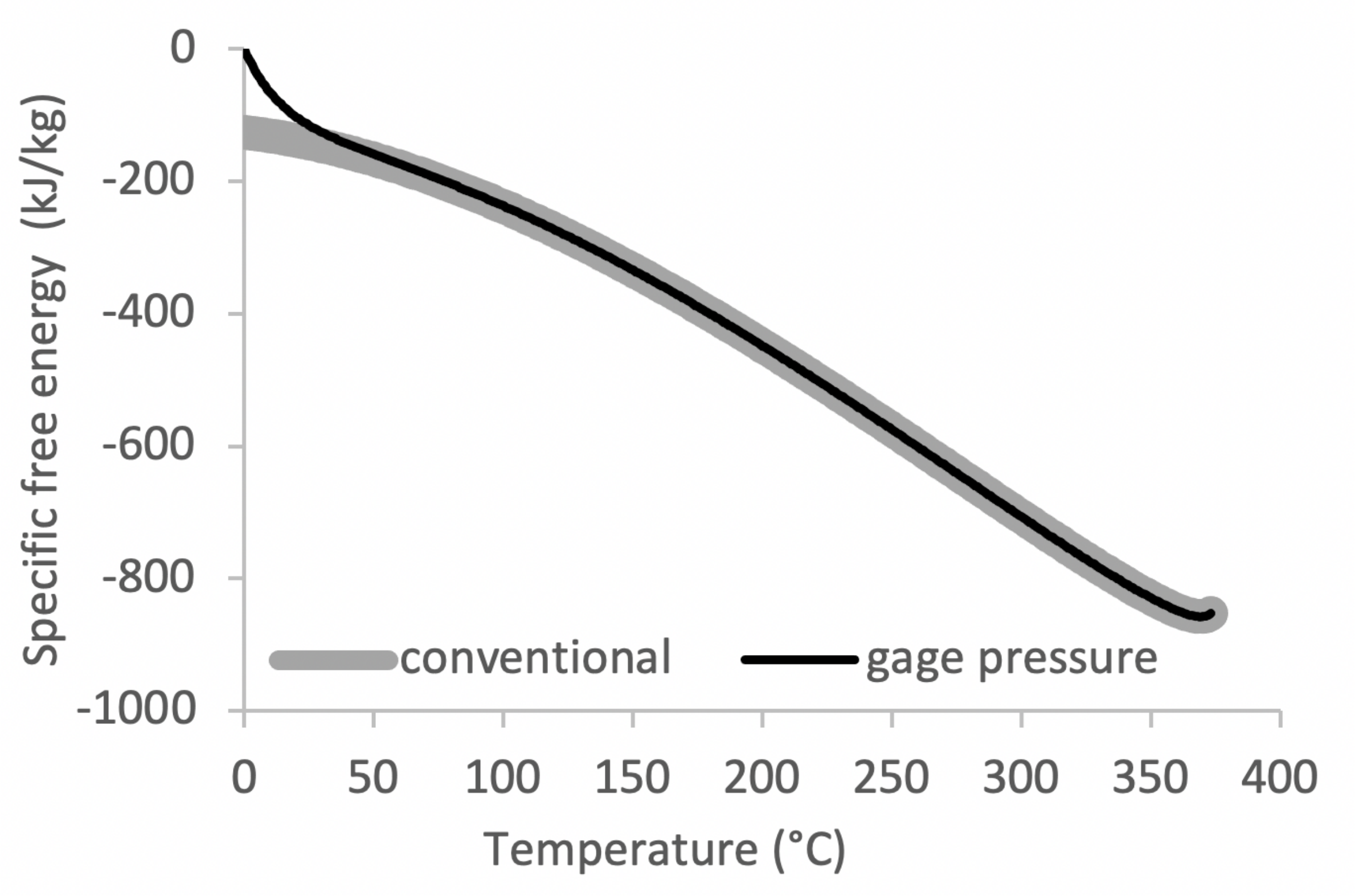}

}
\par\end{centering}
\begin{centering}
\subfloat[]{\includegraphics[width=3.25in]{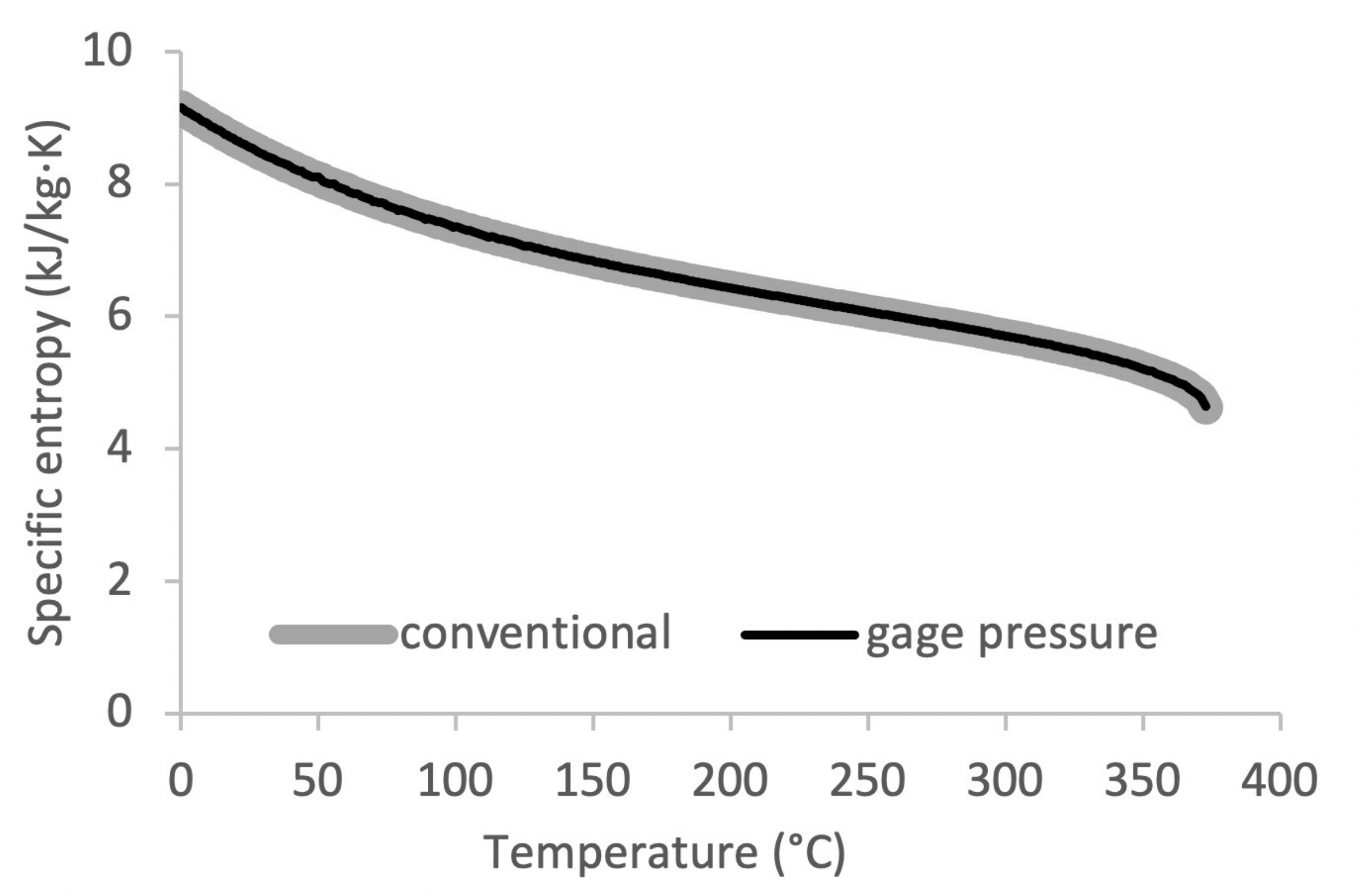}

}
\par\end{centering}
\caption{Thermodynamic properties of water vapor on the saturation curve, comparing
conventional values from standard tables (NIST, https://webbook.nist.gov/chemistry/fluid/)
to those evaluated in this study, where gauge pressure is used instead
of absolute pressure. (a) The specific free energy $a^{v}$ in our
approach starts from zero in the reference configuration (at the triple
point), in contrast to the value from standard tables. (b) The specific
entropy $s^{v}$ is identical in both approaches.\label{fig:H2Ovapor-properties-1}}
\end{figure}
\begin{figure}
\begin{centering}
\subfloat[]{\includegraphics[width=3.25in]{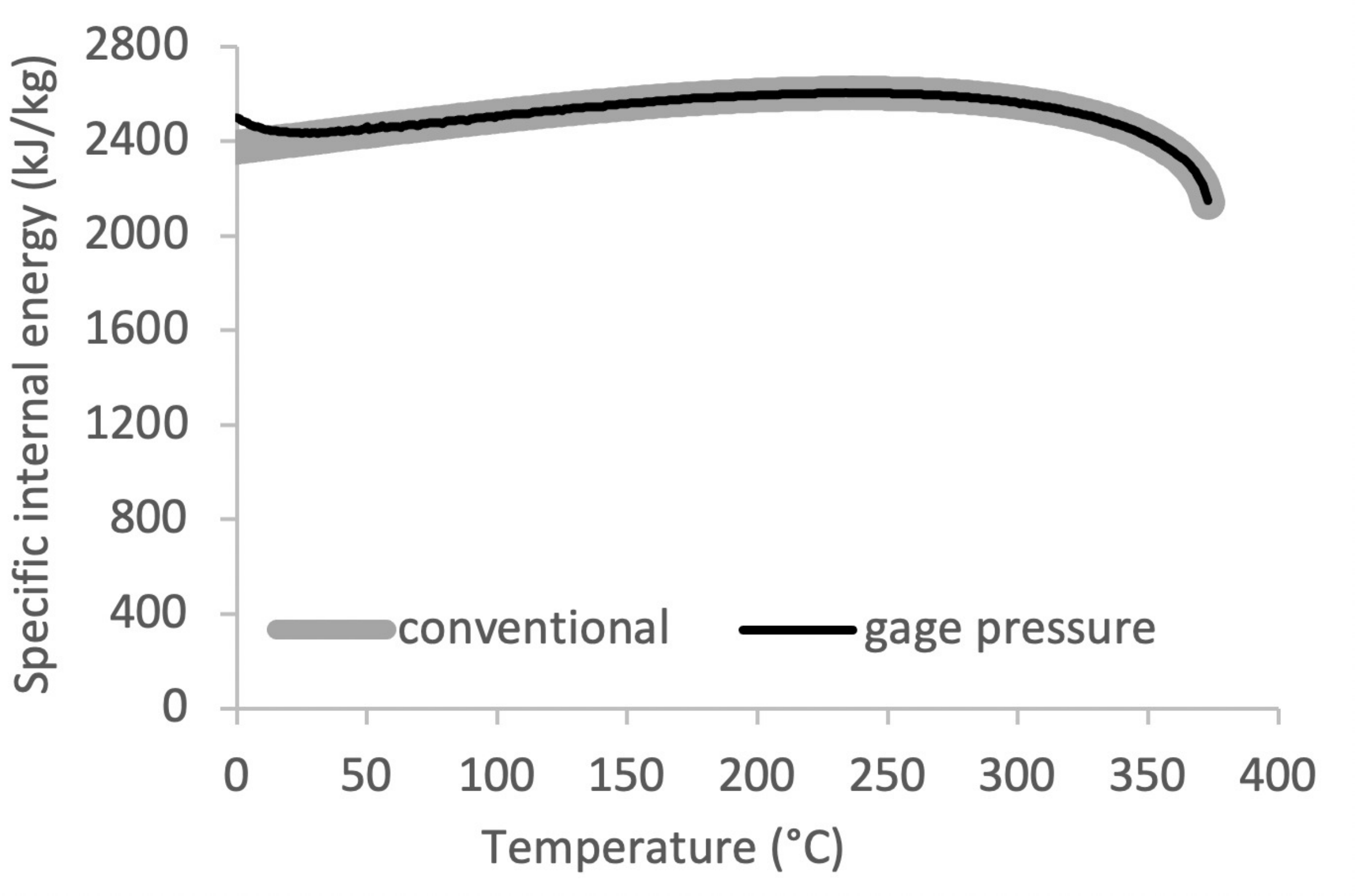}

}
\par\end{centering}
\begin{centering}
\subfloat[]{\includegraphics[width=3.25in]{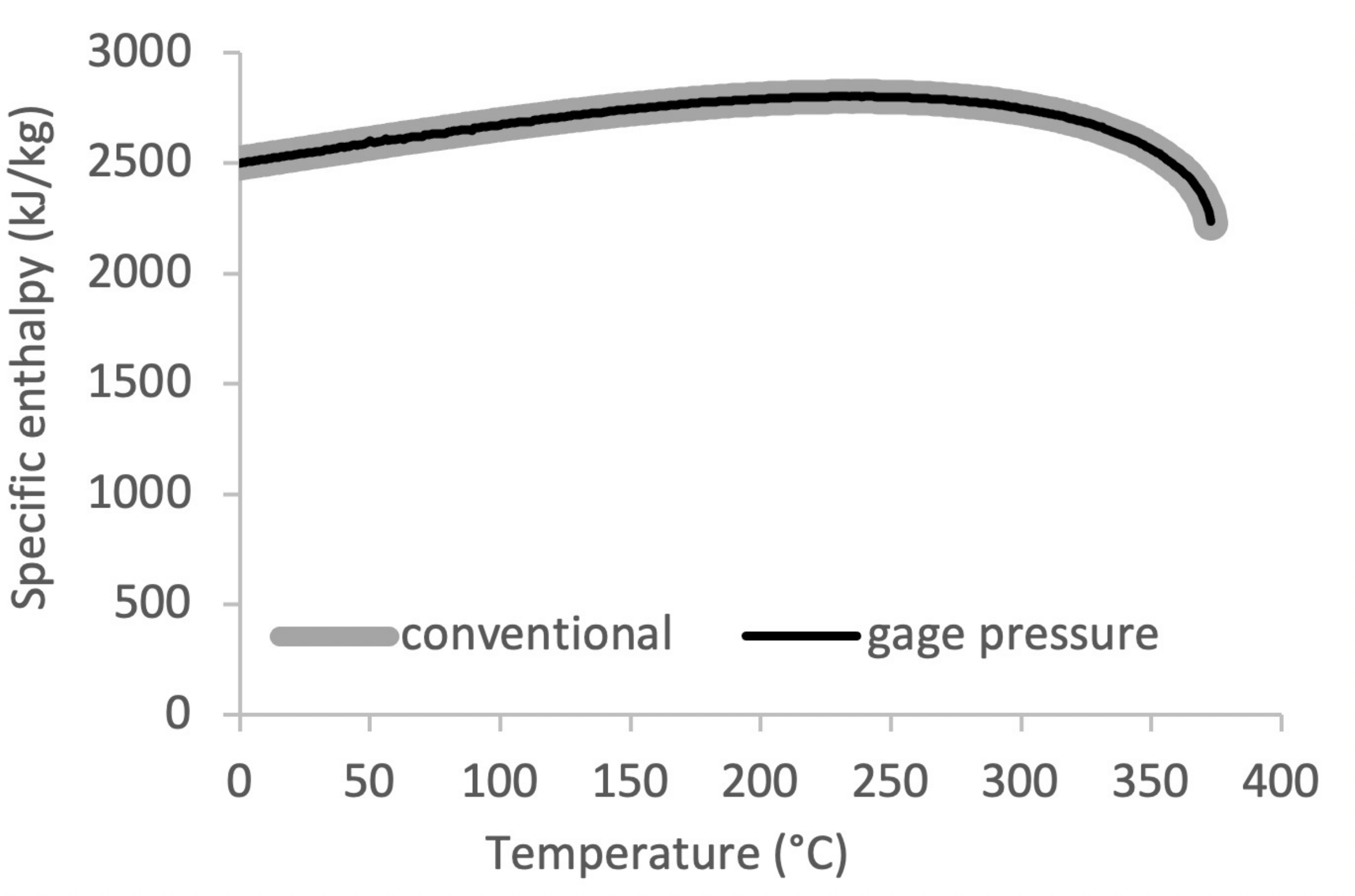}

}
\par\end{centering}
\caption{Thermodynamic properties of water vapor on the saturation curve, comparing
conventional values from standard tables (NIST, https://webbook.nist.gov/chemistry/fluid/)
to those evaluated in this study, where gauge pressure is used instead
of absolute pressure. (a) The specific internal energy $u^{v}$ shows
small differences for temperatures up to approximately $50\,\text{°C}$.
(b) The enthalpy $h^{v}$ in the conventional approach is the same
as the gauge enthalpy in the pressure gauge approach.\label{fig:H2Ovapor-properties-2}}
\end{figure}
\end{example}

To the best of our knowledge, the presentation of this section represents
an original literature contribution. In particular, our emphasis on
using gauge pressure instead of absolute pressure is a distinction
rarely addressed in the continuum mechanics and thermodynamics literature.
Our approach emphasizes the cautionary distinctions that arise from
using one or the other. Though readers may be concerned about our
conclusion that the specific internal energy $u^{v}$ of the vapor
phase of a thermoelastic fluid differs from entries in standard thermodynamic
tables, this should be viewed as a minor concern since those standard
tables also provide the enthalpy $h^{v}$ (whose values are the same
as our gauge enthalpy), along with the absolute pressure $P^{v}$
and mass density $\rho^{v}$. Therefore, one can use the same standard
thermodynamics tables to evaluate $u^{v}=h^{v}-\left(P^{v}-P_{r}\right)/\rho^{v}$
where $P_{r}$ is the triple point pressure, and recover the values
of the specific internal energy $u^{v}$ of our formulation.

\section{Phase Transformation Kinetics\label{sec:Phase-Transformation-Kinetics}}

\subsection{Constitutive Model for Reactive Mass Flux\label{subsec:Constitutive-Model-Mass-Flux}}

To complete the set of available equations, we need to propose a constitutive
relation for the reactive mass flux $M\bar{\zeta}$. As reviewed by
Persad and Ward \citep{Persad16}, the standard models described in
the literature are the Hertz-Knudsen or Hertz-Knudsen-Shrage relations,
based on the kinetic theory of gases, and the statistical rate theory
(SRT) expression for the evaporation flux. These authors also discussed
the application of molecular dynamics to calculate the evaporation
and condensation coefficients, which represent material parameters
needed for the Hertz-Knudsen relation and its modifications. Badam
et al. \citep{Badam07} proposed an alternative constitutive model
for the evaporative flux based on phenomenological equations and Onsager's
reciprocal principle, also relying on the work of Bedeaux and Kjelstrup
\citep{Bedeaux99} which appealed to the concept of chemical potentials
of the liquid and vapor phases at the interface.

Our goal here is to develop an original constitutive model for the
reactive mass flux which is consistent with our continuum thermodynanics
framework, without appealing to the kinetic theory of gases, molecular
mechanisms, statistical approaches, or chemical potentials, since
none of these concepts have emerged naturally from the governing equations
presented so far. This constitutive model should reduce the mass flux
to zero under the conditions that are consistent with phase equilibrium
as outlined in the previous section. It must also satisfy the jump
condition (\ref{eq:entropy-jump-latent-1}) or (\ref{eq:entropy-jump-latent-3})
on the entropy inequality. Based on these requirements, the simplest
option is to let
\begin{equation}
M\bar{\zeta}=-\frac{1}{\left[\left[s\right]\right]}\left[\left[\frac{1}{T}\mathbf{q}\right]\right]\cdot\mathbf{n}^{\Gamma}\,.\label{eq:PT-constitutive-relation}
\end{equation}
In other words, the reactive mass flux for a phase transition on $\Gamma$
is proportional to the normal jump in thermal entropy flux, and inversely
proportional to the jump in entropy. This choice of constitutive relation
would satisfy the first requirement, since
\[
\left[\left[\frac{1}{T}\mathbf{q}\right]\right]\cdot\mathbf{n}^{\Gamma}=\frac{1}{T^{\Gamma}}\underbrace{\left[\left[\mathbf{q}\right]\right]\cdot\mathbf{n}^{\Gamma}}_{=0}+\underbrace{\left[\left[\frac{1}{T}\right]\right]}_{=0}\mathbf{q}^{\Gamma}\cdot\mathbf{n}^{\Gamma}=0\quad\text{at phase equilibrium,}
\]
when we consider phase equilibrium as the limiting condition leading
to $\left[\left[g\right]\right]=0$ and $\left[\left[1/T\right]\right]=0$.
This constitutive relation also satisfies the entropy inequality jump
(\ref{eq:entropy-jump-latent-1}) without any residual dissipation.
Consequently, we would describe the phase boundary as a non-dissipative
interface for this choice of constitutive model. However, since non-zero
heat fluxes $\mathbf{q}$ must exist on either side of $\Gamma$ to
drive the phase transformation, the phase transition process is in
fact irreversible (dissipative) when accounting for the domains of
$\mathcal{S}^{a}$ and $\mathcal{S}^{b}$ across the interface $\Gamma$.

\begin{figure}
\begin{centering}
\includegraphics[width=3.25in]{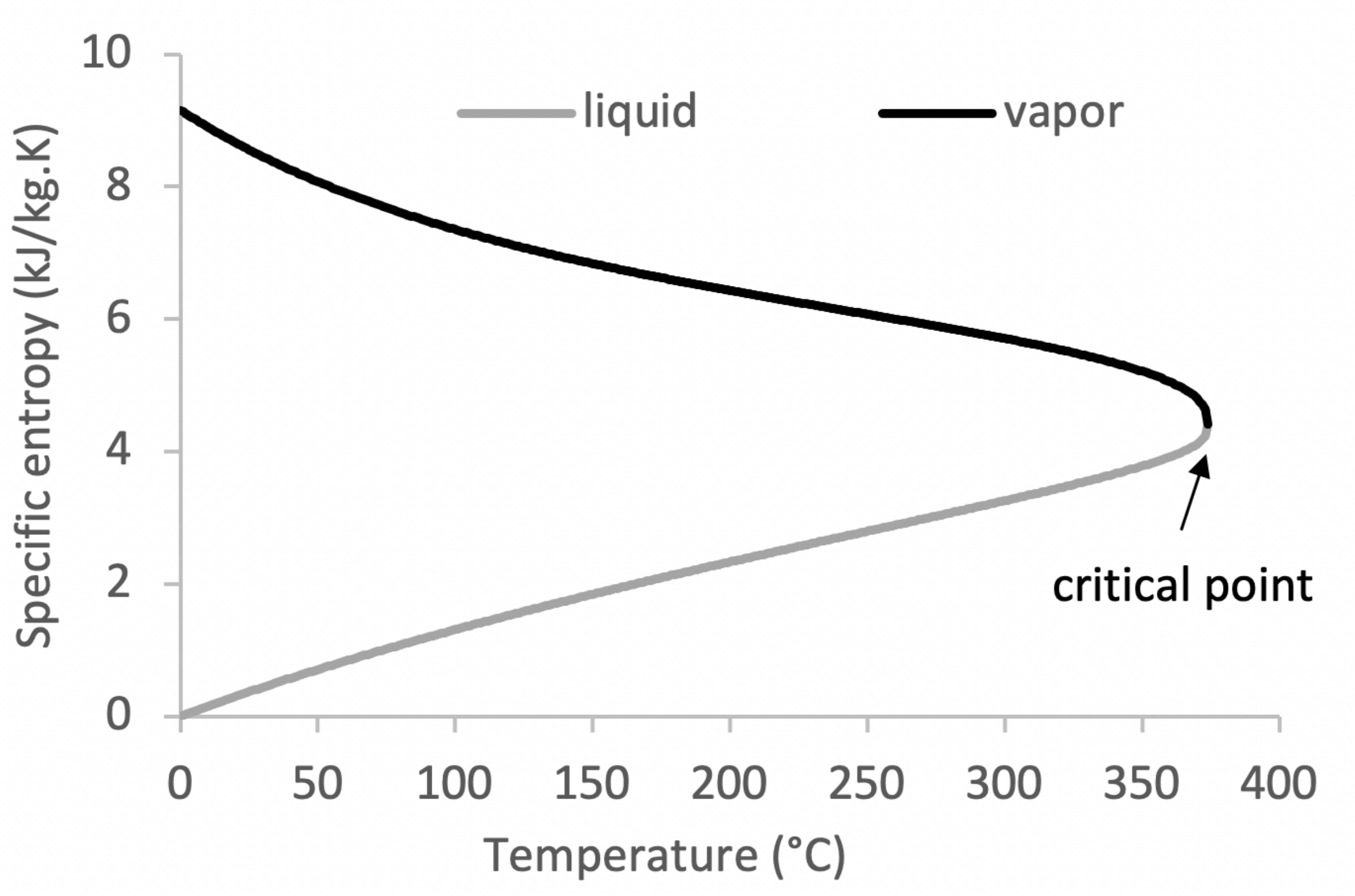}
\par\end{centering}
\caption{The entropies of liquid and vapor water on the saturation curve (NIST,
https://webbook.nist.gov/chemistry/fluid/) converge toward each other
as the temperature rises to the critical point ($373.946\,\text{°C}$),
showing that their difference $\left[\left[s\right]\right]$ reduces
to zero at that point. This convergence produces a singularity in
the constitutive model (\ref{eq:PT-constitutive-relation}) for the
phase transformation mass flux $M\bar{\zeta}$, consistent with the
disappearance of the phase boundary in the supercritical regime.\label{fig:entropy-critical-point}}
\end{figure}

Since the entropy of the vapor phase of a substance is greater than
that of the liquid phase, the term $-1/\left[\left[s\right]\right]$
in (\ref{eq:PT-constitutive-relation}) is always positive whenever
the $+$ side represents the liquid phase and the $-$ side is the
vapor phase. Thus, the sign of the reactive mass flux $M\bar{\zeta}$
depends on the sign of $\left[\left[\mathbf{q}/T\right]\right]\cdot\mathbf{n}^{\Gamma}$;
a positive value implies evaporation while a negative value implies
condensation. It is also interesting to note that (\ref{eq:PT-constitutive-relation})
predicts a singularity in the reactive mass flux as $\left[\left[s\right]\right]\to0$
when the numerator is not zero (recall that $\left[\left[s\right]\right]$
depends only on temperatures and volumetric strains, whereas the numerator
also depends on temperature gradients, implying that numerator and
denominator are independent of each other). This singularity coincides
with the emergence of the critical point in the liquid and vapor phases,
where their entropy values are equal, as shown in a plot of the entropies
of fluid water on the saturation curve (\figurename~\ref{fig:entropy-critical-point}).
Thus, the model of (\ref{eq:PT-constitutive-relation}) is consistent
with the fact that liquid and vapor phases become indistinguishable
in the supercritical regime.

Based on the alternative form of the residual dissipation in (\ref{eq:entropy-jump-latent-3}),
we may also write the constitutive relation (\ref{eq:PT-constitutive-relation})
as
\begin{equation}
M\bar{\zeta}=\frac{\left[\left[\frac{1}{T}\right]\right]T^{\Gamma}\mathbf{q}^{\Gamma}\cdot\mathbf{n}^{\Gamma}}{\left[\left[u\right]\right]+p^{\Gamma}\left[\left[\frac{1}{\rho}\right]\right]-T^{\Gamma}\left[\left[s\right]\right]}\,.\label{eq:PT-constitutive-redux}
\end{equation}
This equivalent alternative form makes it evident that a reactive
mass flux for phase transformation can only occur in the presence
of a non-zero temperature jump, $\left[\left[\frac{1}{T}\right]\right]\ne0$,
and a non-zero average heat flux, $\mathbf{q}^{\Gamma}\cdot\mathbf{n}^{\Gamma}\ne0$,
for this choice of constitutive relation.

Substituting this constitutive model into the expression for the specific
latent heat $L$ in (\ref{eq:PT-latent-heat-def}) produces
\begin{equation}
L=-\left[\left[s\right]\right]\frac{\left[\left[\mathbf{q}\right]\right]\cdot\mathbf{n}^{\Gamma}}{\left[\left[\frac{1}{T}\mathbf{q}\right]\right]\cdot\mathbf{n}^{\Gamma}}=-\left[\left[s\right]\right]\left(\frac{1}{T^{\Gamma}}+\left[\left[\frac{1}{T}\right]\right]\frac{\mathbf{q}^{\Gamma}\cdot\mathbf{n}^{\Gamma}}{\left[\left[\mathbf{q}\right]\right]\cdot\mathbf{n}^{\Gamma}}\right)^{-1}\,.\label{eq:PT-constitutive-L}
\end{equation}
In the limit of phase equilibrium $\left[\left[g\right]\right]=0$
and $\left[\left[1/T\right]\right]=0$, the right-hand-side of this
expression reduces to $L_{0}=-\left[\left[h\right]\right]$ in (\ref{eq:tabulated-latent-heat})
from the fact that $g=h-Ts$, such that phase equilibrium implies
$\left[\left[h\right]\right]=T^{\Gamma}\left[\left[s\right]\right]$.
More generally, the expression of (\ref{eq:PT-constitutive-L}) shows
that the specific latent heat $L$ may deviate substantially from
$L_{0}$ for general phase transition processes.

To determine the temperature jump across $\Gamma$ under general conditions,
we need to solve the field equations (the kinematic constraint (\ref{eq:kinematic-constraint}),
the momentum balance (\ref{eq:momentum-balance}), and the energy
balance in (\ref{eq:energy-balance-fluid})) in both phases across
$\Gamma$, and use the momentum and energy jumps in (\ref{eq:momentum-jump-redux})
and (\ref{eq:energy-jump-redux}) as well as the constitutive relation
of (\ref{eq:PT-constitutive-relation}) as interface jump conditions.
These interface conditions depend on the constitutive relations for
$a\left(T,J\right)$ and $\mathbf{q}\left(T,J,\mathbf{g}\right)$
on either side of $\Gamma$, which have been formulated for each of
the two phases $\mathcal{S}^{a}$ and $\mathcal{S}^{b}$ of the pure
substance $\mathcal{S}$ independently of these interface jump conditions.
Thus, jump conditions can only be satisfied for specific pairs of
state variables $\left(T^{a},J^{a},\mathbf{g}^{a}\right)$ and $\left(T^{b},J^{b},\mathbf{g}^{b}\right)$.

To the best of our knowledge, the prior literature on continuum thermodynamics
has not proposed a constitutive model for the phase transformation
mass flux of the type given in (\ref{eq:PT-constitutive-relation}).
Therefore, all of the material presented in this section represents
an original contribution. This model requires experimental and numerical
validations to establish some measure of confidence in its validity.
Since the model is predicated on the existence of a temperature jump
across the phase boundary, we may reference experimental studies that
have reported such temperature jumps under controlled conditions of
a liquid-vapor phase transformation.

\subsection{Experimental Validation\label{subsec:Experimental-Validation}}

Persad and Ward \citep{Persad16} have tabulated an extensive set
of results for the evaporation of water and ethanol under conditions
that produce a temperature jump. However, these tabulated results
do not include measurements of the temperature gradient $\mathbf{g}$
or heat flux $\mathbf{q}$ on the liquid and vapor sides of the phase
boundary $\Gamma$, preventing us from using them to validate the
model of (\ref{eq:PT-constitutive-relation}). However, Badam et al.
\citep{Badam07} provided a comprehensive set of experimental measurements
relevant to the constitutive model in (\ref{eq:PT-constitutive-relation})
or its equivalent form (\ref{eq:PT-constitutive-redux}). These authors
investigated the steady-state evaporation of water under various temperature
and pressure conditions, where they found a temperature jump as high
as $15.68\,\text{K}$ as they increased the vapor phase heating. In
addition to measurements of the evaporative mass flux $M\bar{\zeta}$,
temperatures across $\Gamma$, and the vapor pressure, they also measured
temperature gradients on the liquid and vapor sides and computed the
heat fluxes using thermal conductivities.

\begin{figure}
\begin{centering}
\includegraphics[width=3.25in]{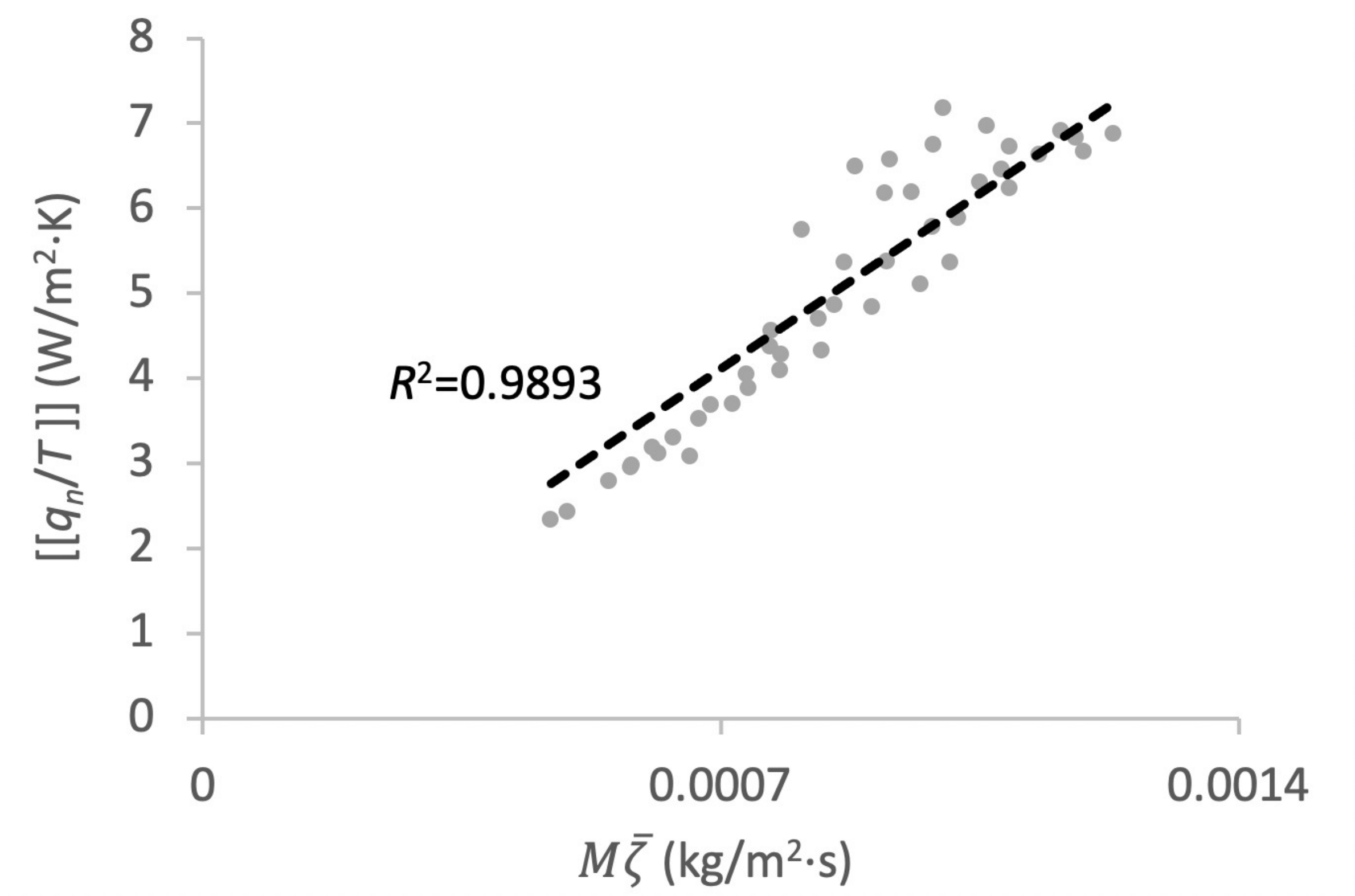}
\par\end{centering}
\caption{Plot of experimental data of Badam et al. \citep{Badam07} in the
form of (\ref{eq:EV-correlation}). A linear regression of the data
produces $\left[\left[q_{n}/T\right]\right]=5876\,M\bar{\zeta}$ with
a coefficient of determination $R^{2}=0.9893$.\label{fig:Badam-correlation}}
\end{figure}

Badam et al.'s experiments were conducted at an average gauge pressure
$p^{v}$ of $-100\,\text{Pa}$, with liquid water temperatures $T^{\ell}$
averaging $269.5\,\text{K}$ and water vapor temperatures $T^{v}$
averaging $277.4\,\text{K}$ \citep{Badam07}. Since their measurements
remained close to the triple point of water, it is reasonable to assume
that the entropy of liquid water was negligible compared to that of
the vapor, such that the entropy jump appearing in (\ref{eq:PT-constitutive-redux})
could be approximated as $\left[\left[s\right]\right]\approx-s^{v}$.
Consequently, our constitutive model (\ref{eq:PT-constitutive-relation})
may be reduced to the simplified form
\begin{equation}
\left[\left[\frac{q_{n}}{T}\right]\right]=\frac{q_{n}^{\ell}}{T^{\ell}}-\frac{q_{n}^{v}}{T^{v}}\approx s^{v}M\bar{\zeta}\,,\label{eq:EV-correlation}
\end{equation}
where $q_{n}^{\ell}=\mathbf{q}^{\ell}\cdot\mathbf{n}^{\Gamma}$ and
$q_{n}^{v}=\mathbf{q}^{v}\cdot\mathbf{n}^{\Gamma}$ denote the heat
flux components normal to the phase boundary in the liquid and vapor
phases, respectively. Thus, as long as relative variations in $s^{v}$
remain small over the range of experimental conditions in the vapor
phase, this constitutive model predicts a nearly linear relationship
between $\left[\left[q_{n}/T\right]\right]$ and the evaporative flux
$M\bar{\zeta}$. When we plot Badam et al.'s reported $\left[\left[q_{n}/T\right]\right]$
versus $M\bar{\zeta}$, we find a very strong linear relationship
with a coefficient of determination $R^{2}=0.9893$ and a slope $s^{v}=5876\,\text{J}/\text{kg}\cdot\text{K}$
with a standard deviation of $92\,\text{J}/\text{kg}\cdot\text{K}$
(\figurename~\ref{fig:Badam-correlation}).

The linear relationship strongly supports our proposed model for $M\bar{\zeta}$;
however, the slope $5876\,\text{J}/\text{kg}\cdot\text{K}$ underestimates
the entropy $s^{v}$ of water vapor in the range of their experimental
conditions, which we estimate to average $9318\,\text{J}/\text{kg}\cdot\text{K}$
based on their reported $p^{v}$ and $T^{v}$ values (ranging from
$9017$ to $9657\,\text{J}/\text{kg}\cdot\text{K}$). In effect, their
experimental data underestimate our hypothesized theoretical value
of the normal jump in thermal entropy flux $\left[\left[q_{n}/T\right]\right]$
by 37\%, most likely due to the oversimplification of their experimental
conditions using a one-dimensional analysis (i.e., using a mass flux
$M\bar{\zeta}$ averaged over a curved $8\,\text{mm}\times23\,\text{mm}$
liquid-vapor interface, while relying on measurements of temperature
along a single centerline normal to that surface, see their Fig. 2).

\section{Discussion\label{sec:Discussion}}

The objective of this study was to formulate the thermodynamics of
phase transformations of a pure thermoelastic fluid using only a continuum
approach. We used gauge pressure instead of absolute pressure in this
framework, since the stress tensor in continuum mechanics is assumed
to reduce to zero in some arbitrary reference configuration. This
seemingly minor recognition had a major influence on our findings,
as became evident in the development presented above. In Section~\ref{subsec:Axioms-of-Conservation},
we presented the familiar differential statements for the axioms of
mass, momentum and energy balance, which include functions of state
for the fluid stress $\boldsymbol{\sigma}$, specific internal energy
$u$, and heat flux $\mathbf{q}$. These functions of state model
the behavior of specific fluids, and therefore they need to be described
by constitutive relations. We proceeded to present the axiom of entropy
inequality and adapted the concept presented by Coleman and Noll \citep{Coleman63},
namely that this inequality statement should be used to place constraints
on the functions of state for arbitrary processes. The entropy inequality
introduced the specific entropy $s$ as an additional function of
state, and its combination with the energy balance, motivated by the
need to eliminate user-specified specific heat supplies $r$ (as further
elaborated upon in our recent study \citep{Ateshian22}), naturally
produced the specific free energy $a$.

In Section~\ref{subsec:State-Variables-Constraints}, we proceeded
to select a list of observable state variables suitable for the analysis
of thermoelastic fluids, which included the temperature, volume ratio
and temperature gradient. Given those state variables, we reviewed
the constraints imposed on the functions of state using the entropy
inequality. Using those constraints, we examined two alternative forms
of the energy balance in Section~\ref{subsec:Implications-Energy-Balance},
involving the material time derivative of temperature and specific
entropy, respectively. The latter form could be used to better understand
the concept of isentropic processes.

The volume ratio $J$ and the fluid velocity $\mathbf{v}$ may be
obtained by solving the kinematic constraint of (\ref{eq:kinematic-constraint}),
concurrently with the differential statement of momentum balance in
(\ref{eq:momentum-balance}). For thermofluid analyses, the energy
balance in (\ref{eq:energy-balance}) is also needed to solve for
the temperature $T$. Though continuum analyses that employ the kinematic
constraint (\ref{eq:kinematic-constraint}) are not common in fluid
mechanics, we have recently demonstrated that they can be used successfully
to develop a general-purpose computational fluid dynamics finite element
solver \citep{Ateshian18,Shim19}. To solve these equations in practice,
we need constitutive relations for the relevant functions of state.
Based on the constitutive restrictions presented in Section~\ref{subsec:State-Variables-Constraints},
the only functions of state that require constitutive relations are
the specific free energy $a$ and the heat flux $\mathbf{q}$. Constitutive
relations for all other functions of state can be derived from $a$.

In Section~\ref{subsec:Constitutive-Relations}, we proceeded to
formulate the constitutive relation for the specific free energy $a$
of ideal gases, real gases and liquids. As reviewed in that section,
the resulting expressions we find for the specific free energy of
ideal (\ref{eq:IG-free-energy}) and real (\ref{eq:rg-a-final}) gases
represent original contributions, since they appear to be the first
to be derived starting from an expression for the gauge pressure $p$,
instead of absolute pressure as in \citep{Muller85}. A virial expansion
was similarly used to formulate the constitutive model for the specific
free energy $a$ of liquids, as shown in (\ref{eq:liquid-free-energy}).
Since the forms of the constitutive models for the specific free energy
of real liquids and gases presented in this study is novel, we demonstrated
that it could reproduce standard thermodynamic properties of liquid
water in Example~\ref{exa:Liquid-water-properties}, and vapor water
in Example~\ref{exa:Vapor-water-properties}.

In Sections~\ref{subsec:Interface-Jump-Conditions} and \ref{sec:Reactive-Mixtures}
we presented interface jump conditions for the axioms of mass, momentum
and energy balance, and the axiom of entropy inequality, allowing
for reactive mass exchanges across the interface. Jump conditions
of this type have been derived previously in the continuum mechanics
literature, including for reactive mixtures \citep{Kelly64,Eringen65,Hutter04,Muller85,Ateshian07}.
The specific free enthalpy function $h$ emerged naturally in the
energy jump.\footnote{It is also possible to rewrite the energy balance equation (\ref{eq:energy-balance})
as $\rho\dot{h}=\dot{p}-\divg\mathbf{q}+\rho r$, which could facilitate
the analysis of isobaric ($\dot{p}=0$) or isenthalpic processes ($\dot{h}=0$).} As mentioned earlier in this presentation, this enthalpy function
uses the gauge pressure $p$, instead of the absolute pressure $P$
as normally done in conventional thermodynamics.

Most significantly, the jump condition (\ref{eq:entropy-jump}) on
the entropy inequality can be used to enforce constraints on the thermodynamic
feasibility of interfacial processes. We illustrated the application
of this jump constraint to the case of a normal shock wave in Section~\ref{subsec:Normal-Shock-Wave},
showing that it led to the common understanding that such a shock
wave can only occur spontaneously (i.e., without the addition of heat)
when supersonic flow transitions to subsonic flow \citep{Liepmann57}.
Moreover, this example served as a useful reminder that temperature
need not be continuous across an immaterial interface, even in the
absence of a phase transformation.

The existence of temperature jumps across interfaces is accepted in
the research literature but remains ambiguously defined in thermodynamics
and heat transfer textbooks. In the thermodynamics of phase transition,
textbooks typically emphasize that temperature and pressure remain
continuous across the phase boundary. In heat transfer textbooks,
the convective heat transfer formula between a solid and fluid assigns
a different temperature to each side of their interface, though the
fluid temperature is often described as the free stream temperature,
generally implying that it is the temperature outside of the boundary
layer located in the vicinity of the solid wall. Whether a temperature
jump may exist on the interface is generally not discussed in those
textbooks. A major objective of the current study was to eliminate
this ambiguity entirely and provide the theoretical foundations that
support the existence of a temperature jump, specifically in the case
of a phase boundary.

In Section~\ref{sec:Phase-Transformations}, we specialized the general
jump conditions to the case of a reactive interface between the liquid
and vapor phases of a pure substance. When applied to mass, momentum
and energy balance, this specialization produced equations consistent
with those of Stefan as he applied them to the investigation of the
ice-liquid water interface \citep{Stefan91,Ward04}. A detailed examination
of the entropy jump led us to the conclusion that continuous temperature
and pressure across the phase boundary is a sufficient condition for
phase equilibrium ($M\bar{\zeta}=0$), though not necessary. By examining
small disturbances in the mass flux $M\bar{\zeta}$ about this equilibrium,
we concluded from (\ref{eq:Gibbs-inequality-constraint}) that phase
equilibrium corresponds to continuity of the specific free enthalpy
(the Gibbs function) across the phase boundary, consistent with classical
concepts. This condition represented the first appearance of the specific
free enthalpy function $g$ in this continuum thermodynamics formulation,
implying that $g$ is only needed to identify phase equilibrium.

Upon establishing this continuity condition in the limiting case of
phase equilibrium, we were able to determine how to relate the reference
values $a_{r}$ for the specific free energy, and $s_{r}$ for the
specific entropy, of liquid and vapor phases (Section~\ref{subsec:Reference-Configurations}).
Whereas the liquid values of $a_{r}$ and $s_{r}$ were arbitrarily
set to zero at the triple point, as done conventionally, the corresponding
values for the vapor were derived from the continuity condition $\left[\left[g\right]\right]=0$.
Contrary to the conventional approach, due to our use of gauge pressure
$p$ instead of absolute pressure $P$, we found that $a_{r}$ for
the vapor phase should also equal zero at the triple point. This led
to a deviation of our thermodynamic values of $a$ from those calculated
from conventional thermodynamic tables, by the value given in (\ref{eq:a-vapor-classical})
at the triple point, though this deviation was not constant and became
negligible with increasing temperature on the saturation curve, as
illustrated for water in \figurename~\ref{fig:H2Ovapor-properties-1}.
For this comparison, we evaluated $a=u-Ts$ from thermodynamic values
of $u$ and $s$ reported in the NIST tables, since they do not report
$a$ directly. As explained in Section~\ref{subsec:Reference-Configurations},
there were no differences in $s$ and $h$ between our values and
standard tables, but the value of the specific internal energy $u$
exhibited the same pattern as $a$.

As a side note, taking the triple point as the reference configuration
for the three phases of a pure substance is the logical choice when
analyzing phase transformations. However, other applications, such
as the analysis of reactive mixtures of liquids or gases, may employ
an arbitrary (e.g., standard) reference configuration common to all
mixture constituents, in order to properly evaluate the heat of reaction.
Therefore, the existence of a triple point is not essential to the
formulation of reactive mixture thermodynamics. It is just a convenient
choice for the special case of phase transformations of a pure substance,
which is the topic of this study.

To analyze more general cases of phase transformation kinetics, where
the reactive mass flux is not negligibly small (i.e., irreversible
thermodynamics of phase transformations), it became necessary to formulate
a constitutive model for $M\bar{\zeta}$, using the entropy inequality
jump (\ref{eq:entropy-jump-latent-1}) as a constraint on this choice
of model. An examination of that jump condition showed that the only
function of state which had not already been constrained by the axiom
of entropy inequality in the liquid and vapor phases was the interfacial
reactive mass flux $M\bar{\zeta}$. Given the simplicity of the expression
in (\ref{eq:entropy-jump-latent-1}), and since the sign of $\left[\left[s\right]\right]$
remains fixed for an interface between liquid and vapor phases of
a substance (e.g., as seen in \figurename~\ref{fig:entropy-critical-point}
at phase equilibrium), whereas $M\bar{\zeta}$ could be positive or
negative, the constitutive relation presented in (\ref{eq:PT-constitutive-relation})
was the self-evident form that would satisfy the entropy inequality,
under the assumption that the interface is non-dissipative. In this
relation, the direction of the mass flux (evaporation versus condensation)
becomes contingent on the sign of the normal jump in thermal entropy
flux, $-\left[\left[\mathbf{q}/T\right]\right]\cdot\mathbf{n}^{\Gamma}$.

To the best of our knowledge, this constitutive relation for the phase
transition mass flux has not been proposed previously. By relying
on temperature gradients on either side of the phase boundary, it
deviates from conventional formulations that depend only on temperature
and mass densities. The physical insight underlying this constitutive
model is that the energy needed to cause evaporation or condensation
(the latent heat of transformation) must be transported to the interface
by the heat flux, (\ref{eq:PT-latent-heat-def}), which must be a
function of the temperature gradient according to the axiom of entropy
inequality as presented in Section~\ref{sec:Heat-Conduction}, thus
it represents heat conduction. In contrast, standard textbook chemistry
formulations for the heat of vaporization do not explicitly address
how that energy is transported in the reacting medium or across the
phase boundary. Textbook formulations typically assume that the temperature
remains uniform in a process; as shown in Section~\ref{sec:Irreversible-and-Reversible},
a state of uniform temperature for a pure thermoelastic fluid necessarily
implies that all processes in that medium are reversible, thus lacking
the generality of irreversible thermodynamics. We expect that the
formulation presented in this study fills this important gap, in the
context of a continuum framework.

Other investigators have previously used the jump conditions from
mixture theory to examine phase transitions. For example, our derivations
in Sections~\ref{sec:Mass-Balance-Jump}, \ref{subsec:Momentum-Balance-Jump-1}
and \ref{subsec:Energy-Balance-Jump-1} show considerable analogy
with the approach of Svendsen and Gray \citep{Svendsen96,Gray97},
however these authors did not discuss the implications of these jump
conditions as we do in Section~\ref{subsec:Entropy-Inequality-Jump-1},
nor did they propose a constitutive model for the phase transition
flux, as done in Section~\ref{subsec:Constitutive-Model-Mass-Flux}.
Buratti et al. \citep{Buratti03} also presented the same jump conditions
across the phase interface as done in this study, but they adopted
the assumption of temperature continuity early in their derivation.
In contrast, Danescu \citep{Danescu04} presented the same result
as our equation (\ref{eq:entropy-jump-latent-1}) in Section~\ref{subsec:Entropy-Inequality-Jump-1},
and based on the work of Fried and Shen \citep{Fried99} they identified
what we called $T^{\Gamma}$ as the ``equivalent temperature of the
interface,'' which they defined as a constitutive choice. Despite
the close analogy of our approach with the prior work of Danescu \citep{Danescu04},
this author did not identify the interfacial phase transition mass
flux as a function of state requiring a constitutive relation, such
as the one we proposed in (\ref{eq:PT-constitutive-relation}) or
its equivalent form in (\ref{eq:PT-constitutive-redux}).

To test the validity of our constitutive model against experimental
data, we had access to the data set from the study of Badam et al.
\citep{Badam07} which was the only one we could find that provided
measurements of temperature gradients across the phase boundary (Gatapova
et al. \citep{Gatapova17} also reported evaporative temperature jumps
that depend on temperature gradients across the interface, but their
experiments were conducted at the interface of liquid water and air).
Using this data set with our model produced encouraging but mixed
results: A very strong linear relation was observed between the model
and experimental data in \figurename~\ref{fig:Badam-correlation},
but the slope of the response underestimated the model by approximately
37\%. While this outcome may tempt us to adjust the constitutive relation
in (\ref{eq:PT-constitutive-relation}) by including a positive scale
factor less than or equal to unity, perhaps serving as some kind of
material property for the phase transition mass flux, it is easy to
show that this type of scaling would not satisfy the entropy inequality
(\ref{eq:entropy-jump-latent-1}) under general conditions. It is
more likely that the experimental conditions of Badam et al. \citep{Badam07}
could not be reduced to a one-dimensional analysis. Indeed, a closer
examination of their experimental results (not reported here) shows
that they would also not satisfy the general energy jump (\ref{eq:PT-energy-redux}),
nor the classical Stefan energy jump condition (\ref{eq:PT-slow-mass-flux}),
satisfactorily when assuming that one-dimensional conditions prevail.
Therefore, more evidence is needed to support or reject the validity
of our proposed constitutive model for the phase transformation mass
flux.

If this constitutive relation is found to be valid, it would represent
a powerful tool for analyzing phase transition kinetics under arbitrary
conditions, including those that deviate significantly from phase
equilibrium. As reviewed in Section~\ref{subsec:Constitutive-Model-Mass-Flux},
this constitutive model clearly establishes that phase transitions
can only occur in the presence of a temperature jump, $\left[\left[1/T\right]\right]\ne0$,
where the heat flux entering the interface from one side differs from
the heat flux leaving the interface on the opposite side, $\mathbf{q}^{\Gamma}\cdot\mathbf{n}^{\Gamma}\ne0$.
Evidently, the difference in heat flux across $\Gamma$ contributes
to the specific latent heat of the phase transformation, as shown
in (\ref{eq:PT-constitutive-L}). Thus, phase transformations always
require a non-zero heat flux on either side of $\Gamma$, implying
that this process is dissipative (irreversible), regardless of the
fact that the immaterial phase interface $\Gamma$ itself is modeled
as non-dissipative.

In summary, we have revisited fundamental concepts in the thermodynamics
of pure fluids and their phase transformation, using only a continuum
approach, with no reference to alternative frameworks such as statistical
thermodynamics. We cited some of the classical and recent literature
in the development of well-known fundamental principles and highlighted
our novel contributions in the relevant sections, with further comparisons
reported in the above paragraphs. While we recovered almost all the
fundamental relations of classical thermodynamics, we found that a
self-consistent framework should employ gauge pressure relative to
a suitable reference configuration, instead of absolute pressure,
when evaluating the specific free energy and its related functions
of state. Using gauge pressure, we were also able to formulate a novel
closed-form expression for the specific free energy of ideal gases,
which served as the basis for related formulations for real gases
and liquids using virial expansions. To examine phase transformation
kinetics, we placed a special emphasis on the jump condition for the
axiom of entropy inequality, which allowed us to recover the conventional
result that phase equilibrium coincides with continuity of temperature,
pressure, and free enthalpy across the phase boundary. These results
were found to be consistent with the prior classical literature going
back to the work of Stefan \citep{Stefan91} and Gibbs \citep{Gibbs06},
as well as the more recent continuum thermodynamic literature \citep{Hutter04,Fried95,Svendsen96,Gray97,Fried99}.
Moreover, this jump condition allowed us to formulate an original
constitutive relation for arbitrary phase transition kinetics, showing
that phase transformations must be accompanied by a jump in temperature
across the phase boundary. Further experimental and computational
evidence is needed to conclusively validate (or reject) this proposed
constitutive model. This continuum framework is well suited for implementation
in a computational framework, such as the finite element method \citep{Ateshian18,Shim19}
and the phase field method \citep{Chen22}. Though not addressed explicitly
in this study, the analysis of solid-fluid phase transformations may
be easily extended from the presentation given here, by allowing the
Cauchy stress tensor $\boldsymbol{\sigma}$ of the solid phase to
depend on a suitable strain tensor, instead of just $J$.

\section*{Acknowledgment}

The authors thank Prof. Arvind Narayanaswamy of Columbia University
for his insightful comments.

\setcounter{section}{0}
\renewcommand{\thesection}{\Alph{section}}
\renewcommand{\theequation}{\thesection\arabic{equation}}
\renewcommand{\thefigure}{\thesection\arabic{figure}}
\numberwithin{equation}{section}

\section{Appendix\label{sec:Appendix}}

\subsection{Derivation of Interface Jump Conditions}

\begin{figure}
\begin{centering}
\includegraphics[width=3.13in]{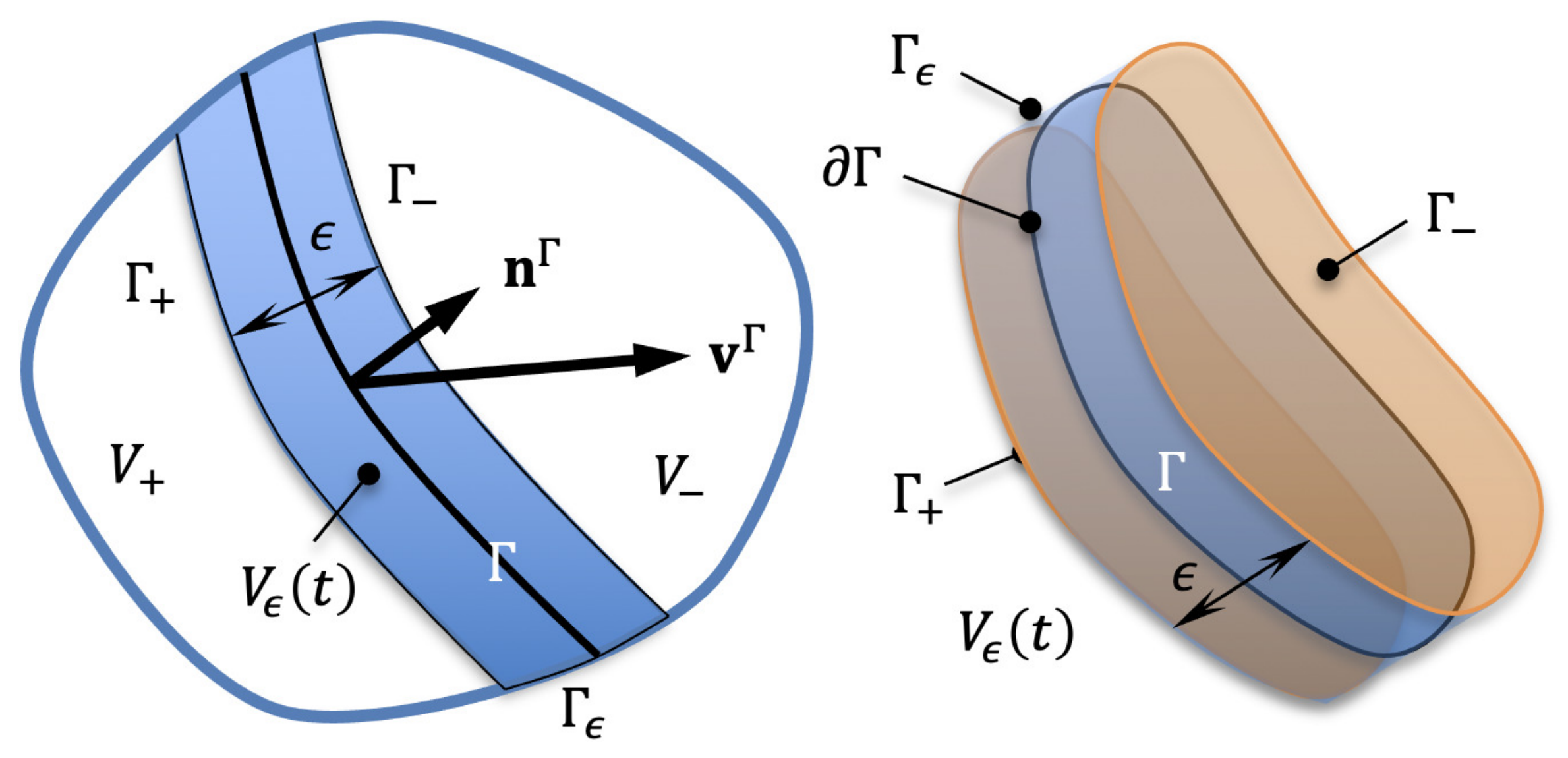}
\par\end{centering}
\caption{Interface jump conditions are defined across an interfacial surface
$\Gamma$ and evaluated using axioms of conservation over a thin region
$V_{\epsilon}\left(t\right)$ of thickness $\epsilon$ within the
material, whose mid-surface is $\Gamma$. The interface $\Gamma$
moves with a velocity $\mathbf{v}^{\Gamma}$ and its surface normal
is $\mathbf{n}^{\Gamma}$. The boundary $\partial V_{\epsilon}\left(t\right)$
is the union of the front surface $\Gamma_{+}$, the back surface
$\Gamma_{-}$, and the side surface $\Gamma_{\epsilon}$. As $\epsilon\to0$,
$\Gamma_{+}$ and $\Gamma_{-}$ collapse onto $\Gamma$, whereas $\Gamma_{\epsilon}$
collapses onto the boundary contour curve $\partial\Gamma$ of $\Gamma$.\label{fig:Interface-jump-conditions}}
\end{figure}

Let the interface $\Gamma$ move at a velocity $\mathbf{v}^{\Gamma}$
through the control volume $V$ and let the unit normal to $\Gamma$
be denoted by $\mathbf{n}^{\Gamma}$ (\figurename~\ref{fig:Interface-jump-conditions}).
Our goal is to formulate the jump conditions for mass, momentum, energy
and entropy across $\Gamma$, in terms of the values of state variables
and functions of state on both sides of $\Gamma$. To achieve this
goal, we define a region $V_{\epsilon}\left(t\right)$ about $\Gamma$
as the thin volume extending along $\pm\mathbf{n}^{\Gamma}$ by a
total thickness $\epsilon$, with the remaining regions of $V$ denoted
by $V_{+}$ and $V_{-}$ ($V=V_{+}\cup V_{-}\cup V_{\epsilon}$).
The boundary surface between $V_{+}$ and $V_{\epsilon}$ is $\Gamma_{+}$,
with its outward normal relative to $V_{+}$ given by $\mathbf{n}_{+}=\mathbf{n}^{\Gamma}$,
and that between $V_{-}$ and $V_{\epsilon}$ is $\Gamma_{-}$ with
outward normal relative to $V_{-}$ given by $\mathbf{n}_{-}=-\mathbf{n}^{\Gamma}$;
the side surface of $V_{\epsilon}\left(t\right)$ is denoted by $\Gamma_{\epsilon}$,
so that $\partial V_{\epsilon}=\Gamma_{+}\cup\Gamma_{-}\cup\Gamma_{\epsilon}$
represents the entire boundary of $V_{\epsilon}\left(t\right)$.

Let $f\left(\mathbf{x},t\right)$ represent a material density function
defined in $V_{\epsilon}$; for example, $f$ may represent mass,
momentum or internal and kinetic energy density of the material in
$V_{\epsilon}$. An integral statement of $f$ over $V_{\epsilon}\left(t\right)$
may be evaluated in the limit as $\epsilon\to0$ as
\begin{equation}
\lim_{\epsilon\to0}\int_{V_{\epsilon}\left(t\right)}f\,dV=\lim_{\epsilon\to0}\int_{\Gamma}f\,\epsilon\,dA=0\,,\label{eq:volume-integral-thin}
\end{equation}
since the elemental mass in $V_{\epsilon}$ goes to zero in that limit.

Now consider an integral statement of a function $\mathbf{f}\cdot\mathbf{n}_{\epsilon}$
on $\partial V_{\epsilon}$, where $\mathbf{n}_{\epsilon}$ is the
outward normal to $\partial V_{\epsilon}$. For example, $\mathbf{f}\cdot\mathbf{n}_{\epsilon}$
may represent mass, momentum or internal and kinetic energy flux across
the surface $\partial V_{\epsilon}$, positive in the outward direction
to $\partial V_{\epsilon}$; it may also represent surface traction
or its rate of work, or heat flux across $\partial V_{\epsilon}$.
Note that $\mathbf{n}_{\epsilon}=-\mathbf{n}_{+}$ on $\Gamma_{+}$
and $\mathbf{n}_{\epsilon}=-\mathbf{n}_{-}$ on $\Gamma_{-}$. In
that case,
\begin{equation}
\int_{\partial V_{\epsilon}}\mathbf{f}\cdot\mathbf{n}_{\epsilon}\,dA=-\int_{\Gamma_{+}}\mathbf{f}_{+}\cdot\mathbf{n}_{+}\,dA-\int_{\Gamma_{-}}\mathbf{f}_{-}\cdot\mathbf{n}_{-}\,dA+\int_{\Gamma_{\epsilon}}\mathbf{f}_{\epsilon}\cdot\mathbf{n}_{\epsilon}\,dA\,,\label{eq:integral-thin-surface}
\end{equation}
where $\mathbf{f}_{+}$, $\mathbf{f}_{-}$ and $\mathbf{f}_{\epsilon}$
are the values of $\mathbf{f}$ on $\Gamma_{+}$, $\Gamma_{-}$ and
$\Gamma_{\epsilon}$, respectively. In the limit as $\epsilon\to0$
we note that the surfaces $\Gamma_{+}$ and $\Gamma_{-}$ collapse
onto $\Gamma$, so that $\mathbf{f}_{+}\cdot\mathbf{n}_{+}dA\to\mathbf{f}_{+}\cdot\mathbf{n}^{\Gamma}\,dA$
and $\mathbf{f}_{-}\cdot\mathbf{n}_{-}dA\to-\mathbf{f}_{-}\cdot\mathbf{n}^{\Gamma}\,dA$.
On the lateral surface $\Gamma_{\epsilon}$, we note that the elemental
area may be written as $dA=\epsilon\,ds$, where $ds$ is a path variable
along the contour curve $\partial\Gamma$. Therefore, as $\epsilon\to0$,
the lateral surface shrinks to zero. For quantities proportional to
mass flux (including momentum or internal and kinetic energy flux),
this integral thus reduces to zero,
\begin{equation}
\lim_{\epsilon\to0}\int_{\Gamma_{\epsilon}}\mathbf{f}_{\epsilon}\cdot\mathbf{n}_{\epsilon}\,dA=\lim_{\epsilon\to0}\int_{\partial\Gamma}\epsilon\mathbf{f}_{\epsilon}\cdot\mathbf{n}_{\epsilon}\,ds=0\,.\label{eq:area-integral-thin}
\end{equation}
For other quantities, such as surface traction $-p_{\epsilon}\mathbf{n}_{\epsilon}$,
its rate of work $-p_{\epsilon}\mathbf{v}_{\epsilon}\cdot\mathbf{n}_{\epsilon}$,
or heat flux $\mathbf{q}_{\epsilon}\cdot\mathbf{n}_{\epsilon}$ normal
to $\Gamma_{\epsilon}$, we choose to neglect those contributions
in our treatment here (which is equivalent to neglecting surface tension,
surface energy, and heat leakage tangential to $\Gamma$). We may
thus write the integral statement (\ref{eq:integral-thin-surface})
in the limit as $\epsilon\to0$ as
\begin{equation}
\lim_{\epsilon\to0}\int_{\Gamma_{\epsilon}}\mathbf{f}\cdot\mathbf{n}_{\epsilon}\,dA=-\int_{\Gamma}\left[\left[\mathbf{f}\right]\right]\cdot\mathbf{n}^{\Gamma}\,dA\,,\label{eq:integral-surface-jump}
\end{equation}
where
\begin{equation}
\left[\left[\mathbf{f}\right]\right]\equiv\mathbf{f}_{+}-\mathbf{f}_{-}\label{eq:jump-definition}
\end{equation}
represents the jump in $\mathbf{f}$ across $\Gamma$.

\subsection{Mass Balance Jump\label{sec:Mass-Balance-Jump}}

We now apply these identities to evaluate the mass balance jump across
the interface $\Gamma$. We write the integral form of the axiom of
mass balance over the thin domain $V_{\epsilon}\left(t\right)$, recognizing
that $\partial V_{\epsilon}\left(t\right)$ moves with a velocity
$\mathbf{w}_{\epsilon}$ such that the velocity of the material crossing
$\partial V_{\epsilon}\left(t\right)$ is $\left(\mathbf{v}_{\epsilon}-\mathbf{w}_{\epsilon}\right)\cdot\mathbf{n}_{\epsilon}$,
\[
\begin{aligned}\frac{d}{dt}\int_{V_{\epsilon}\left(t\right)}\rho\,dV & =-\int_{\partial V_{\epsilon}\left(t\right)}\rho\left(\mathbf{v}-\mathbf{w}\right)\cdot\mathbf{n}_{\epsilon}\,dA\\
 & =\int_{\Gamma_{+}}\rho_{+}\left(\mathbf{v}_{+}-\mathbf{w}_{+}\right)\cdot\mathbf{n}_{+}\,dA+\int_{\Gamma_{-}}\rho_{-}\left(\mathbf{v}_{-}-\mathbf{w}_{-}\right)\cdot\mathbf{n}_{-}\,dA\\
 & -\oint_{\partial\Gamma}\rho_{\epsilon}\left(\mathbf{v}_{\epsilon}-\mathbf{w}_{\epsilon}\right)\cdot\mathbf{n}_{\epsilon}\,\epsilon ds
\end{aligned}
\,,
\]
Taking the limit as $\epsilon\to0$, we find that $\mathbf{w}_{+}\to\mathbf{v}^{\Gamma}$
on $\Gamma_{+}\to\Gamma$ and $\mathbf{w}_{-}\to\mathbf{v}^{\Gamma}$
on $\Gamma_{-}\to\Gamma$, whereas the contour integral over $\partial\Gamma$
reduces to zero since the mass flux goes to zero, so that
\begin{equation}
0=\int_{\Gamma}\left[\left[\rho\left(\mathbf{v}-\mathbf{v}^{\Gamma}\right)\right]\right]\cdot\mathbf{n}^{\Gamma}\,d\Gamma\,.\label{eq:mass-jump-integral}
\end{equation}
Since the above condition must hold for arbitrary interfaces $\Gamma$,
we conclude that the jump condition derived from the mass balance
is satisfied pointwise on $\Gamma$ as given in (\ref{eq:mass-balance-jump}).

\subsection{Linear Momentum Balance Jump\label{subsec:Momentum-Balance-Jump}}

Following the same procedure, we may write the integral form of the
linear momentum balance over $V_{\epsilon}\left(t\right)$ as
\begin{equation}
\frac{d}{dt}\int_{V_{\epsilon}\left(t\right)}\rho\mathbf{v}\,dV=-\int_{\partial V_{\epsilon}\left(t\right)}\rho\mathbf{v}\otimes\left(\mathbf{v}-\mathbf{w}\right)\cdot\mathbf{n}_{\epsilon}\,dA-\int_{\partial V_{\epsilon}\left(t\right)}p\mathbf{n}_{\epsilon}\,dA+\int_{V_{\epsilon}\left(t\right)}\rho\mathbf{b}\,dV\,.\label{eq:momentum-jump-formulation}
\end{equation}
Taking the limit as $\epsilon\to0$ produces
\begin{equation}
0=\int_{\Gamma}\left[\left[\rho\mathbf{v}\otimes\left(\mathbf{v}-\mathbf{v}^{\Gamma}\right)\right]\right]\cdot\mathbf{n}^{\Gamma}\,dA+\int_{\Gamma}\left[\left[p\right]\right]\mathbf{n}^{\Gamma}\,dA+0\,.\label{eq:momentum-jump-integral}
\end{equation}
Since this integral statement must remain valid for arbitrary $\Gamma$,
we may write
\begin{equation}
\left[\left[p\mathbf{I}+\rho\mathbf{v}\otimes\mathbf{u}^{\Gamma}\right]\right]\cdot\mathbf{n}^{\Gamma}=\mathbf{0}\,.\label{eq:linear-momentum-jump}
\end{equation}
If we take the dyadic product of the mass balance jump in (\ref{eq:mass-balance-jump})
with $\mathbf{v}^{\Gamma}$ and subtract it from the momentum jump
above, we may rewrite it as shown in (\ref{eq:momentum-jump-redux}).

\subsection{Energy Balance Jump\label{subsec:Energy-Balance-Jump}}

The integral statement of the axiom of energy balance over $V_{\epsilon}\left(t\right)$
is
\[
\begin{aligned}\frac{d}{dt}\int_{V_{\epsilon}\left(t\right)}\rho\left(u+\frac{1}{2}\mathbf{v}\cdot\mathbf{v}\right)\,dV & =-\int_{\partial V_{\epsilon}\left(t\right)}\rho\left(u+\frac{1}{2}\mathbf{v}\cdot\mathbf{v}\right)\left(\mathbf{v}-\mathbf{w}\right)\cdot\mathbf{n}_{\epsilon}\,dV\\
 & -\int_{\partial V_{\epsilon}\left(t\right)}p\mathbf{v}\cdot\mathbf{n}_{\epsilon}\,dA\\
 & +\int_{V_{\epsilon}\left(t\right)}\rho\mathbf{b}\cdot\mathbf{v}\,dV-\int_{\partial V_{\epsilon}\left(t\right)}\mathbf{q}\cdot\mathbf{n}_{\epsilon}\,dA+\int_{V_{\epsilon}\left(t\right)}\rho r\,dV
\end{aligned}
\,.
\]
In the limit as $\epsilon\to0$ this expression reduces to
\[
\begin{aligned}0 & =\int_{\Gamma}\left[\left[\rho\left(u+\frac{1}{2}\mathbf{v}\cdot\mathbf{v}\right)\left(\mathbf{v}-\mathbf{v}^{\Gamma}\right)\right]\right]\cdot\mathbf{n}^{\Gamma}\,dA\\
 & +\int_{\Gamma}\left[\left[p\mathbf{v}\right]\right]\cdot\mathbf{n}^{\Gamma}\,dA\\
 & +0+\int_{\Gamma}\left(\left[\left[\mathbf{q}\right]\right]\cdot\mathbf{n}^{\Gamma}\right)\,dA+0
\end{aligned}
\,.
\]
This energy jump may now be rewritten as
\begin{equation}
\left[\left[\rho\left(u+\frac{1}{2}\mathbf{v}\cdot\mathbf{v}\right)\mathbf{u}^{\Gamma}+p\mathbf{v}+\mathbf{q}\right]\right]\cdot\mathbf{n}^{\Gamma}=0\,.\label{eq:energy-jump}
\end{equation}
Multiply the mass balance jump in (\ref{eq:mass-balance-jump}) by
$\frac{1}{2}\mathbf{v}^{\Gamma}\cdot\mathbf{v}^{\Gamma}$ and subtract
it from this expression; also subtract the dot product of the momentum
balance jump in (\ref{eq:momentum-jump-redux}) with $\mathbf{v}^{\Gamma}$
to produce the relation of (\ref{eq:energy-jump-redux}).

\subsection{Entropy Inequality Jump\label{subsec:Entropy-Inequality-Jump}}

The integral statement of the axiom of entropy inequality over $V_{\epsilon}\left(t\right)$
is
\[
\frac{d}{dt}\int_{V_{\epsilon}\left(t\right)}\rho s\,dV+\int_{\partial V_{\epsilon}\left(t\right)}\rho s\left(\mathbf{v}-\mathbf{w}\right)\cdot\mathbf{n}_{\epsilon}\,dA+\int_{\partial V_{\epsilon}\left(t\right)}\frac{1}{T}\mathbf{q}\cdot\mathbf{n}_{\epsilon}\,dA-\int_{V_{\epsilon}\left(t\right)}\rho\frac{r}{T}\,dV\ge0\,.
\]
In the limit as $\epsilon\to0$ this expression reduces to
\[
\begin{aligned} & 0-\int_{\Gamma}\left[\left[\rho s\mathbf{u}^{\Gamma}\right]\right]\cdot\mathbf{n}^{\Gamma}\,dA\\
 & -\int_{\Gamma}\left[\left[\frac{1}{T}\mathbf{q}\right]\right]\cdot\mathbf{n}^{\Gamma}\,dA-0\ge0
\end{aligned}
\,.
\]
This relation is valid for arbitrary $\Gamma$, thus it reduces to
the expression in (\ref{eq:entropy-jump}).

\bibliographystyle{amsplain}

\providecommand{\bysame}{\leavevmode\hbox to3em{\hrulefill}\thinspace}
\providecommand{\MR}{\relax\ifhmode\unskip\space\fi MR }
\providecommand{\MRhref}[2]{%
  \href{http://www.ams.org/mathscinet-getitem?mr=#1}{#2}
}
\providecommand{\href}[2]{#2}

\end{document}